\documentclass{jfm}

\usepackage{multirow}
\usepackage{booktabs}
\usepackage{graphicx}
\usepackage{xcolor}
\usepackage{caption}
\usepackage{subcaption}
\usepackage{newtxtext}
\usepackage{newtxmath}
\usepackage{natbib}
\usepackage{textcomp}
\usepackage{manyfoot}
\usepackage{algorithm}
\usepackage{algorithmicx}
\usepackage{algpseudocode}
\usepackage{listings}
\usepackage[percent]{overpic}
\usepackage[normalem]{ulem}

\definecolor{green}{RGB}{0, 128, 0}
\newcommand{\red}[1]{\textcolor{black}{#1}}
\newcommand{\blue}[1]{\textcolor{black}{#1}}
\newcommand{\green}[1]{\textcolor{black}{#1}}
\newcommand{\ac}[1]{}
\newcommand{\pp}{^{\scriptscriptstyle\prime\mkern-1.5mu\prime}}

\title{Mechanisms of transonic buffet over supercritical airfoils revealed by direct numerical simulations} 

\author{Giulio Soldati\aff{1}
  \corresp{\email{giulio.soldati@uniroma1.it}},
  Hang Song\aff{2}, 
  Sergio Pirozzoli\aff{1}
  and Sanjiva K. Lele\aff{2}}

\affiliation{
\aff{1} 
Dipartimento di Ingegneria Meccanica e Aerospaziale, 
Sapienza Universit\`a di Roma,
Rome 00184, Italy
\aff{2} 
Center for Turbulence Research, 
Stanford University, 
Stanford, CA, USA
}

\begin{document}
\maketitle
%\linenumbers

\begin{abstract}
\green{
We use direct numerical simulations to study the 
turbulent transonic flow over a supercritical airfoil, 
spanning the transition from stable conditions to fully developed buffet 
at chord-based Reynolds numbers of $3\times10^5$ and $6\times10^5$. 
The Reynolds number determines the incoming boundary-layer state, thereby affecting the separation behaviour.
Spectral proper orthogonal decomposition reveals a single dominant
mode that emerges below onset and persists across the transition 
to buffet. 
The energy of this mode increases with the angle of attack 
while its spatial structure and fundamental frequency 
remain nearly unchanged, consistent with a global instability 
that saturates into a nonlinear limit cycle. Geometrical acoustics 
provides a first-order description 
of the global mode propagation 
and identifies two paths connecting the acoustic source, 
located in the near wake, to the shock. 
A shorter path impinges on the shock from behind, while 
a longer one circumvents the shock tip before reaching it from the front. 
Linearised shock--acoustic interaction theory predicts a stronger 
shock response to low-frequency, front-impinging acoustic disturbances. 
Space--time correlations link the establishment of a downstream 
path to complete boundary-layer separation, which is induced 
intermittently by the upstream motion of the shock.
The sum of the upstream acoustic time along the front-impinging path 
and the downstream convective time is compatible with the measured 
buffet period at both Reynolds numbers. 
Together, these results are consistent with a unified picture of 
buffet, where an aeroacoustic feedback mechanism provides 
the physical pathways through which perturbations 
can sustain the global dynamics.
}
\end{abstract}


%==============================================
\section{Introduction}\label{sec:intro}
%==============================================

In the typical cruise regime of modern transport aircraft, supercritical
airfoils generate a local supersonic region terminated by a shock wave.
At a fixed free-stream Mach number ($M$), increasing the angle of attack
($\alpha$) strengthens the shock and may provoke shock-induced
boundary-layer separation.
\red{
Beyond a critical incidence, the flow transitions to a buffet state 
characterised by low-frequency shock excursions associated with a periodic 
alternation between large-scale boundary-layer separation and reattachment.
}
These shock oscillations are self-sustained and persist in the absence of
external forcing.
The intense wall-pressure fluctuations associated with the shock motion
induce large variations in aerodynamic loads on time scales significantly
longer than those typical of wall-bounded turbulence.
This low-frequency forcing poses a severe operational hazard, as it can
excite structural eigenmodes, leading to aeroelastic vibrations known as
\emph{buffeting}.
Such coupling degrades aerodynamic performance, accelerates fatigue
failure, and compromises passenger comfort.
Consequently, conservative safety margins are imposed that restrict the
usable flight envelope \citep{sansica2022system}.

Shock buffet instability occurs for specific combinations of Mach number
and angle of attack, defining a bounded region in the $(M,\alpha)$
parameter space \citep{lee2001self}.
\citet{stanewsky1990experimental} originally defined the buffet boundary
as the locus in the $(M,\alpha)$ plane beyond which the flow downstream of
the shock is fully separated.
Several mechanisms have been proposed to explain the onset of shock
unsteadiness, including: bursting of the separation bubble at the
shock foot \citep{pearcey1958method}; merging of the shock-induced
separation bubble with trailing-edge separation, leading to a fully
separated boundary layer \citep{pearcey1962simple}; and attainment
of a critical Mach number immediately upstream of the shock
\citep{mabey1981oscillatory}.
Subsequent experimental investigations have consistently identified flow
separation as a necessary condition for buffet development
\citep{lee1990oscillatory, jacquin2009experimental, d2021experimental}.
Depending on the airfoil geometry and operating conditions, however,
separation may first appear at the shock foot or near the trailing edge
\citep{lee1990transonic}.

Unsteady Reynolds-averaged Navier--Stokes (URANS) simulations by
\citet{raghunathan1998transonic} underlined that the shock must be
sufficiently strong to generate a separation bubble for buffet to grow.
In this picture, the periodic shock motion is sustained by the
alternating expansion and collapse of the separation bubble, which in
turn alters the effective camber of the airfoil.
Nevertheless, several numerical studies did not observe a clear
connection between buffet onset and flow separation
\citep{crouch2009origin, iovnovich2012reynolds, sartor2015stability}.
A possible explanation is that these studies relied on
Reynolds-averaged Navier--Stokes (RANS) closures with turbulence models, 
which often struggle to predict separated boundary layers accurately.
Indeed, significant sensitivity to numerical schemes and turbulence-model
choices has been reported, leading to large discrepancies in the
prediction of the shock-induced separated region
\citep{goncalves2004turbulence}.
On the other hand, URANS simulations capture the global dynamics of shock
buffet reasonably well, suggesting that they may also capture its
dominant physical mechanisms.

\green{
Using steady RANS simulations, \citet{nitzsche2019fluid} identified 
three distinct flow regimes within the $(M,\alpha)$ parameter space: 
1) shock-free subsonic flow, 2) {\em regular} shock motion, where the shock moves 
downstream with increasing $\alpha$, and 3) {\em inverse} shock 
motion, where the shock moves upstream with increasing $\alpha$. 
This inverse motion indicates a `pre-buffet' flow state 
\citep{nitzsche2009numerical}, characterised by shock-induced 
flow separation. 
Subsequently, \citet{accorinti2022experimental} and \citet{schauerte2023experimental}
confirmed experimentally that this reversal in shock trajectory 
is a necessary (but not sufficient) condition for buffet onset, 
which ultimately occurs at a slightly higher angle of attack.
}

As $\alpha$ increases, the flow accelerates on the suction side and the
upstream pressure is reduced, which intensifies the shock.
Shock strength may be quantified by the downstream-to-upstream pressure
ratio and, in a first stage, this intensification tends to shift the
shock downstream.
At a critical $\alpha$, trailing-edge separation forms and the shock
motion is concurrently reversed.
One possible explanation is that trailing-edge separation raises the
downstream pressure and increases the shock strength, which forces the
shock towards the leading edge in order to raise the upstream relative
Mach number and re-establish the Rankine--Hugoniot relations.
As this upstream shift progresses, the steady solution eventually
becomes unstable and the shock begins to oscillate.
However, the mechanism by which the steady shock configuration evolves
into a self-sustained unsteady oscillation at buffet onset remains an
open question.

A physical explanation for the self-sustained nature of shock buffet was
first proposed by \citet{lee1990oscillatory}, who cast the phenomenon
as an acoustic feedback loop. 
Unsteady shock motion generates pressure disturbances that 
amplify while convecting downstream through the separated shear layer.
Upon reaching the trailing edge, these amplified disturbances excite
upstream-travelling pressure waves that propagate through the subsonic
region above the separated flow and interact with the shock, exchanging
energy and closing the loop.
In this picture, the spatial growth of disturbances in the separated
shear layer, together with the upstream and downstream propagation
speeds, determine the buffet frequency.

Subsequent computational and experimental investigations have provided
mixed quantitative support for Lee's model.
Good agreement between modelled and measured reduced frequencies was
reported for the BGK No.\,1 airfoil in several studies, and comparable
buffet frequencies were obtained using URANS \citep{xiao2006numerical}
and zonal DES \citep{deck2005numerical} simulations.
By contrast, poor frequency agreement has also been reported in the 
literature \citep{garnier2010large, jacquin2009experimental}. 
To address this, \citet{jacquin2009experimental} proposed an extension of the 
feedback picture in which
upstream waves also travel along the lower surface and around the
leading edge, thereby impinging on the shock from both sides.
\citet{hartmann2013interaction} argued that the
characteristic length traversed by upstream-travelling disturbances is
the distance between the trailing edge and the shock tip, where the
shock is weaker.
Substituting this length into the original model yields good agreement
between modelled and measured Strouhal numbers.
A more robust estimation of the upstream travel time was obtained by
\citet{memmolo2018scrutiny}, who employed geometrical acoustics to trace
the path of upstream-travelling acoustic rays and showed that these rays
can circumvent the shock tip and impact the shock from the upstream
region. In their analysis, the predicted buffet frequency was virtually
coincident with the computed one.

\red{
On the other hand, \citet{paladini2019various} used URANS simulations 
to demonstrate that buffet can be suppressed by locally applying selective 
frequency damping in specific flow regions. These areas are the shock wave 
and the separated zone, which therefore play an active role 
in sustaining the instability. In contrast, the subsonic region above 
the boundary layer, often invoked in acoustic feedback models, 
was found to play a marginal role. 
In support of this finding, \citet{moise2022large} showed  
via large-eddy simulations (LES) that the oscillation period 
predicted using Lee's model is substantially shorter 
than the buffet period obtained from LES. 
Subsequent research by \citet{moise2024connecting} revealed that 
transonic buffet is associated with incompressible 
low-frequency oscillations, indicating that shock waves are 
not essential for buffet and that flow separation is a more 
significant factor in the underlying mechanism.
}

An alternative perspective on the mechanism of transonic buffet was
proposed by \citet{crouch2009origin}. Using global linear stability analysis, 
they associated buffet onset with the emergence of an unstable aerodynamic mode.
In this view, shock oscillations appear once the angle of attack exceeds
a critical threshold, in analogy with the onset of vortex shedding in a
cylinder wake when the Reynolds number crosses a critical value
\citep{jackson1987finite}. When the unstable mode is triggered, pressure 
perturbations originating at the shock foot propagate vertically along the 
shock and through the boundary layer, rather than within the separated shear 
layer as in Lee's model. Upon reaching the shock tip, these perturbations
propagate forward and dissipate into the incoming flow.
Recently, \citet{crouch2024weakly} extended this 
perspective by characterising the instability as a Hopf bifurcation, 
employing the Landau equation to model the transition from a steady 
base flow to limit-cycle oscillations. 

The interpretation of buffet as a global flow instability has since 
been widely employed to compute buffet boundaries. 
\blue{
\citet{sartor2015stability} 
used resolvent analysis to show that the shock--boundary-layer 
interaction (SBLI) acts as a low-pass filter, while the mixing 
layers above the separated region amplify locally 
medium-frequency disturbances.
Through eigenvalue analysis, the authors found that the 
buffet frequency depends primarily on the Mach number rather 
than on the angle of attack. This suggests that the size of the 
separation region, which increases with $\alpha$, is not the 
dominant parameter governing frequency selection, and indicates 
a stronger connection with the acoustic timescale than with a 
purely convective timescale. Furthermore, 
their adjoint analysis revealed that flow receptivity 
is concentrated along a specific characteristic line in the 
supersonic region, which maps how upstream perturbations 
travel down to impact the shock foot from the front side.  
}
 
\blue{
The shock foot was later identified by \citet{kojima2020resolvent} 
as the origin of transonic buffet. Using resolvent analysis, 
they showed that perturbations at the shock foot drive the 
instability amplification, producing low-frequency shock 
oscillations even at a chord-based Reynolds number of 2000. 
\citet{iwatani2023identifying} performed resolvent analysis using wall-resolved 
LES data to isolate two distinct self-sustaining feedback loops from the forcing 
and response modes. The first mechanism involves the trailing-edge separation, 
where periodic variations in the separation height alter the post-shock pressure 
and modulate the shock-jump condition. The second loop constitutes a feedback model 
where pressure disturbances are generated downstream of the shock foot, travel 
around the sonic line, and propagate along the characteristic line to impinge on the 
shock foot from the upstream supersonic region. The spatial structure of this upstream 
forcing mode matches the oblique adjoint mode of \citet{sartor2015stability}, 
suggesting that the characteristic line is the path through which the information 
reaches the shock foot and sustains the oscillations. 
}

\blue{
The transonic buffet phenomenon on three-dimensional (3D) swept wings has
also received attention in recent literature. \citet{molton2013control} showed
that, compared to two-dimensional (2D) airfoils, the amplitude of shock oscillations
on wings is an order of magnitude smaller and the dominant frequency is four to seven
times higher. Subsequent numerical \citep{iovnovich2015numerical} and experimental
\citep{dandois2016experimental} investigations revealed the emergence of 
3D flow structures, referred to as \textit{buffet cells}, 
which travel along the span of the swept wing.
Global stability analyses have identified unstable modes associated with
large-scale unsteady flow phenomena on both infinite \citep{crouch2019global,
paladini2019transonic, he2021triglobal} and finite-span \citep{sansica2023global,
timme2020global, plante2021link} swept wings. These studies identify the coexistence
of a travelling-wave mode alongside the 2D oscillatory mode typical of airfoil buffet.
This travelling-wave mode convects outboard along the wing, with a dominant spanwise
wavelength approximately equal to the chord.
}

\red{
Additional instabilities associated with transonic buffet on swept 
wings are decoupled from the two-dimensional buffet mode and thus require 
separate consideration when targeting more realistic flight conditions. 
The airfoil configuration is a fundamental problem in its own right 
and serves as the baseline against which swept-wing cases can be compared. 
Despite decades of intensive research, characterising this 2D instability 
remains challenging, and the underlying mechanisms are still the subject 
of ongoing debate.
}

\blue{
The present investigation employs the V2C supercritical airfoil developed 
by Dassault Aviation. This profile has been extensively documented in the literature 
through wind tunnel experiments \citep{davidson2016influence, placek2016wind, 
placek2018flow} and numerical simulations \citep{szubert2016numerical, memmolo2018scrutiny, 
zauner2019direct} within the European TFAST project. Using LES and RANS methods, 
\citet{moise2022large, moise2023transonic, moise2024connecting} carried out an 
extensive parametric study for the V2C airfoil across a wide range of angles of 
attack, Mach numbers and Reynolds numbers. 
These recent studies have focused primarily on the laminar buffet, showing that 
its principal distinction from turbulent buffet is the emergence of a multiple-shock 
structure on the suction surface. Conversely, tripping the boundary layer or 
increasing the Reynolds number restores a single, well-defined shock. In addition, 
laminar flow configurations feature intense vortex shedding at the trailing edge, 
which is attenuated in the fully turbulent regime. 
} 

Against this background, we present the first fully-resolved direct numerical 
simulation (DNS) database of turbulent transonic buffet over an airfoil. 
We gradually increase the angle of attack from stable conditions into the 
buffet regime to characterise the flow reorganisation as the critical 
threshold is crossed. 
\green{
Additionally, we assess the impact of the Reynolds 
number on this process by conducting two distinct DNS sets, 
one at $\Rey=3\times10^5$ and another at $6\times10^5$. 
These values of Reynolds numbers are selected 
to span distinct incoming boundary-layer states. 
The lower value provides a limiting Reynolds number 
at which sustained turbulence can be maintained, 
resulting in an incoming boundary layer that lacks a 
well-defined logarithmic region, while 
the higher value yields a fully developed turbulent state. 
}
The resulting high-fidelity dataset is used to investigate the mechanisms 
underlying transonic buffet. In addition, it can serve as a benchmark for
lower-order modelling approaches, such as RANS closures, which
remain among the most widely used tools in both industrial and academic
research on transonic turbulent buffet.

%==============================================
\section{Methodology}\label{sec:method}
%==============================================

%==============================================
\subsection{Numerical setup}
\label{sec:method-setup}
%==============================================
The DNS campaign was performed using the in-house solver STREAmS-2.0
\citep{salvadore2025streams} on the EuroHPC LUMI supercomputer.
The solver integrates the curvilinear formulation of the legacy code
FLEW \citep{soldati2024flew}, extending GPU architecture support to AMD
accelerators.
The compressible Navier--Stokes equations in generalised curvilinear
coordinates are solved for an ideal gas with heat capacity ratio
$\gamma = 1.4$ and Prandtl number $\Pra = 0.72$.
Spatial discretisation is based on high-order finite-difference schemes.
Convective terms are approximated using sixth-order central schemes
expanded in skew-symmetric form, ensuring numerical stability and
discrete preservation of kinetic energy in the inviscid limit
\citep{pirozzoli2011stabilized}. 

\blue{
A fifth-order weighted essentially 
non-oscillatory (WENO) scheme is activated adaptively when the shock 
sensor $\Theta$ exceeds a threshold value, 
}
\begin{equation}
\Theta=\mathrm{max}\left\{-\nabla \cdot \mathbf{u}\left[{(\nabla \cdot \mathbf{u})^2+(\nabla \times \mathbf{u})^2
+\left(u_0 / c\right)^2}\right]^{-1/2}, \, 0\right\} > \Theta_0, 
\end{equation}
\blue{
where $\mathbf{u}$ is the velocity vector. The threshold is set to 
$\Theta_0 = 0.25$ in the present DNS, which is selected to limit the 
activation of WENO schemes only near the shock wave.
}
The dynamic viscosity $\mu$ follows a power-law temperature dependence,
$\mu = \mu_0 (T/T_0)^{0.76}$, where the subscript $0$ denotes freestream
quantities. Further details on the solver and its validation can be found in 
the aforementioned references. In addition, the present DNS setup is further 
validated against reference LES data for the V2C airfoil in appendix \ref{app:val}.

The flow is characterised by three parameters: the angle of attack
($\alpha$), the freestream Mach number ($M$), and the chord-based Reynolds
number ($\Rey = c u_0 / \nu_0$), where $c$ is the chord length, $u_0$ the
freestream velocity, and $\nu_0$ the kinematic viscosity.
In all cases, $M$ is fixed at $0.7$, representative of typical cruise
conditions for commercial transport aircraft. The angle of attack spans 
from stable conditions ($\alpha = 4^\circ,\,5^\circ$) to buffet 
conditions ($\alpha = 6^\circ,\,7^\circ$). For each value of $\alpha$, 
two Reynolds numbers are considered, $\Rey = 3 \times 10^{5}$ and $6 \times 10^{5}$.
The complete set of flow parameters for each simulation is reported in
table~\ref{tab:1}.

\begin{table}
\begin{center}
\renewcommand{\arraystretch}{1.0}
\begin{tabular}{cccccccc}
\toprule
$\alpha$ & $M$ & $\Rey$ & $N_s \times N_n \times N_z$ &
$\overline{x}_s / c$ & $\Delta x_{s,\mathrm{max}}/c$ & \green{$\sigma_s/c$} & $\St_B$ \\
\midrule
$4^\circ$ & $0.7$ & $3\times 10^5$ &
 $5536 \times 491 \times 296$ & $0.460$ & $0.031$ & \green{$0.005$} & $0.09$ \\
$5^\circ$ & $0.7$ & $3\times 10^5$ &
 $5536 \times 491 \times 296$ & $0.486$ & $0.016$ & \green{$0.003$} & $0.09$ \\
$6^\circ$ & $0.7$ & $3\times 10^5$ &
 $5536 \times 491 \times 296$ & $0.483$ & $0.042$ & \green{$0.009$} & $0.09$ \\
$7^\circ$ & $0.7$ & $3\times 10^5$ &
 $5536 \times 491 \times 296$ & $0.436$ & $0.179$ & \green{$0.044$} & $0.09$ \\
$4^\circ$ & $0.7$ & $6\times 10^5$ &
$10496 \times 652 \times 556$ & $0.469$ & $0.026$ & \green{$0.004$} & $0.12$ \\
$5^\circ$ & $0.7$ & $6\times 10^5$ &
$10496 \times 652 \times 556$ & $0.473$ & $0.018$ & \green{$0.003$} & $0.08$ \\
$6^\circ$ & $0.7$ & $6\times 10^5$ &
$10496 \times 652 \times 556$ & $0.450$ & $0.050$ & \green{$0.010$} & $0.09$ \\
$7^\circ$ & $0.7$ & $6\times 10^5$ &
$10496 \times 652 \times 556$ & $0.403$ & $0.177$ & \green{$0.049$} & $0.11$ \\
\bottomrule
\end{tabular}
\renewcommand{\arraystretch}{1.}
\caption{Flow parameters: angle of attack, freestream Mach number,
chord-based Reynolds number, number of grid points in the
$s$, $n$, and $z$ directions, mean shock position, 
maximum shock excursion, standard deviation of shock excursion, 
buffet Strouhal number.}
\label{tab:1}
\end{center}
\end{table}

\begin{figure}
\centering
\includegraphics[width=\textwidth]{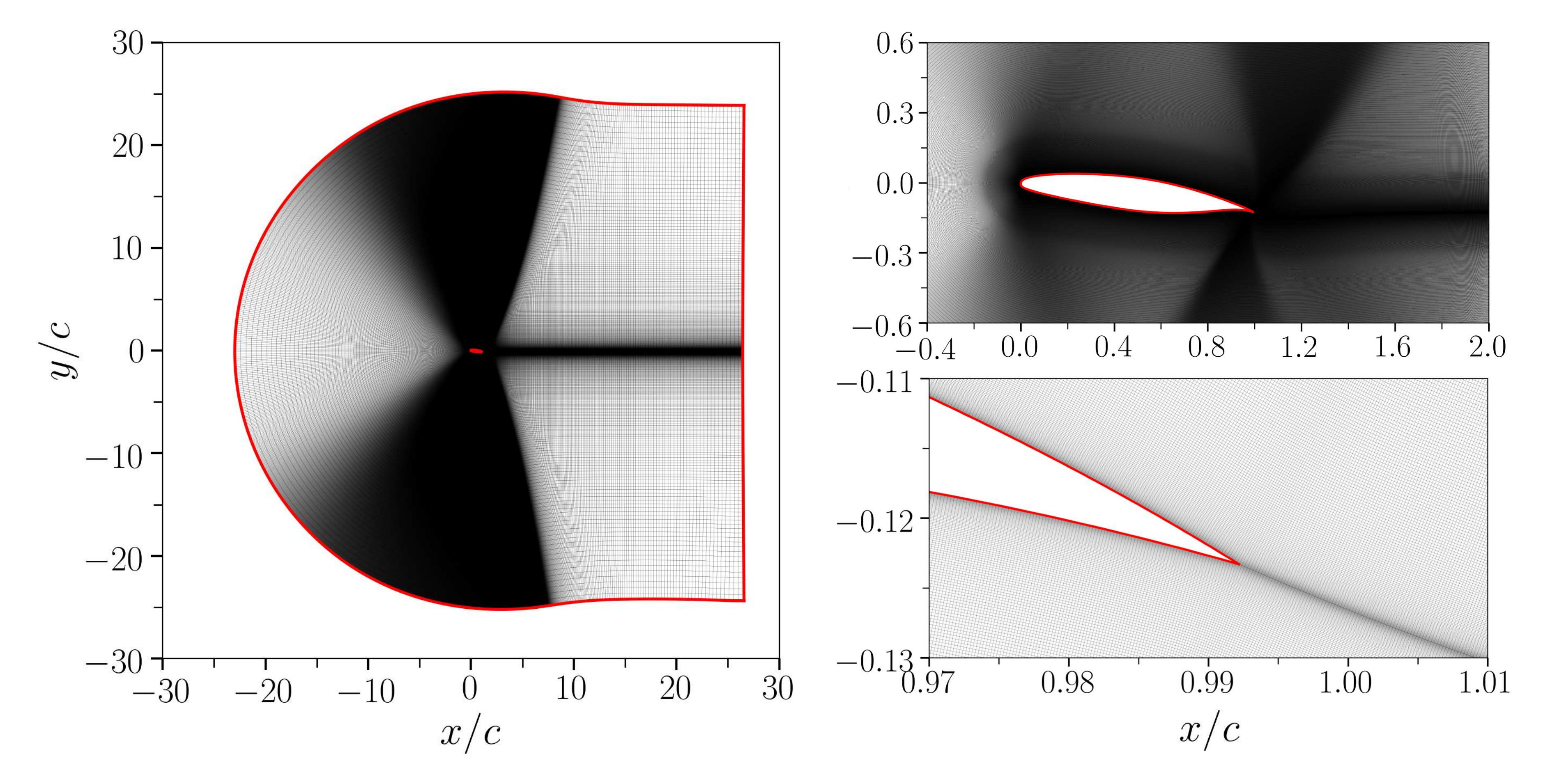}
\caption{Computational grid for the V2C supercritical airfoil at $M=0.7$,
$\Rey = 6 \times 10^{5}$, $\alpha = 7^\circ$.
Left: global view of the computational domain in the $(x,y)$ plane.
Top right: close-up of the near field.
Bottom right: further zoom on the trailing-edge region.}
\label{fig:airgrid}
\end{figure}

The numerical setup follows the procedures adopted in previous
studies by our group \citep{memmolo2018scrutiny, quadrio2022drag,
berizzi2025aerodynamic}.
The computational domain, sketched in the left panel of
Figure \ref{fig:airgrid}, extends radially to approximately $26\,c$, where
$c$ is the chord length.
In the following, both Cartesian ($x,y,z$) and curvilinear
($s,n,z$) coordinate systems are employed.
The origin of the Cartesian system is located at the leading edge; the
$x$-axis is aligned with the chord and directed downstream, the $y$-axis
is normal to the chord and oriented upward, and the $z$-axis is oriented
in the spanwise direction. The curvilinear coordinates $s$ and $n$ 
denote the directions tangent and normal to the airfoil wall.
Cartesian velocity components are denoted by $(u_x, u_y, u_z)$, 
whereas $(u, v, w)$ denotes the velocity components along curvilinear directions.

\red{
The sensitivity of the results to the spanwise 
domain extent was studied systematically via implicit large-eddy simulations 
(ILES) by \citet{zauner2020wide} for the laminar buffet regime, and 
by \citet{lusher2024effect} for turbulent flow over the NASA CRM airfoil. 
Covering a range of incidences at $M = 0.72$ and $\Rey = 5 \times 10^5$, 
\citet{lusher2024effect} assessed domain widths spanning $L_z = 0.025c$ to 
$0.5c$ and established that $L_z = 0.1c$ ensures flow statistics independent 
of $L_z$. This conclusion was recently confirmed by \citet{lusher2025implicit}, 
who extended the independence study to much wider computational domains. Their analysis 
showed an excellent collapse of the chordwise distributions of pressure and friction 
coefficients between the $L_z = 0.1c$ and $L_z = 3c$ configurations. Furthermore, 
the discrepancy in the mean lift coefficient was only $0.1\%$, and the characteristic 
buffet frequencies remained virtually identical. 
Therefore, the spanwise extent of the domain is set to $L_z = 0.1c$ for the present DNS. 
The adequacy of this choice is assessed via two-point correlations in appendix \ref{app:span}. 
}

%==============================================
\subsection{Computational grid}
\label{sec:method-grid}
%==============================================
\green{
A body-fitted curvilinear mesh is employed to conform to the airfoil geometry. 
The mesh features a C-type topology, wherein the computational domain is bounded 
by the airfoil surface, an outer curvilinear perimeter (denoted as the C-boundary), 
and a vertical downstream outlet. A branch cut, referred to as the wake cut, 
originates at the trailing edge and extends to the downstream outlet, partitioning the 
wake domain. These geometric features are illustrated in figure \ref{fig:airgrid}. 
Within this framework, the grid lines are nearly aligned with the curvilinear coordinates 
$s$ and $n$. The $s$-coordinate lines originate at the downstream outlet, propagate along 
one side of the wake cut, wrap clockwise around the airfoil surface, and return along the 
opposite side of the wake cut to terminate at the outlet. The $n$-coordinate 
lines extend in the normal direction, originating from either the airfoil surface or the wake 
cut and terminating at the outer C-boundary. This arrangement avoids the trailing-edge 
singularity characteristic of O-grids while maintaining excellent resolution across the wake. 
}

The mesh in the $(x,y)$ plane is generated using an in-house hyperbolic grid generator based 
on the open-source Construct2D library \citep{construct2d}. The algorithm is customised to minimise 
cell non-orthogonality, which reaches a maximum value of approximately $4^\circ$ in the trailing-edge 
region. The wake cut follows an exponentially decaying profile, ensuring tangential continuity at 
the trailing edge and a smooth relaxation of the wake boundary from the trailing-edge bisector to 
the horizontal direction. Each simulation employs a case-specific grid obtained by rotating the 
airfoil by $\alpha$ degrees, rather than the incoming velocity vector. Consequently, the region 
of maximum grid refinement clustered along the wake cut remains aligned with the physical 
wake behind the airfoil, thereby optimising spatial resolution in the wake region.

A key challenge in DNS of inhomogeneous flows is resolving the local
Kolmogorov length scale $\eta$, which varies substantially
across the computational domain.
To address this issue, a two-step strategy is adopted.
First, the near-wall grid distribution is designed under the assumption
of local boundary-layer equilibrium.
Under this assumption, the streamwise and spanwise spacings are selected
according to standard wall-unit constraints, namely $\Delta s^+<10$ and
$\Delta z^+<4.5$.
Wall-normal clustering is prescribed using the stretching function of
\citet{pirozzoli2021natural}, which guarantees $n^+<1$ at the wall and
provides adequate resolution of both the buffer and logarithmic layers.
Second, a precursor RANS computation at identical flow conditions is used
to estimate the local turbulent kinetic energy production, from which
the local dissipation rate $\varepsilon$ is inferred assuming a local
production–dissipation balance. The Kolmogorov scale is then obtained as
$\smash{\eta = \left({\nu^3}/{\varepsilon}\right)^{1/4}}$, 
allowing explicit sizing of the computational mesh.
The two approaches yield similar target resolutions over the airfoil
surface.
However, only the RANS-based estimate enables a practical prescription
of the wake resolution, which is enforced by requiring
$\Delta s < 6\,\eta$ \citep{pope2001turbulent} for at least one
chord length downstream of the trailing edge, and
$\Delta n < 6\,\eta$ over a distance of $0.2\,c$ in the
$n$ direction.
The adequacy of the selected grid spacings is confirmed
\emph{a posteriori}.

The computational meshes comprise $5536\times491\times296$ grid points
for the low-Reynolds-number cases ($\Rey=3\times10^{5}$) and
$10496\times652\times556$ points for the high-Reynolds-number cases
($\Rey=6\times10^{5}$).
The mesh for the case at $\Rey=6\times10^{5}$ and $\alpha=7^\circ$, shown in
figure \ref{fig:airgrid}, is representative of the grid distribution.
Along the $s$ direction, $5598$ points lie on the airfoil surface,
$2085$ points are allocated to the near wake ($1 < s/c \le 2$), and $364$
points cover the far wake ($2 < s/c \le 26$), where the grid spacing
increases according to a geometric progression.
Along the $n$ direction, $153$ points are located within the
boundary-layer thickness ($n/c < 0.025$), $233$ points populate the
intermediate region ($0.025 \le n/c < 0.2$) to resolve wake structures, and
the remaining $266$ grid planes extend towards the far-field boundary.
The most refined mesh regions are visible as the dark patches in
figure \ref{fig:airgrid}.

%==============================================
\subsection{Boundary conditions}
\label{sec:method-bc}
%==============================================
Spanwise periodicity is prescribed for the flow field. At the downstream 
rectilinear boundary, a subsonic extrapolation with a specified reference 
pressure is enforced alongside non-reflecting characteristic-based boundary 
conditions. At the upstream curvilinear boundary, characteristic-based 
conditions are coupled with a narrow sponge layer to smoothly relax the 
flow variables toward a reference far-field state. This far-field target 
is obtained from case-specific precursor RANS simulations. 
\blue{
The detailed 
formulation of the non-reflecting characteristic conditions employed here 
can be found in \citet{pirozzoli2013generalized}. 
}

The C-mesh entails a branch cut extending from the trailing edge
to the downstream boundary, separating the upper and lower wake regions
in the computational domain. While these boundaries are computationally 
disjoint, they represent a physically continuous fluid region.
To enforce this continuity, a point-to-point data exchange procedure is
implemented.
The ghost layers (grid points at $j = 0, -1, -2$) are updated by enforcing
the solution computed at the topologically coincident interior points
($j = 2, 3, 4$).
The streamwise correspondence is handled by offsetting the index $i$ by
the mesh periodicity $N_s - 1$.
This exact reconstruction guarantees strict continuity across the wake
interface, avoiding the artificial dissipation introduced by averaging
the solution along the boundary \citep{fukushima2018wall}.

Boundary-layer transition is forced on both airfoil surfaces at
$x = 0.1\,c$ by applying a time-varying wall-normal body force that
emulates experimental roughness elements. The time and length scales 
of the forcing are defined based on the local boundary-layer displacement 
thickness, namely $\delta^* \approx 10^{-3}\,c$ and $\delta^* \approx 7 
\times 10^{-4}\,c$ for the two Reynolds numbers considered.
Following \citet{schlatter2012turbulent}, the spatial extent of the 
forcing region is defined by a Gaussian blob centred at $x = 0.1\,c$, 
with attenuation lengths of $4\,\delta^*$ and $\delta^*$ in $s$ and $n$ directions. 
The spanwise cutoff scale is $1.7\,\delta^*$, the time scale is $4\,\delta^*/u_0$, 
and the tripping amplitude is $150\,u_0^2/c$. 

%==============================================
\subsection{Time advancement}
\label{sec:method-time}
%==============================================
Time advancement is performed using a low-storage third-order
Runge--Kutta scheme, with the maximum Courant--Friedrichs--Lewy number set
to $\mathrm{CFL} = 0.9$.
The time step, which is updated dynamically, has average values of
$\Delta t \approx 1.6 \times 10^{-5}\,c/u_0$ and
$\Delta t \approx 8.5 \times 10^{-6}\,c/u_0$, depending on the Reynolds
number.

All simulations are initialised from case-specific RANS precursor
solutions, the same as those used to prescribe the
far-field boundary conditions and to inform the preliminary assessment
of the grid resolution required for the DNS.
The RANS initialisation provides a physically consistent mean flow and
boundary-layer development, thereby avoiding an impulsive start from
uniform flow conditions and the associated generation of strong spurious
acoustic disturbances.
As a result, the duration of the initial transient is significantly
reduced.

Each DNS is first advanced to $t_1 = 20\,c/u_0$, which is sufficient to
reach a statistically stationary state, and then continued to a final
time $t_2 = 200\,c/u_0$.
Throughout the paper, flow quantities averaged in the spanwise direction 
and in time are denoted by an overbar, $\overline{q}$. 
Favre averages are denoted by a tilde, 
$\tilde{q} = \overline{\rho q}/\bar{\rho}$, 
and the corresponding Favre fluctuations are defined as 
$q\pp = q-\tilde{q}$.

The adopted averaging interval, $t_2 - t_1 = 180\,c/u_0$, contains
approximately 20 buffet periods
($T_B \approx 9.1\,c/u_0$ at high $\Rey$) and ensures statistical
convergence for all the flow cases considered.
The same averaging interval is employed for frequency-domain analyses,
including spectral proper orthogonal decomposition (SPOD).

Wall-parallel and cross-stream snapshots of the flow field are stored at
regular intervals of
$\Delta t_s \approx 8 \times 10^{-3}\,c/u_0$, yielding approximately
$1125$ snapshots per buffet cycle.
Time signals are analysed in the frequency domain using the Welch method.
As a compromise between spectral resolution and spectral smoothness,
nine segments with $75\%$ overlap are employed, such that each segment
spans $60\,c/u_0$.

%==============================================
\section{Flow regimes}\label{sec:flow}
%==============================================

%==============================================
\subsection{Boundary-layer state}\label{sec:bl}
%==============================================
\blue{
Recent literature has focused on the role of the upstream 
boundary-layer state on normal shock interaction 
\citep{davidson2018influence} and particularly on shock buffet 
\citep{dandois2018large, moise2022large, moise2023transonic, zauner2023co, 
lusher2024effect}. In wind tunnel experiments, transition is 
typically forced using carborundum strips \citep{jacquin2009experimental, 
d2021experimental}, roughness elements \citep{accorinti2022experimental}, 
or aluminium zig-zag tapes \citep{schauerte2025influences}, 
which ensure that the boundary layer reaches a fully turbulent state 
within a short development length. 
}

\begin{figure}
\centering
(a)\includegraphics[width=\textwidth]{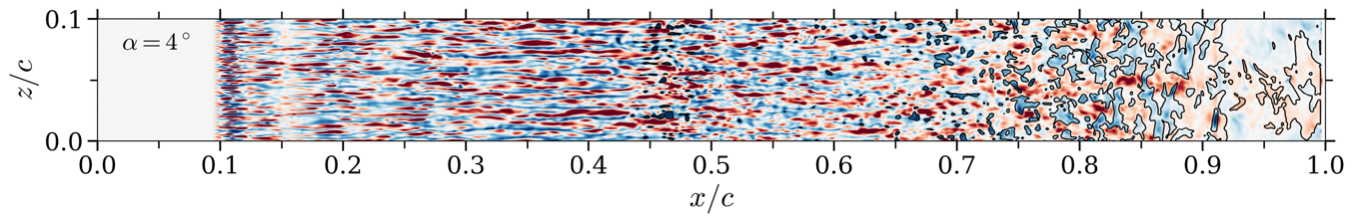}\\[-0.5cm]
(b)\includegraphics[width=\textwidth]{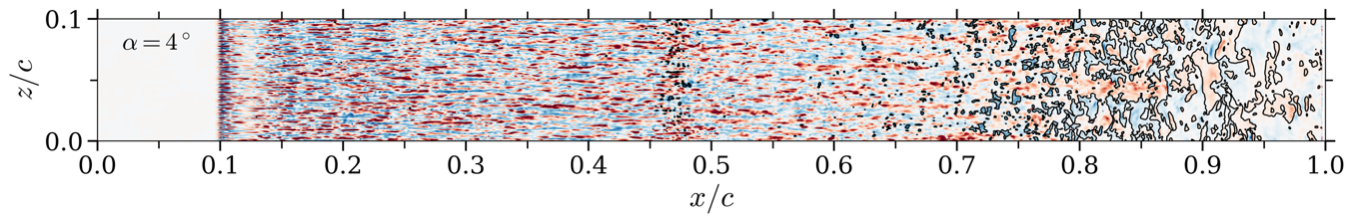}\\[-0.5cm]
(c)\includegraphics[width=\textwidth]{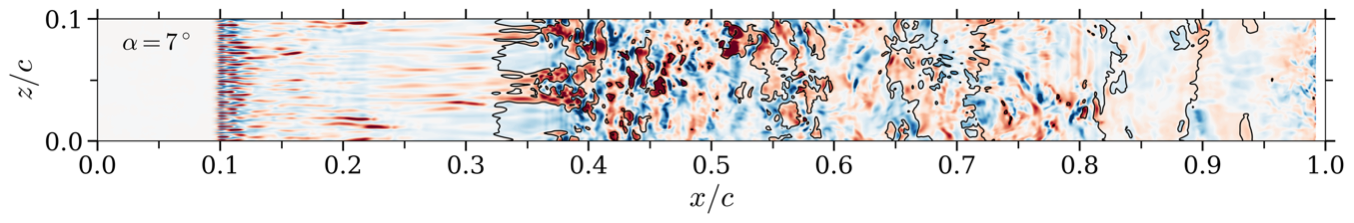}\\[-0.5cm]
(d)\includegraphics[width=\textwidth]{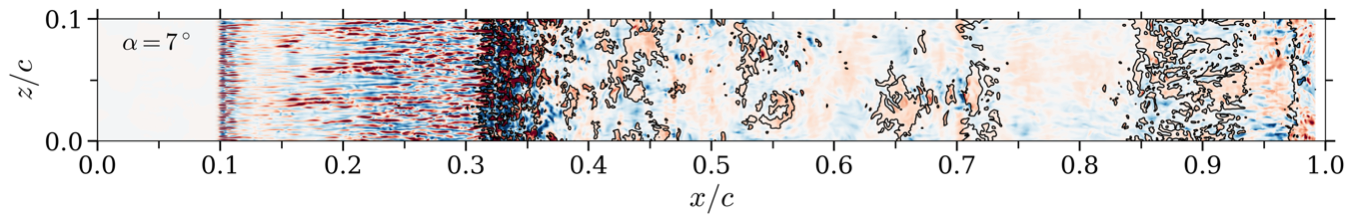}
\caption{
Instantaneous spanwise fluctuations of wall shear stress 
\eqref{eq:tauw} on the suction surface at 
$\Rey = 3 \times 10^5$ (a, c) and $6 \times 10^5$ (b, d), at 
$\alpha = 4^\circ$ (a, b) and $7^\circ$ (c, d). 
Snapshots (c, d) correspond to the upstream shock-motion phase. 
Black contours denote $\tau_w = 0$ isolines, indicating regions 
of instantaneous reverse flow.
}
\label{fig:tauw}
\end{figure}

\blue{
From the numerical perspective, \citet{schlatter2012turbulent} studied 
systematically the effects of tripping on turbulent boundary layers 
at moderate Reynolds numbers. \citet{hosseini2016direct} performed a DNS 
of incompressible flow over a wing section at $\Rey = 4\times10^5$, 
where transition was forced by applying a localised volume forcing 
to achieve fully developed turbulence conditions. 
In the compressible regime, turbulent buffet has been recently simulated 
using LES at $\Rey = 5\times10^5$ \citep{moise2023transonic, lusher2024effect}. 
In these studies, transition was triggered by time-varying blowing and 
suction strips, designed to mimic experimental arrays of tripping dots. 
\citet{lusher2024effect} demonstrated that a sufficiently large tripping 
amplitude ensures a fully developed turbulent state prior to shock interaction, 
yielding consistent buffet dynamics.
}
\red{
In this work, the tripping implementation follows the works by 
\citet{schlatter2012turbulent} and \citet{hosseini2016direct}, 
as detailed in \S\ref{sec:method-bc}.
}
\red{
To visualise the boundary layer transition, we show in figure 
\ref{fig:tauw} the instantaneous spanwise fluctuations of 
the suction-side shear stress, 
}
\begin{equation}
\tau_w' = \tau_w - \langle \tau_w \rangle_z, 
\quad \tau_w = \mu_w \frac{ \partial u }{\partial n}\bigg|_w, 
\label{eq:tauw}
\end{equation}
\red{
where $\langle \rangle_z$ denotes spanwise average. 
}

\blue{
In all cases, the flow downstream of the leading edge is laminar and remains 
spanwise-uniform. At $x = 0.1\,c$, the trip forcing generates a regular 
spanwise footprint of artificial disturbances. The transition process depends
on both the Reynolds number and the angle of attack. The non-buffet 
cases (figure~\ref{fig:tauw}a--b) exhibit a concentrated breakdown, namely 
the transition is induced at a nearly constant streamwise distance from the 
tripping strip across the span. Alternating wall-shear fluctuations mark the 
formation of streamwise-oriented streaks, which break down into turbulence 
within a development length of approximately $0.1$ ($\Rey = 3\times 10^5$ in 
figure \ref{fig:tauw}a) and $0.05$ ($\Rey = 6\times 10^5$ in figure \ref{fig:tauw}b).
}

\blue{
Figures~\ref{fig:tauw}(c--d) show fully developed buffet cases. The snapshots are 
captured at the maximum upstream shock excursion, which is the stage of the cycle 
when the distance between the tripping strip and the shock reaches its minimum. 
Under these conditions, the Reynolds number dependence becomes more pronounced. 
At $\Rey = 6\times 10^5$ (figure~\ref{fig:tauw}d), the boundary layer transitions 
to a fully turbulent state before interacting with the shock, as evidenced by the broadband, spanwise-homogeneous near-wall structures. 
At $\Rey = 3\times 10^5$ (figure~\ref{fig:tauw}c), the streaky structures are 
distributed non-uniformly across the span, indicating incomplete turbulent 
breakdown upstream of the shock.
}

\blue{
The state of the incoming boundary layer can be inspected through the 
`incompressible' Reynolds number ($\Rey_{\theta,i}$) based on the 
momentum thickness ($\theta$), defined as 
}
\begin{equation}
\Rey_{\theta,i} = \frac{ \tilde{u}_e \bar{\rho}_e \theta}{\bar{\mu}_w}, 
\quad \theta = \int_0^\infty \frac{ \bar{\rho}\tilde{u} }{ \bar{\rho}_e \tilde{u}_e} 
\left(1 - \frac{ \tilde{u} }{ \tilde{u}_e }\right) \mathrm{d}n. 
\label{eq:reytheta}
\end{equation} 
\blue{
where the subscripts $w$ and $e$ denote quantities evaluated
at the wall and at the boundary-layer edge, respectively.  
Following \citet{spalart1993experimental}, the boundary-layer edge
is defined as the wall-normal location at which $|\omega_z / \omega_{z,w}| < 10^{-4}$,
where $\omega_z = {\partial \tilde{v}}/{\partial s} - {\partial \tilde{u}}/{\partial n}$.
}

\begin{figure}
\centering
(a)\includegraphics[width=.47\textwidth]{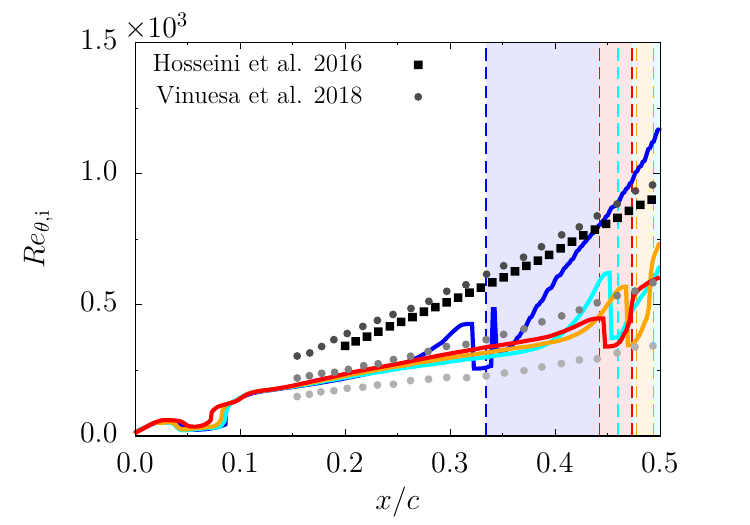}
(b)\includegraphics[width=.47\textwidth]{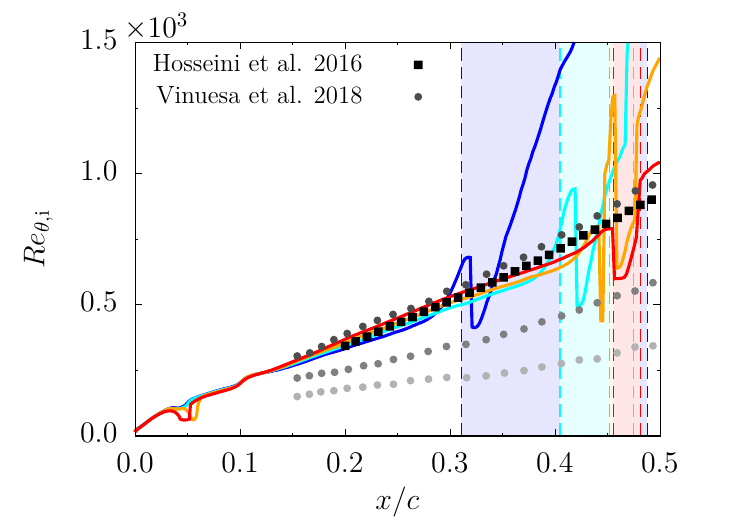}
(c)\includegraphics[width=.47\textwidth]{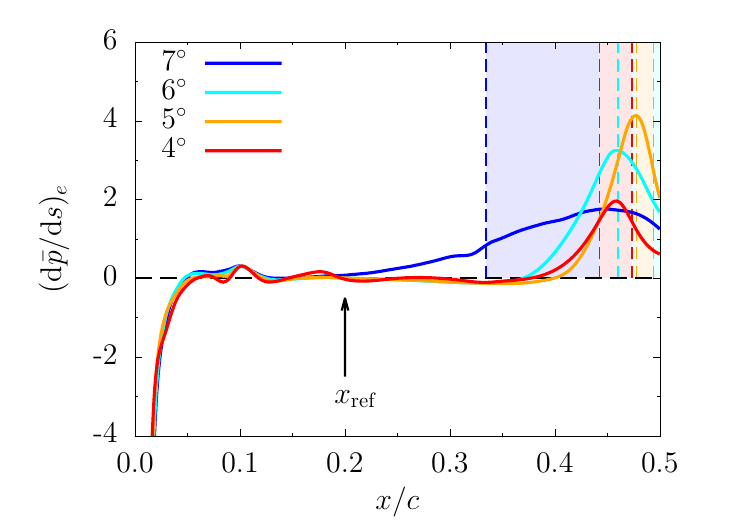}
(d)\includegraphics[width=.47\textwidth]{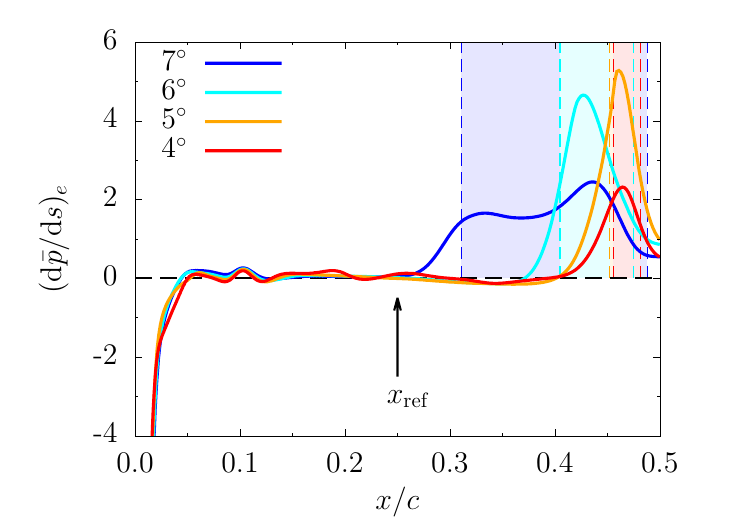}
\caption{
Distributions along the $x$ direction of the incompressible Reynolds number 
based on the momentum thickness (a, b) and streamwise mean-pressure gradient 
(c, d) at $\smash{\Rey = 3 \times 10^5}$ (a, c) and 
$\smash{6 \times 10^5}$ (b, d). Vertical dashed lines denote the maximum upstream 
and downstream shock positions. 
Results are compared with incompressible data of turbulent flow over 
the NACA4412 at $\Rey=4\times10^5$ \citep{hosseini2016direct} and 
$\Rey=10^5$, $2\times10^5$, $4\times10^5$ \citep{vinuesa2018turbulent}. 
}
\label{fig:bl0}
\end{figure}

\blue{
The Reynolds number is reduced to its incompressible value to facilitate a 
comparison with existing turbulent boundary-layer data. Figure~\ref{fig:bl0}(a--b) 
shows the streamwise distribution of $\Rey_{\theta,i}$ along the suction side 
for the investigated cases, alongside the incompressible benchmarks of 
\citet{hosseini2016direct} and \citet{vinuesa2018turbulent}. In those studies,  
DNS and well-resolved LES of turbulent flow over the NACA4412 airfoil 
were performed at Reynolds numbers ranging from $\Rey = 1\times10^5$ 
to $1\times10^6$ (results up to $\Rey = 4\times10^5$ are plotted). 
The shaded region bounded by vertical dashed lines indicates 
the envelope of the shock excursion. 
The present low-$\Rey$ data (figure \ref{fig:bl0}a) fall within 
the range bounded by the LES results at $\Rey=1\times10^5$ and $2\times10^5$. 
The high-$\Rey$ data in figure \ref{fig:bl0}(b) align in the upstream 
region with the DNS benchmark of \citet{hosseini2016direct} at $\Rey = 4\times10^5$. 
}

\blue{
To further assess the state of the incoming boundary layer, 
we identify an upstream reference location using the streamwise 
mean-pressure gradient at the boundary-layer edge 
$(\partial \bar{p}/\partial s)_e$, shown in figure~\ref{fig:bl0}(c--d). 
Downstream of the leading-edge acceleration, the flow 
exhibits a nearly zero-pressure-gradient region, 
characteristic of supercritical airfoils, 
followed by an adverse-pressure-gradient peak induced by the shock. 
While this peak is localised in the non-buffet cases, 
its upstream influence expands significantly at $\alpha = 7^\circ$. 
The reference location is defined immediately upstream of this 
influence region, as indicated by the black arrows 
in figure~\ref{fig:bl0}(c--d). 
}

\blue{
At this station, the incompressible Reynolds number 
is $\Rey_{\theta,i} \approx 250\text{ -- }275$ for the low-$\Rey$ cases 
(figure~\ref{fig:bl0}a) and $\Rey_{\theta,i} \approx 400\text{ -- }450$ 
for the high-$\Rey$ cases (figure~\ref{fig:bl0}b), depending on $\alpha$. \citet{preston1958the} 
established that the lowest Reynolds number at which 
fully developed turbulent 
flow can occur is $\Rey_{\theta,i} \approx 320$, below which the logarithmic 
region vanishes. The low-$\Rey$ cases fall below this threshold, indicating 
that the turbulence is not fully developed. However, sustained turbulence can 
exist in this regime, as demonstrated by \citet{spalart1988direct} 
and \citet{bandyopadhyay1987resonant}, who documented turbulent boundary layers 
at $\Rey_{\theta,i} = 225$ and $285$, respectively. 
}

\begin{figure}
\centering
(a)\includegraphics[width=.47\textwidth]{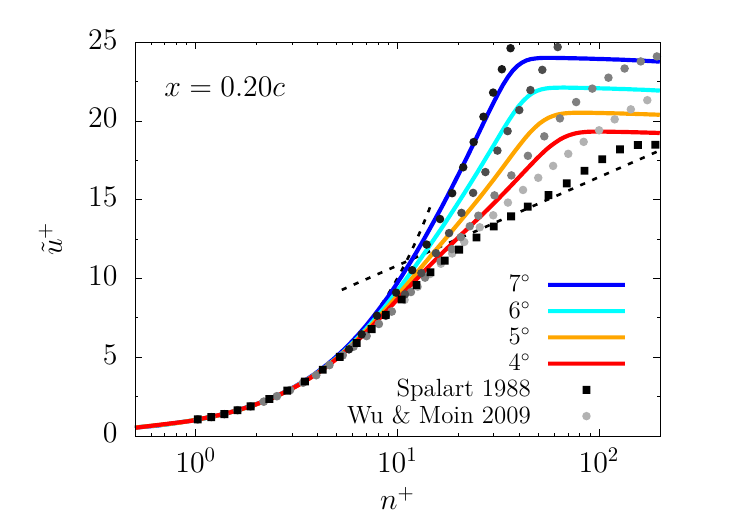}
(b)\includegraphics[width=.47\textwidth]{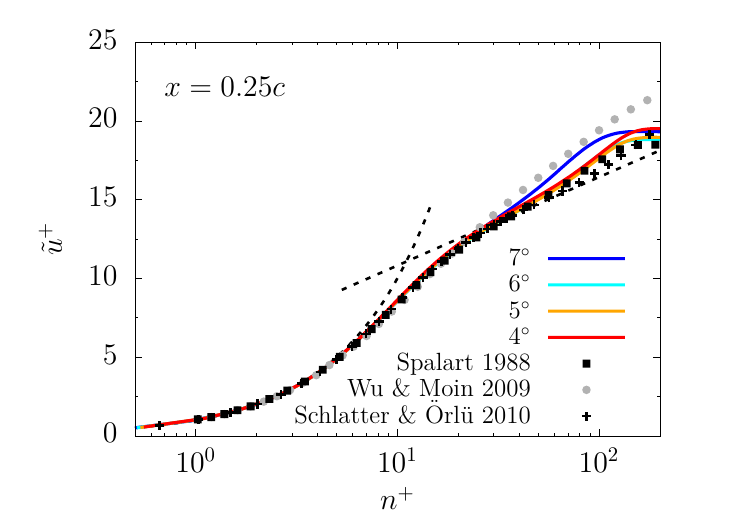}
(c)\includegraphics[width=.47\textwidth]{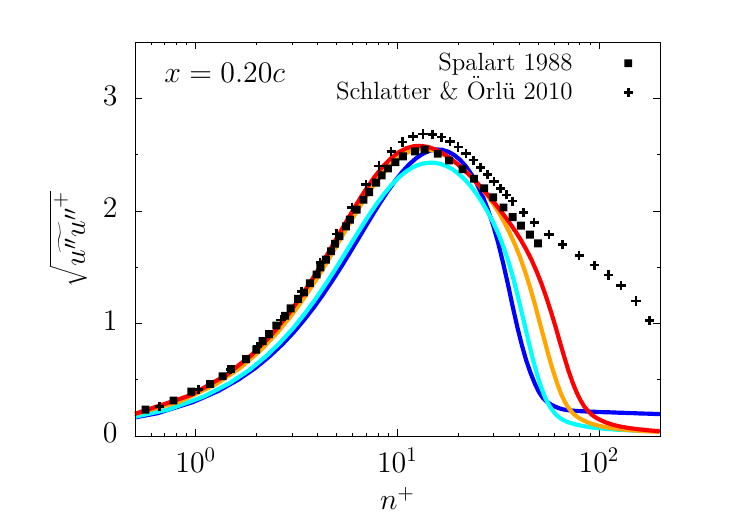}
(d)\includegraphics[width=.47\textwidth]{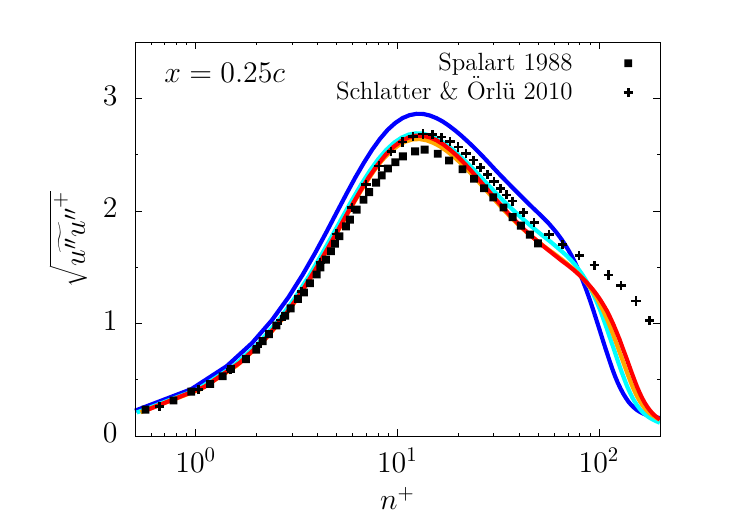}
\caption{
Wall-normal distributions of the mean streamwise velocity 
(a,b) and root-mean-square streamwise velocity (c, d) at the reference 
locations $x=0.2\,c$ (a, c) and $x=0.25\,c$ (b, d). 
Left panels refer to flow cases at $\smash{\Rey = 3 \times 10^5}$, right panels 
to $\smash{\Rey = 6 \times 10^5}$. The dashed line denotes the law of the wall. 
Results are compared with incompressible DNS data of zero-pressure-gradient 
turbulent boundary-layers at $\Rey_{\theta,i}=300$ \citep{spalart1988direct} and 
$\Rey_{\theta,i}=677$ \citep{schlatter2010assessment}, as well as 
transitional boundary-layers at $\Rey_{\theta,i}=300\text{ -- }600$ \citep{wu2009direct}. 
}
\label{fig:uu}
\end{figure}

\blue{
Velocity statistics at the reference locations are compared 
against the incompressible datasets of \citet{spalart1988direct} and 
\citet{schlatter2010assessment} for zero-pressure-gradient turbulent 
boundary layers at $\Rey_{\theta,i} = 300$ and $677$, respectively. 
Transitional boundary-layer data from \citet{wu2009direct} at $\Rey_{\theta,i} 
= 300\text{ -- }600$ are also included. Figure~\ref{fig:uu}(a--b) shows 
the wall-normal profiles of the Van Driest transformed mean streamwise 
velocity ($\tilde{u}^+$). In the low-$\Rey$ cases (figure~\ref{fig:uu}a), 
the inner-scaled velocity profiles lack a logarithmic region and progressively 
deviate from the law of the wall as $\alpha$ increases, aligning with the 
transitional boundary-layer profiles of \citet{wu2009direct}. 
Conversely, velocity profiles at 
the higher Reynolds number (figure~\ref{fig:uu}b) collapse onto the data 
of \citet{spalart1988direct}, except for the $\alpha=7^\circ$ case 
which deviates towards the transitional profile at the highest Reynolds number ($\Rey_{\theta,i} = 600$). 
}

\blue{
Wall-normal profiles of the root-mean-square streamwise velocity, 
$\smash{(\widetilde{u\pp u\pp}^+})^{1/2}$, are shown in figure~\ref{fig:uu}(c--d). 
At low Reynolds number (figure~\ref{fig:uu}c), the results align with the 
reference data only at low angles of attack, exhibiting the characteristic 
streamwise turbulence intensity peak at $n^+ \approx 12$. At higher incidence, 
this peak is attenuated and shifts away from the wall, reflecting the attenuation 
of the near-wall turbulence regeneration cycle. 
A better collapse is achieved at high Reynolds number (figure~\ref{fig:uu}d) in 
terms of both peak location and intensity. 
}

\begin{figure}
\centering
(a)\includegraphics[width=.47\textwidth]{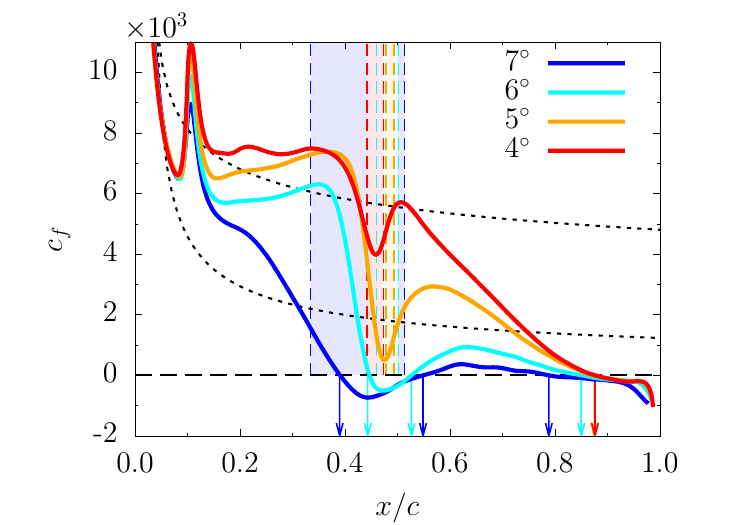}
(b)\includegraphics[width=.47\textwidth]{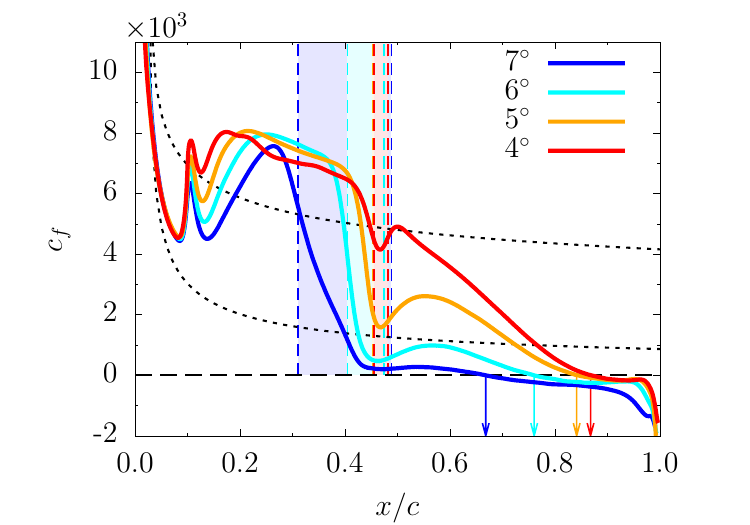}
(c)\includegraphics[width=.47\textwidth]{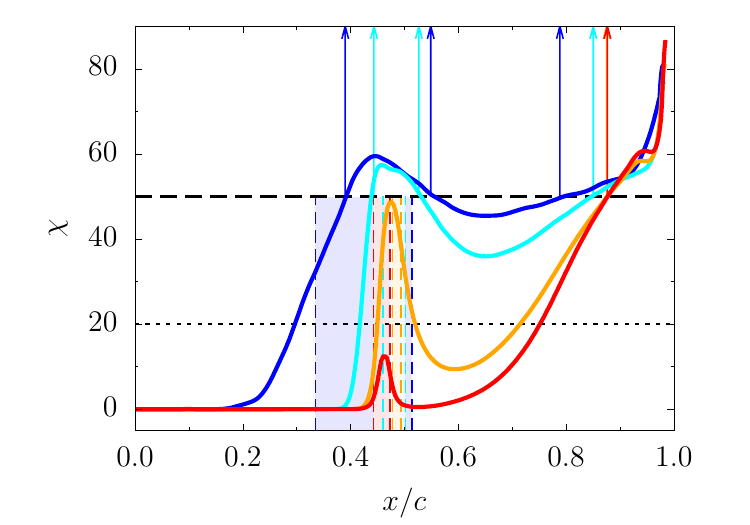}
(d)\includegraphics[width=.47\textwidth]{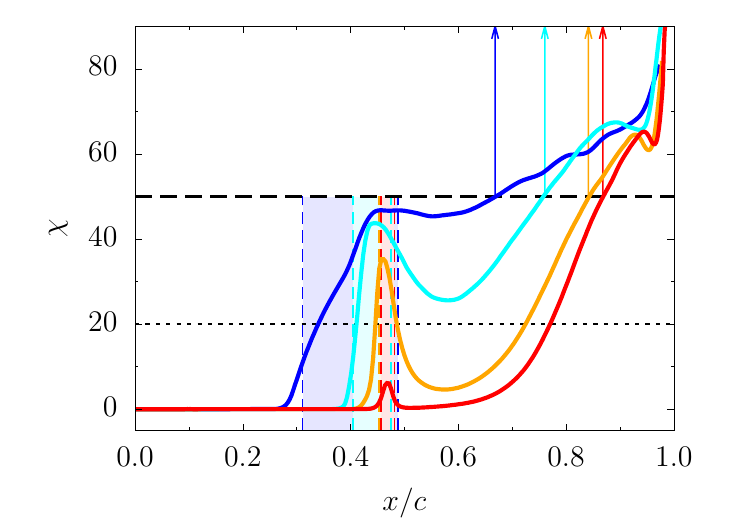}
\caption{
Distributions along the suction side of the friction 
coefficient (a, b) and backflow probability (c, d) 
at $\Rey = 3 \times 10^5$ (a, c) and $6 \times 10^5$ (b, d). 
Vertical dashed lines denote the maximum upstream and downstream 
shock positions, while solid arrows indicate the mean separation 
points. Dotted lines in (a, b) indicate equations \eqref{eq:friction}.
}
\label{fig:bl1}
\end{figure}

\green{
The distributions of friction coefficient ($c_f$) 
are presented in figure~\ref{fig:bl1}(a,b) to examine the 
development of the boundary-layer along the suction side. 
The black dotted curves denote the laminar Blasius 
solution, $c_{f,\mathrm{lam}}$, and the turbulent power-law 
correlation, $c_{f,\mathrm{trb}}$ \citep{young1989boundary}, 
}
\begin{equation}
c_f = \frac{2\tau_w}{\rho_0 u_0^2}, \quad  
c_{f,\mathrm{lam}} = 0.664\,\Rey_x^{-1/2}, \quad
c_{f,\mathrm{trb}} = 0.059\,\Rey_x^{-1/5}, 
\label{eq:friction}
\end{equation}
\green{
which are included as qualitative references. 
The numerical tripping induces a localised peak at $x = 0.1\,c$, 
causing the friction profiles to depart from the laminar branch and 
overshoot the turbulent reference. In the low-$\Rey$ cases 
(figure~\ref{fig:bl1}a) at $\alpha < 7^\circ$, the $c_f$
continues to rise downstream of the trip, confirming that the boundary 
layer has not yet relaxed to a fully developed turbulent state. 
At $\alpha = 7^\circ$, the friction settles at values intermediate to 
the laminar and turbulent limits, indicating an incomplete transition.
}
\green{
In contrast, the high-$\Rey$ cases (figure~\ref{fig:bl1}b) exhibit a faster 
recovery from the forced transition, as anticipated from the flow 
visualisations in figure \ref{fig:tauw}. In this regime, the downstream 
decay of the friction coefficient is consistent with the establishment 
of a fully developed 
turbulent boundary layer upstream of the shock interaction.
}

\green{
The mean separation point is identified by the vanishing 
of the friction coefficient ($c_f = 0$), indicated by solid vertical arrows 
in figure \ref{fig:bl1}. Separation can also be quantified using the backflow 
probability, defined as
}
\begin{equation}
\chi = \int_{-\infty}^0 \mathrm{PDF}(\tau_w) \mathrm{d}\tau_w,
\end{equation} 
\green{
where $\mathrm{PDF}(\tau_w)$ is the probability density function of the 
instantaneous wall shear stress. Following \citet{simpson1989turbulent}, 
thresholds of $\chi = 20\%$ and $\chi = 50\%$ denote intermittent and mean 
separation, respectively. The location of intermittent separation is associated 
with an {\em unrelaxed} velocity profile \citep{sandborn1974boundary}. The streamwise 
distributions of $\chi$, shown in figure~\ref{fig:bl1}(c--d), highlight distinct 
regimes between the investigated cases. In non-buffet conditions, backflow 
probability remains well below $20\%$ across the suction side, with local peaks 
beneath the shock and near the trailing edge. Conversely, in buffet regimes $\chi$ exceeds $20\%$ continuously from the shock 
interaction zone to the trailing edge. 
}

\green{
Both the shock-interaction and trailing-edge regions are sensitive to the 
Reynolds number. A shock-induced separation bubble forms 
exclusively in the low-$\Rey$ cases (figure~\ref{fig:bl1}c) at high incidence.
This separated region is characterised by a single local minimum in 
$c_f$, a signature of transitional separation 
\citep{sansica2016instability, lusher2020shock}, 
contrasting with the double local minimum typical of 
laminar separation \citep{katzer1989lengthscales, lusher2020effect}. 
}

\green{
The pronounced aft camber of the V2C airfoil induces 
trailing-edge separation across all investigated cases, 
with the separation point moving upstream as $\alpha$ increases. 
At $\alpha=7^\circ$, the low-$\Rey$ cases exhibit a trailing-edge 
separation point located further downstream relative 
to the high-$\Rey$ cases. 
This behaviour aligns with the results 
of \citet{lusher2024effect}, who reported that 
transitional boundary layers under buffet conditions, 
although affected by shock-induced 
separation, remain attached further downstream 
towards the trailing edge compared to 
their fully turbulent counterparts. 
}

\begin{figure}
\centering
(a)\includegraphics[width=.47\textwidth]{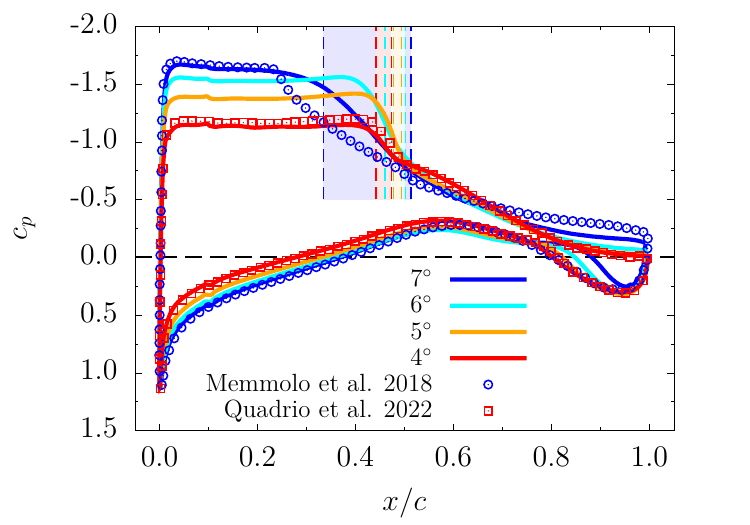}
(b)\includegraphics[width=.47\textwidth]{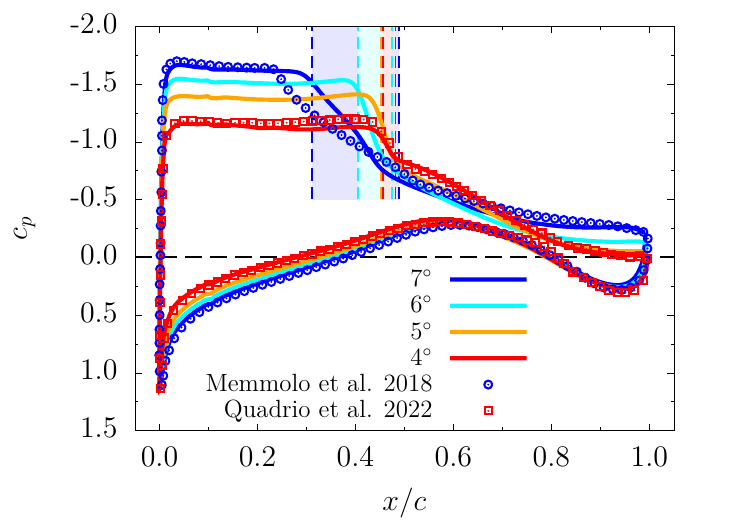}
(c)\includegraphics[width=.47\textwidth]{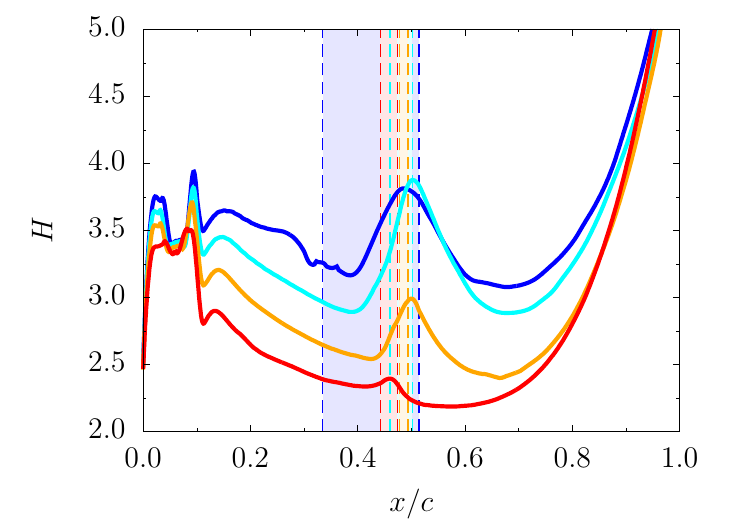}
(d)\includegraphics[width=.47\textwidth]{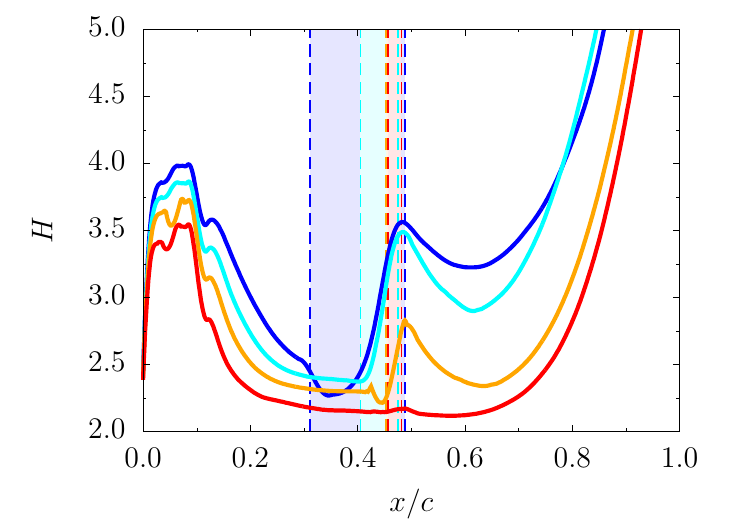}
\caption{
Distributions along the $x$ direction of the pressure
coefficient (a, b) and suction-side shape factor (c, d) 
at $\Rey = 3 \times 10^5$ (a, c) and
$6 \times 10^5$ (b, d). 
Vertical dashed lines denote the maximum upstream and downstream shock
positions, while solid arrows indicate the mean separation points.
}
\label{fig:bl2}
\end{figure}

Figure~\ref{fig:bl2}(a--b) displays the streamwise distribution of the mean 
pressure coefficient, 
\begin{equation}
c_p = \frac{2(\bar{p}_w - p_0)}{\rho_0 u_0^2},
\end{equation} 
where the vertical axis is inverted to show the suction side at the top.  
\red{
Data from \citet{memmolo2018scrutiny} and \citet{quadrio2022drag} are 
included as references. Both studies investigated the same airfoil geometry, 
the former at $M = 0.7$, $\alpha = 7^\circ$, $\Rey=3\times10^6$ using URANS, 
the latter at $M = 0.7$, $\alpha = 4^\circ$, $\Rey=3\times10^5$ using DNS. 
Despite differences in Reynolds number, the present results are in good 
agreement with the reference data. 
In particular, the lower-fidelity URANS model effectively captures 
the mean pressure distribution in the fully developed buffet regime. 
}

\green{
The wall-pressure distributions delineate 
the three flow regions determining the shock position: the upstream supersonic 
plateau, the pressure rise across the shock, and the downstream subsonic region 
\citep{holder1964transonic}. Increasing $\alpha$ intensifies the suction on the 
upper surface, leading to a stronger expansion and a larger pressure jump across 
the shock. 
The local gradient of the pressure coefficient within the interaction 
region provides insight into the shock dynamics. 
The steep pressure rise at $\alpha = 5^\circ$ 
reflects a stable shock configuration (see also figure \ref{fig:bl0}c--d). 
Conversely, the milder pressure gradient observed at $7^\circ$ 
stems from significant shock oscillations, 
which smear the mean pressure profile.
}

\green{
The pressure recovery from the post-shock value to the trailing edge 
is determined by the downstream boundary-layer development. 
At $\alpha = 7^\circ$, the pressure recovery is incomplete, 
as evidenced by the negative pressure coefficient at the trailing edge.
The $c_p$ value at the trailing edge decreases with increasing Reynolds 
number, consistent with the stronger trailing-edge separation observed 
at high $\Rey$ (see figure \ref{fig:bl1}c--d).
}

\blue{
The effects of the Reynolds number are further illustrated 
in figure \ref{fig:bl2}(c--d) through the streamwise 
evolution of the kinematic shape factor, 
$H = \delta^* / \theta$, where $\theta$ is the momentum thickness 
\eqref{eq:reytheta} and 
}
\begin{equation}
\delta^* = \int_0^\infty \left(1 - \frac{ \bar{\rho}\tilde{u} }{ \bar{\rho}_e \tilde{u}_e }\right) \mathrm{d}n 
\end{equation}
is the displacement thickness. 
\blue{
Following the laminar state near the leading edge ($H \approx 2.5$), 
the numerical trip induces a localised distortion of 
the velocity profile, causing a spike in the shape factor ($H \approx 4.0$). 
}

The subsequent relaxation towards equilibrium is governed by the combined 
effects of Reynolds number and angle of attack. 
At low $\Rey$ (figure~\ref{fig:bl2}c), increasing $\alpha$ 
progressively inhibits this relaxation process. While the flow 
at $\alpha = 4^\circ$ approaches a turbulent equilibrium ($H < 2.5$), 
the mean profiles at $\alpha = 7^\circ$ remain highly distorted upstream 
of the shock. 
\green{
This behaviour is attributed to a reduced near-wall momentum 
transport in the under-developed boundary layer, which maintains a 
pronounced velocity defect in the outer layer. 
}
Conversely, the high-$\Rey$ boundary layers (figure \ref{fig:bl2}d) exhibit
greater robustness, relaxing to $H < 2.5$ upstream of the shock even at
high incidence.

The shock interaction delineates clearly the flow regimes. 
Under non-buffet conditions ($\alpha \le 5^\circ$), the shape factor 
is weakly affected by the shock, 
while in the buffet regime ($\alpha \ge 6^\circ$), the shape 
factor peaks at the shock location. 
\green{
This spike reflects a growth in displacement thickness induced by the abrupt 
deceleration of the near-wall flow, which correlates with the high backflow 
probabilities highlighted in figure \ref{fig:bl1}(c--d).
}
Close to the trailing edge, the shape factor increases to values 
characteristic of separated flow ($H > 5$) across all cases.

%==============================================
\subsection{Flow organisation}\label{sec:visua}
%==============================================
%
\begin{figure}
(a)\includegraphics[width=.98\textwidth]{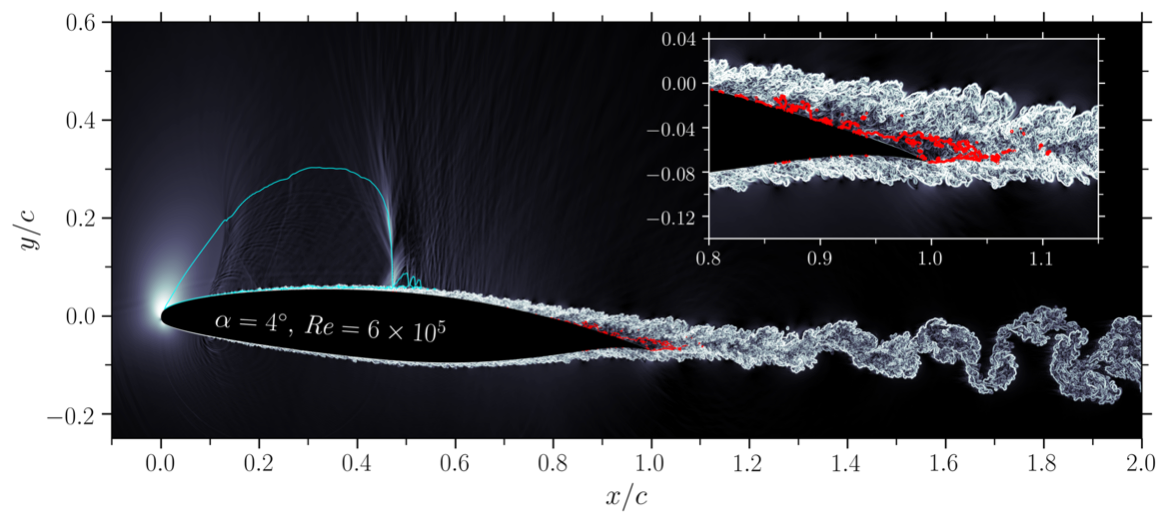}\\[-0.52cm]
(b)\includegraphics[width=.98\textwidth]{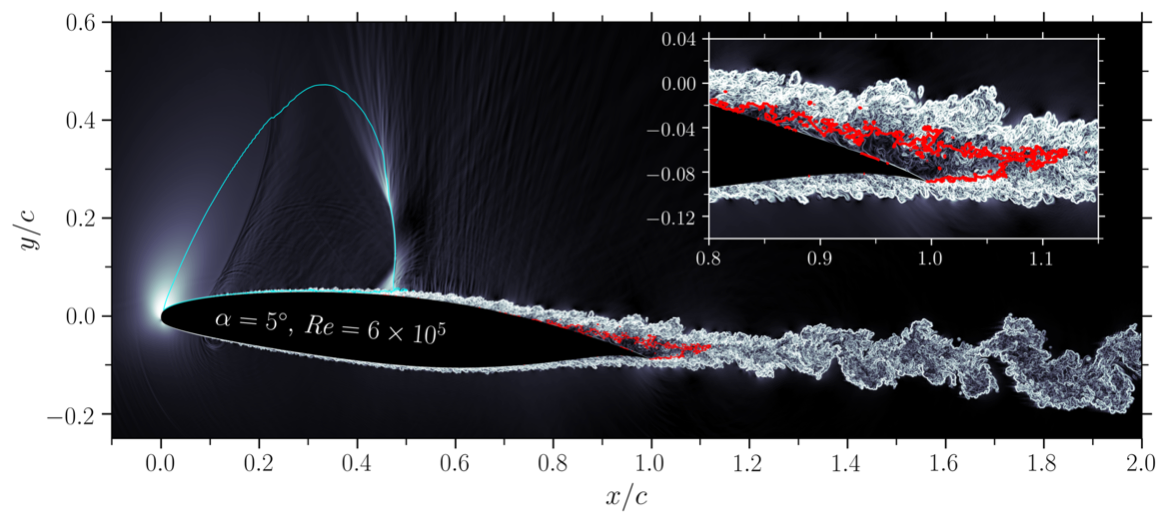}\\[-0.52cm]
(c)\includegraphics[width=.98\textwidth]{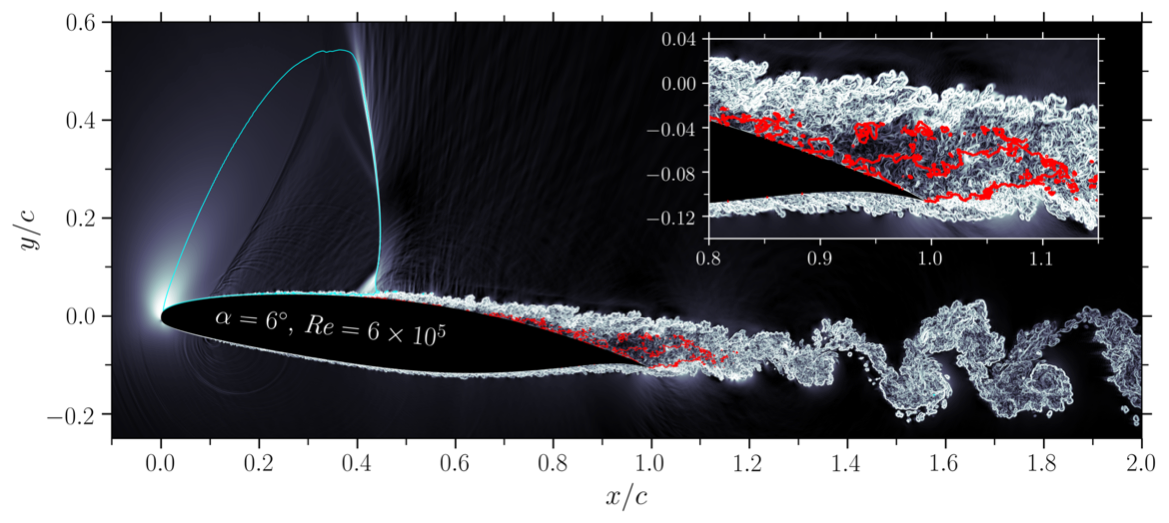}
\caption{
Numerical Schlieren snapshots for flow cases at 
$\alpha = 4^\circ$ (a), $\alpha = 5^\circ$ (b), $\alpha = 6^\circ$ (c). 
The sonic line ($M = 1$) is marked in cyan and the flow reversal line 
($u = 0$) in red. 
}
\label{fig:sch1}
\end{figure}
Figure~\ref{fig:sch1} shows snapshots of the instantaneous 
flow field at $\Rey = 6 \times 10^{5}$. The numerical Schlieren contours, 
computed as $\exp\!\left(-0.5\,\|\nabla\rho\|\,c/\rho_0\right)$, 
visualise the shock wave, turbulent structures, and the wake.
The supersonic region is highlighted by the sonic line (cyan curve), 
and the flow reversal line ($u=0$) is marked in red. 

\red{
At $\alpha = 4^\circ$ (figure \ref{fig:sch1}a), 
the shock wave is relatively weak, allowing the downstream 
flow to re-expand and form a secondary supersonic region 
\citep{babinsky2011shock}. This supersonic tongue is terminated 
by progressively weaker shocklets, which interact with upstream-travelling 
disturbances and intermittently merge with the primary shock.
}

\green{
Increasing the angle of attack to $\alpha = 5^\circ$ (figure~\ref{fig:sch1}b) 
expands the supersonic region and strengthens the shock, 
as shown in \S\ref{sec:bl}. The flow decelerates 
more abruptly to subsonic speeds, preventing the formation of a supersonic tongue. 
The suppression of unsteady shocklets, combined with an intensified shock, 
stabilises the shock position (see $\sigma_s$ in table \ref{tab:1}). 
}
\red{
This behaviour is consistent with \citet{moise2023transonic}, who 
reported a nearly steady lift coefficient for the same airfoil geometry 
under similar conditions ($M=0.7$, $\alpha=5^\circ$, $\Rey=5\times10^5$, 
tripped boundary layer). 
}
At $\alpha = 6^\circ$ (figure \ref{fig:sch1}c), the shock becomes unsteady 
and low-frequency oscillations develop, although the excursions remain
limited. In general, increasing the angle of attack results in more 
pronounced trailing-edge separation (as indicated by the flow reversal line) 
and greater wall-normal extent of the shock.

\begin{figure}
\centering
(a)\includegraphics[width=.46\textwidth]{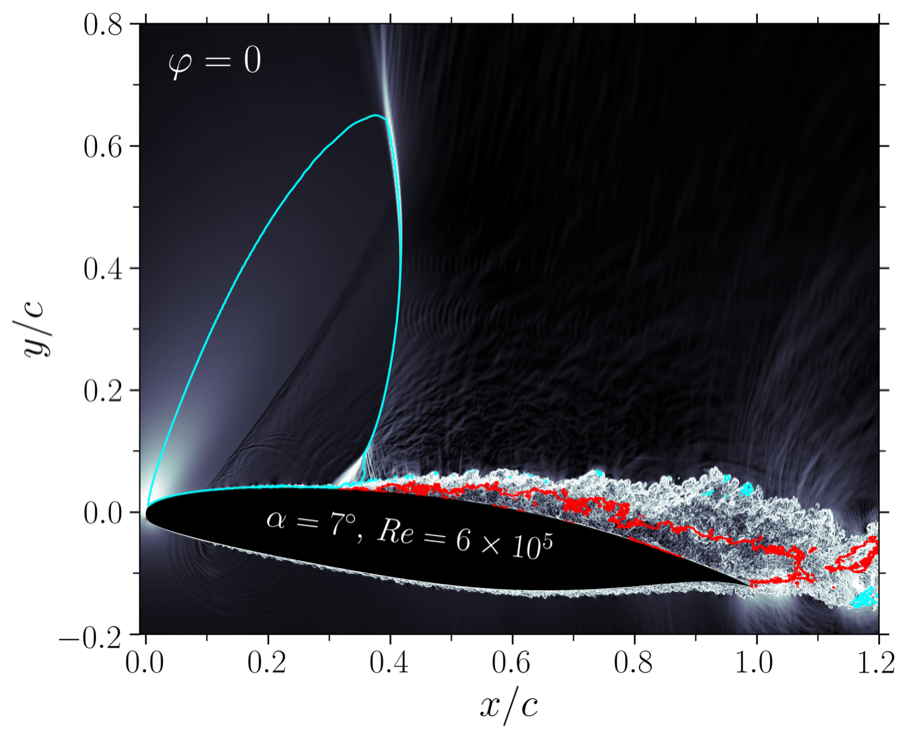}
(b)\includegraphics[width=.46\textwidth]{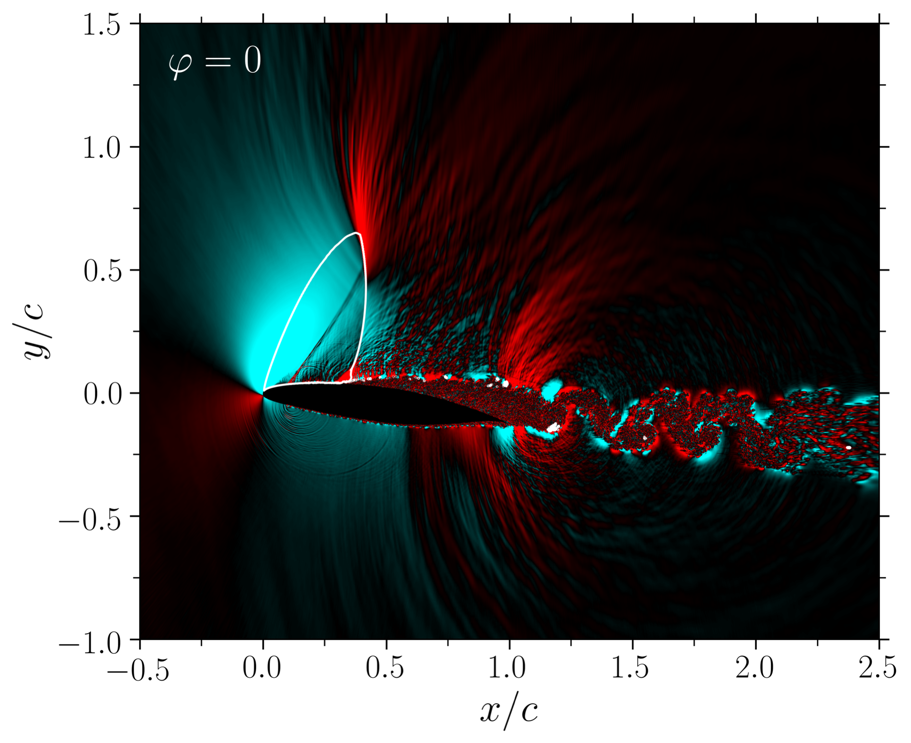}\\[-0.5cm]
(c)\includegraphics[width=.46\textwidth]{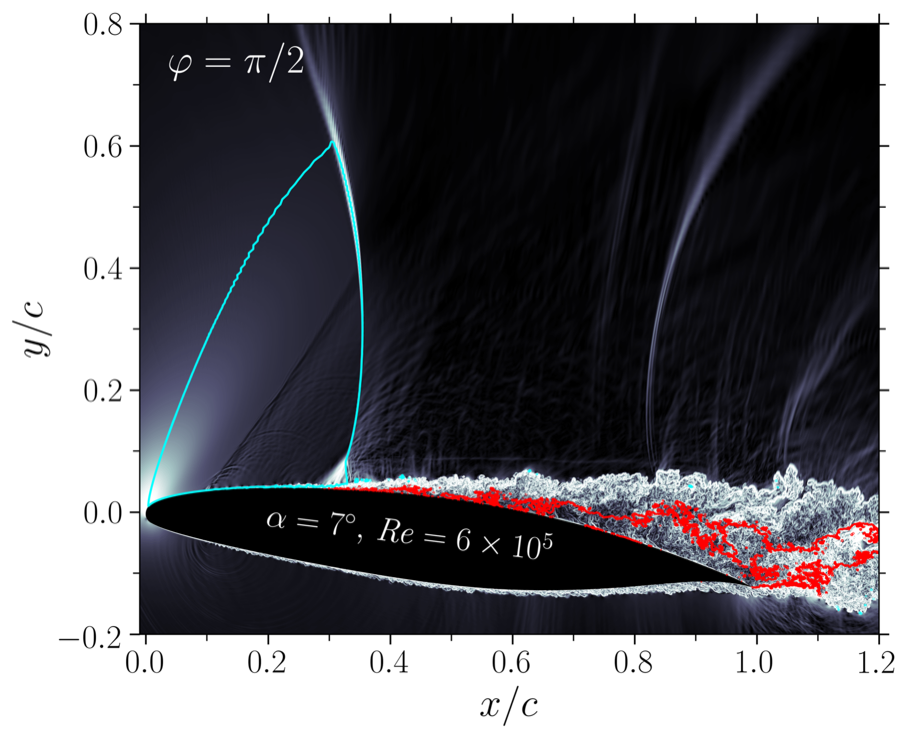}
(d)\includegraphics[width=.46\textwidth]{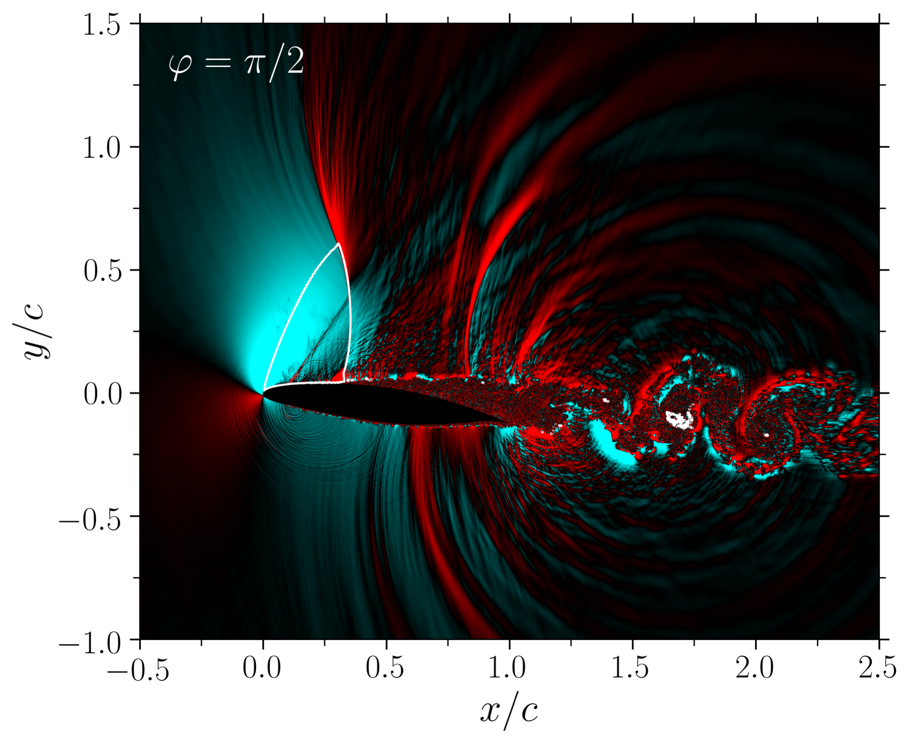}\\[-0.5cm]
(e)\includegraphics[width=.46\textwidth]{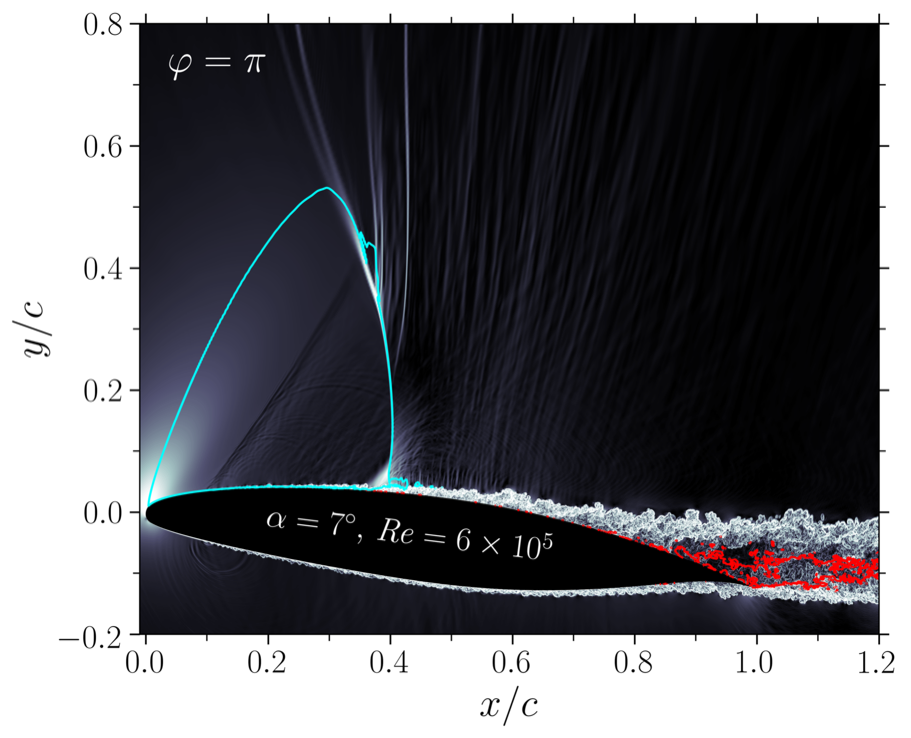}
(f)\includegraphics[width=.46\textwidth]{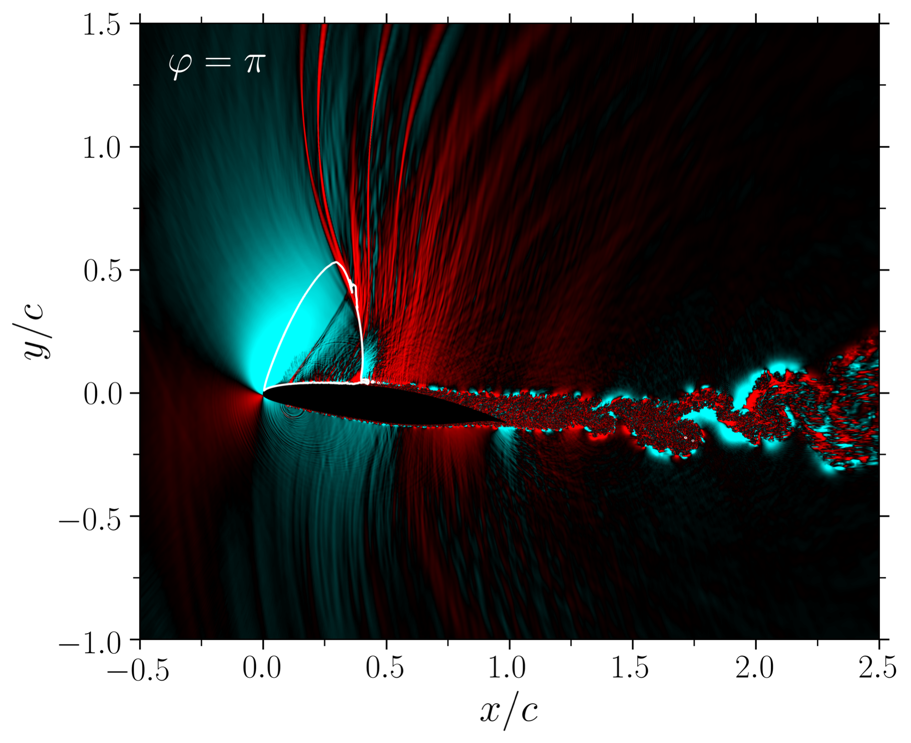}\\[-0.5cm]
(g)\includegraphics[width=.46\textwidth]{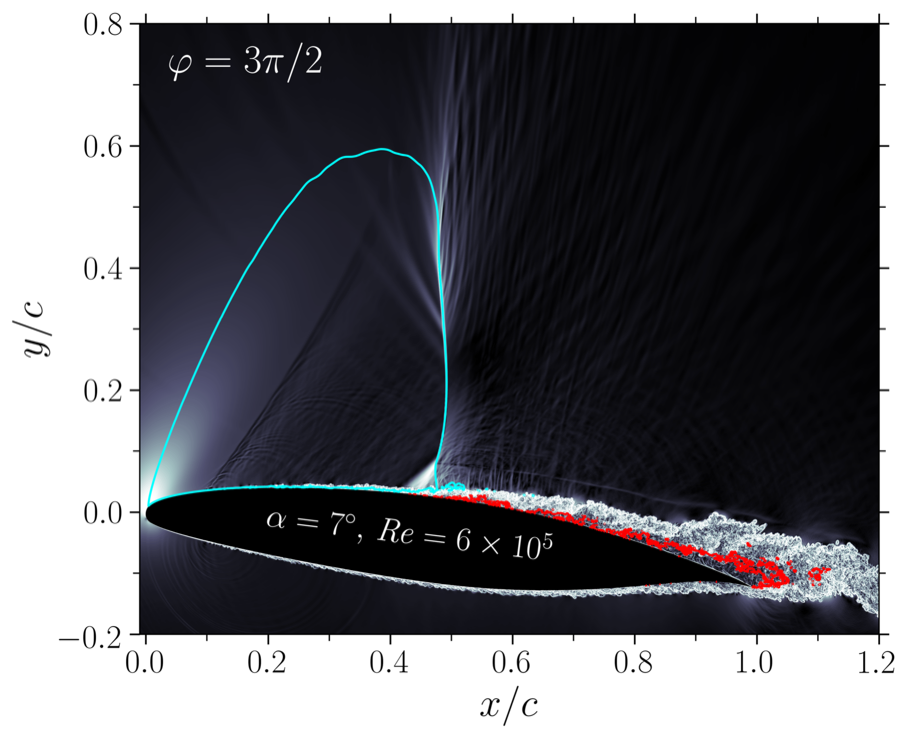}
(h)\includegraphics[width=.46\textwidth]{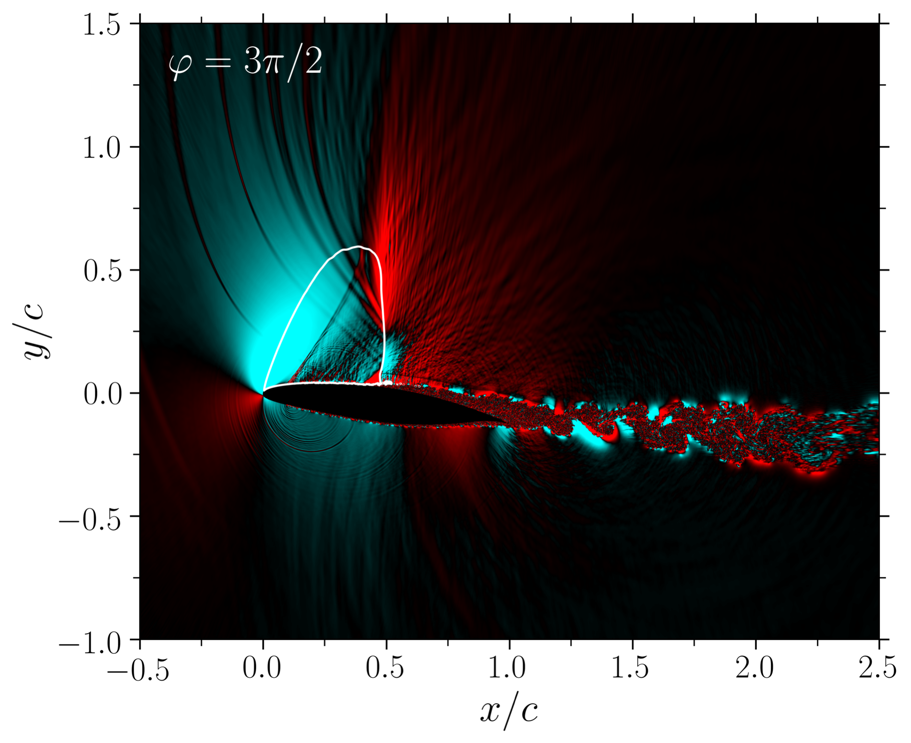}
\caption{
Snapshots of numerical Schlieren (left panels) and modified 
Ducros sensor \eqref{eq:sign_duc} (right panels) for the 
flow case at $\alpha = 7^\circ$. 
The sequence shows four phases of the buffet cycle: 
upstream shock motion ($\varphi = 0$), 
maximum upstream shock excursion ($\varphi = \pi/2$), 
downstream shock motion ($\varphi = \pi$), 
maximum downstream shock excursion ($\varphi = 3\pi/2$). 
}
\label{fig:sch2}
\end{figure}

Further increasing the angle of attack to $\alpha = 7^\circ$ yields 
a fully developed buffet state. 
\blue{
The kinematics of the oscillating shock is characterised by 
the relative shock Mach number, defined as 
$M_s = (\mathrm{d}x_s/\mathrm{d}t)/a_s$,
where $x_s$ is the instantaneous shock position and $a_s$ is the local 
speed of sound immediately upstream of the shock. 
By this definition, $M_s < 0$ denotes a shock moving upstream. 
}

\green{
In contrast to the pre-buffet cases, the flow field varies strongly with
the phase of the buffet cycle. We therefore present four distinct phases at 
$\varphi = 0, \pi/2, \pi, 3\pi/2$, where the phase angle is defined as 
$\varphi = 2\pi \St_B \, t u_0 / c$. As explained in \S\ref{sec:intro}, transonic 
buffet is a global instability, whereby instability waves propagate throughout 
the entire flow field. 
For this reason, alongside the numerical Schlieren of the near field (left panels 
of figure \ref{fig:sch2}), we present a sign-preserving variant of the Ducros sensor, 
}
\begin{equation}
\Theta_s = -\nabla \cdot \mathbf{u} \left[(\nabla \cdot \mathbf{u})^2 + 
(\nabla \times \mathbf{u})^2 +\left(u_0 / c\right)^2 \right]^{-1/2}
\label{eq:sign_duc}
\end{equation}
\green{
(right panels of figure \ref{fig:sch2}), to visualise the acoustic field around the airfoil.
}
 
\green{
Moving upstream, the shock opposes the incoming flow, 
thereby increasing the effective local Mach number. This phase produces 
the strongest shock of the cycle, leading to a fully separated boundary 
layer, as evidenced by the flow reversal line in figure \ref{fig:sch2}(a). 
The resulting shear-layer unsteadiness is accompanied by intense 
acoustic activity near the trailing edge, clearly captured by the 
$\Theta_s$ snapshot in figure \ref{fig:sch2}(b). 
}
\green{
At its most upstream position (figure \ref{fig:sch2}c), the shock bounds 
a minimal supersonic region. The extensive flow separation creates a thick, 
unsteady wake, with acoustic disturbances appearing as upstream-propagating, 
wall-normal density gradients above the shear layer. The $\Theta_s$ contours 
in figure \ref{fig:sch2}(d) reveal sharp, thin bands originating near the 
trailing edge that radiate as near-circular fronts. This alternating pattern of 
compression (red) and expansion (cyan) is characteristic of acoustic waves 
propagating through a non-uniform convective field.
}

\blue{
Figure \ref{fig:sch2}(e) shows the downstream shock motion ($M_s > 0$). 
Large-scale separation is suppressed at this stage, resulting in a narrower 
wake. As the shock travels in the same direction as the incoming flow, the 
effective upstream Mach number decreases and the shock is in its weakest state. 
As the previously emitted acoustic disturbances travel upstream, they distort 
into quasi-wall-normal fronts, with compressive waves coalescing near the shock 
due to wave steepening (figure \ref{fig:sch2}f). 
}

\green{
Lastly, figure \ref{fig:sch2}(g) captures the maximum downstream position 
of the shock. While flow separation begins to re-expand, trailing-edge 
unsteadiness is still mild. Figure \ref{fig:sch2}(h) shows that acoustic 
disturbances continue propagating upstream and persist beyond the shock interaction. 
The wave fronts bend slightly due to spatial gradients in mean velocity and 
speed of sound. Inside the supersonic pocket, this bending 
further intensifies as the fronts tend to align with the local characteristic 
lines.
}

%==============================================
\subsection{Shock motion}\label{sec:shock}
\begin{figure}
\centering
(a)\includegraphics[width=.95\textwidth]{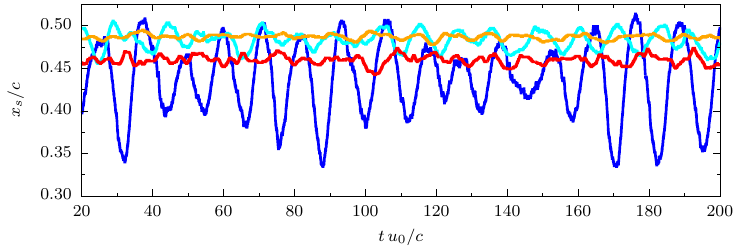}
(b)\includegraphics[width=.95\textwidth]{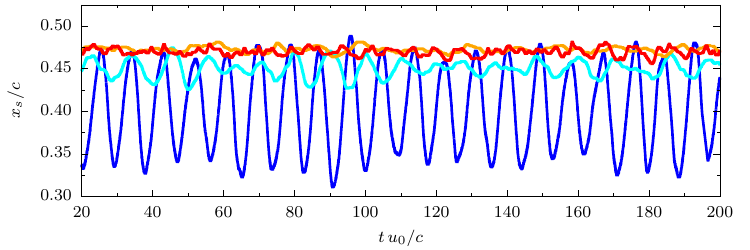}
\caption{Time history of the shock position $x_s$ for flow cases at
$\Rey = 3 \times 10^5$ (a) and $6 \times 10^5$ (b) at various
angles of attack $\alpha$ (see next figure for colour coding).}
\label{fig:xshock}
\end{figure}

%==============================================
The shock motion is compared quantitatively across the considered flow cases 
in figure \ref{fig:xshock}. The instantaneous shock position $x_s$ is 
identified as the location of the maximum wall-parallel pressure gradient
($\mathrm{d}p/\mathrm{d}s$), evaluated at the wall-normal coordinate
$n \approx 6 \times 10^{-2}\,c$, immediately above the triple point 
of the lambda-shock structure.
\green{
To identify fully developed buffet conditions, we use the four criteria 
proposed by \citet{schauerte2023experimental}, namely 1) the amplitude 
of shock oscillations, 2) the occurrence of a sharp frequency peak 
in the power spectral density (PSD) spectrum, 3) the standard deviation 
of the shock position, and 4) the direction of the shock motion with 
increasing incidence.
}

\begin{figure}
\centering
(a)\includegraphics[width=.47\textwidth]{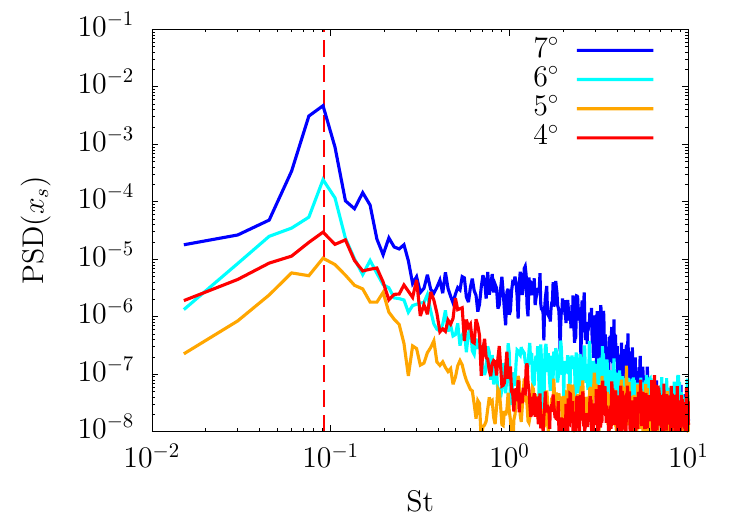}
(b)\includegraphics[width=.47\textwidth]{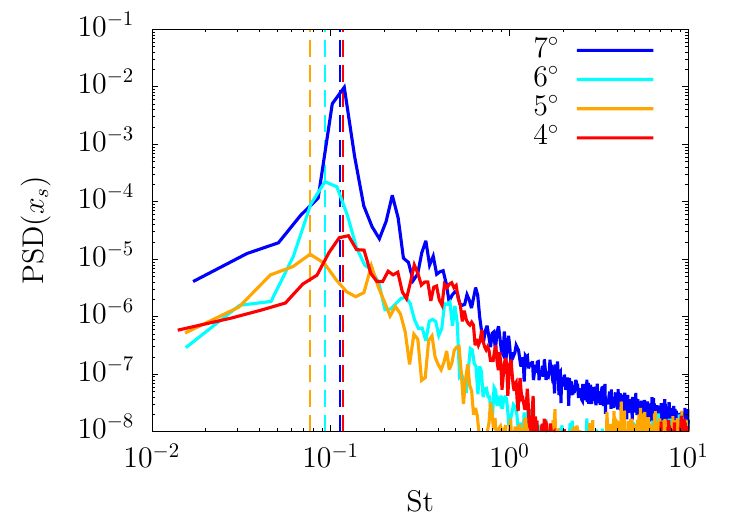}
\caption{
Power spectral density of the shock position $x_s$ as a function
of the Strouhal number $\St$ for flow cases at
$\Rey = 3 \times 10^5$ (a) and $6 \times 10^5$ (b) 
at various $\alpha$. 
}
\label{fig:psdxs}
\end{figure}

\green{
Increasing $\alpha$ from $4^\circ$ to $5^\circ$ results in a downstream 
shift in the mean shock position $\overline{x}_s$, while a further 
increase from $5^\circ$ to $6^\circ$ produces an upstream shift (see table 
\ref{tab:1}). This trend is observed for both Reynolds numbers examined, 
confirming that the reversal of shock motion is a necessary condition for 
buffet onset \citep{nitzsche2019fluid, 
accorinti2022experimental, schauerte2023experimental}.
}
\green{
The fourth criterion for buffet identification requires an onset 
angle greater than $5^\circ$ in the current configuration.
The results listed in table \ref{tab:1} show that at an incidence of 
$\alpha = 6^\circ$, the maximum oscillation amplitude ($\Delta x_{s,\mathrm{max}}$) 
and the standard deviation of the shock position ($\sigma_s$) increase by a 
factor two relative to the lower-incidence regime. This 
amplification characterises the $\alpha = 6^\circ$ condition as an {\em incipient} 
buffet state. 
}
Under these conditions, the buffet cycles are irregular, with both the
amplitude and the period exhibiting cycle-to-cycle variability.

\red{
The flow case at $\alpha = 7^\circ$ represents a fully developed buffet 
state, consistent with \citet{moise2023transonic}. 
}
\green{
The shock oscillation amplitude increases substantially, with $\sigma_s$ 
approximately five times greater than that of the incipient buffet case. 
A comparison of figures \ref{fig:xshock}(a) and \ref{fig:xshock}(b) indicates 
that the maximum oscillation amplitude remains comparable between the two Reynolds 
numbers across the investigated incidence angles. In the fully developed buffet 
regime, a cycle-to-cycle variability is observed at $\Rey = 3\times10^5$, 
while at $\Rey = 6\times10^5$ the time series approaches a periodic state.
}

Figure \ref{fig:psdxs} presents the PSD of the shock-position signal as 
a function of the Strouhal number, defined as $\St = f c / u_0$. 
\green{
A low-frequency peak is observed in all flow cases, including those in a 
pre-buffet state. This indicates that precursor signatures of the instability 
are present well below the buffet threshold. 
The growth of these precursor 
signatures into the fully developed buffet state is non-monotonic. 
In line with regular shock motion, the mean 
shock position at $\alpha = 5^\circ$ relative to $4^\circ$ is further 
downstream, a position associated with a more stationary shock. Consequently, 
the PSD peak at the buffet frequency is more pronounced at $\alpha = 4^\circ$ 
than at $5^\circ$. Beyond this temporary stabilisation, the instability grows 
into incipient buffet at $\alpha = 6^\circ$, where the peak magnitude increases 
by one order of magnitude compared to non-buffet conditions. Ultimately, 
the spectral energy increases by a further two orders of magnitude as the flow 
transitions into the fully developed buffet regime, which is
consistent with the second criterion for buffet identification. 
}

At the lower Reynolds number 
(figure \ref{fig:psdxs}a), the peak frequency collapses to 
$\St \approx 0.09$ across all examined angles of attack. 
Conversely, at the higher Reynolds number (figure \ref{fig:psdxs}b), 
the buffet frequency increases from $\St \approx 0.09$ at 
$\alpha = 6^\circ$ to $\St \approx 0.11$ at $\alpha = 7^\circ$. 
\red{
This dominant frequency is close to the value documented by 
\citet{zauner2019direct} for laminar transonic buffet on the V2C airfoil, 
which was subsequently shown to remain nearly invariant across a range of 
Reynolds numbers \citep{moise2022large}. 
The slight difference in frequency observed at $\alpha = 7^\circ$, 
depending on the Reynolds number, is associated with the state of the 
incoming boundary layer, which only becomes fully turbulent at the higher 
Reynolds numbers. The origin of this difference is discussed in \S\ref{sec:loop}.
}

\red{
While \citet{jacquin2009experimental} reported dominant Strouhal numbers
independent of the angle of attack, subsequent studies observed an
increase in buffet frequency with incidence \citep{iovnovich2012reynolds,
accorinti2022experimental, schauerte2023experimental, schauerte2025influences,
lusher2025implicit}. As evidenced by the flow visualisations and quantified in
\S\ref{sec:bl}, the separated region enlarges significantly with $\alpha$.
If the buffet frequency were primarily set by a characteristic 
time scale associated with the separation size, 
it would be expected to decrease as the separated region grows. 
The slight increase in the peak Strouhal number across incidences
suggests instead that the separation size does not provide the primary
time scale of the buffet instability.
}

%==============================================
\section{Flow unsteadiness}\label{sec:unsteady}
%==============================================

%==============================================
\subsection{Wall-pressure spectra}
\begin{figure}
\centering
(a)\includegraphics[width=.31\textwidth]{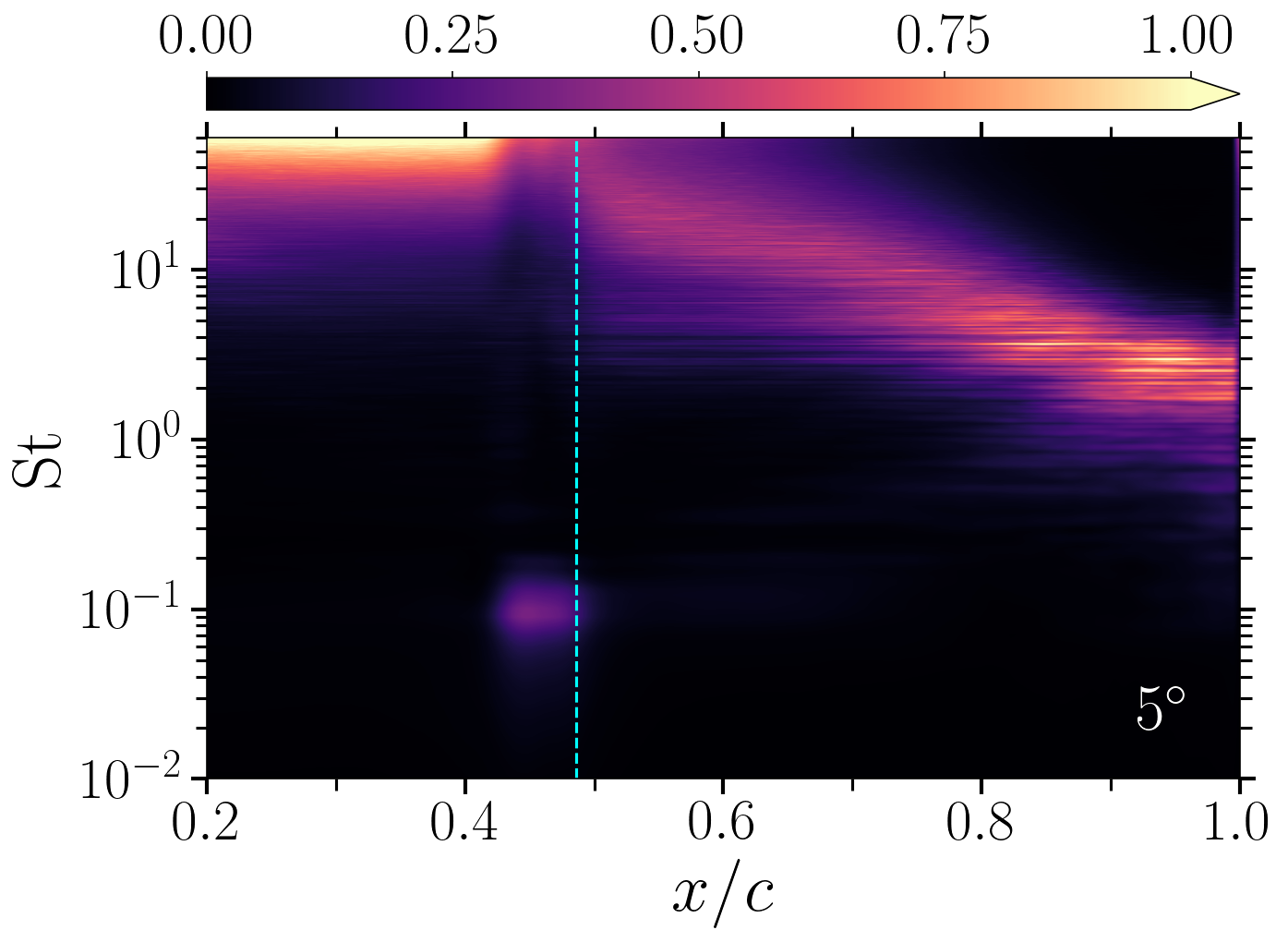}\hspace{-4mm}
(b)\includegraphics[width=.31\textwidth]{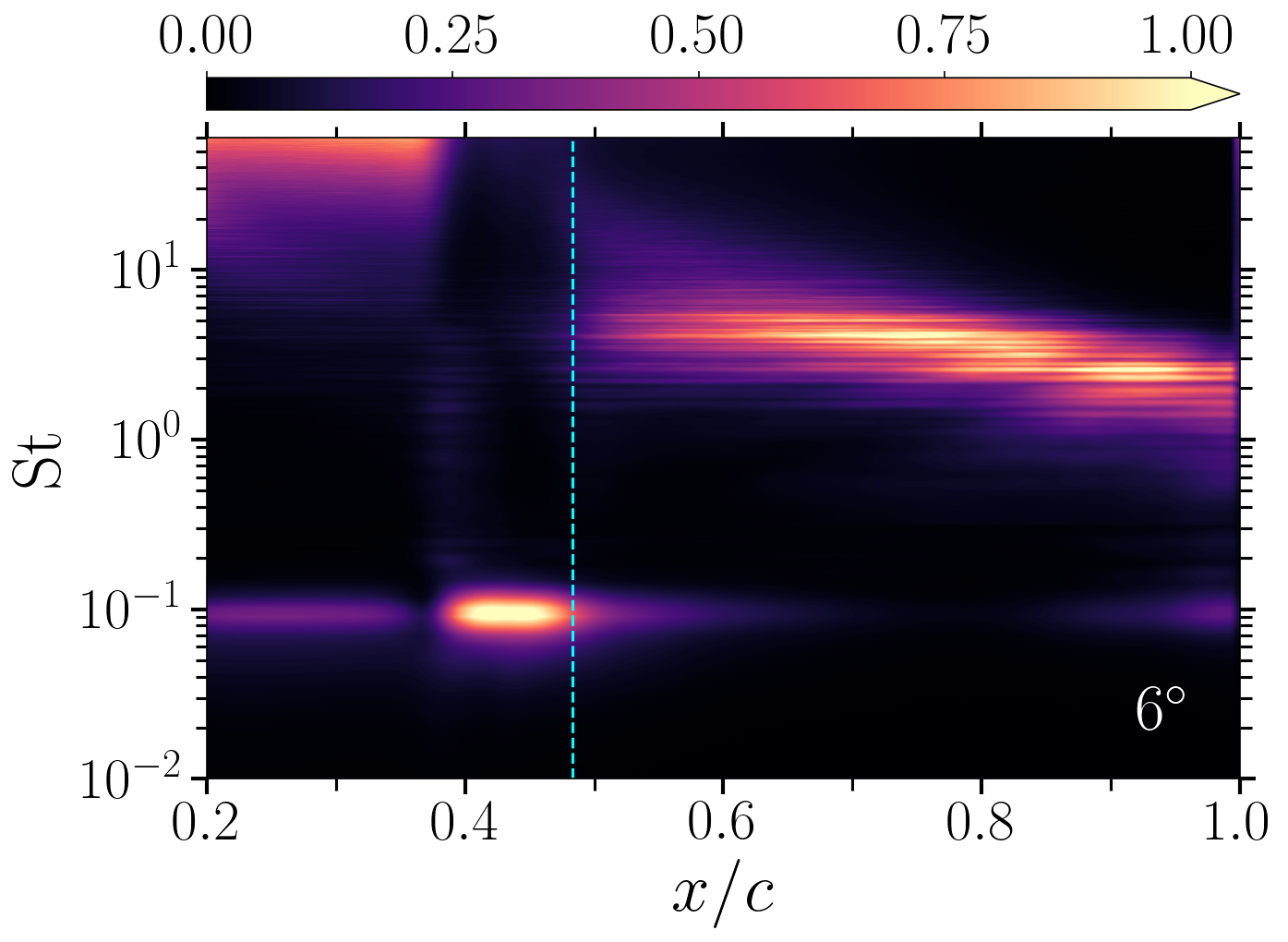}\hspace{-4mm}
(c)\includegraphics[width=.31\textwidth]{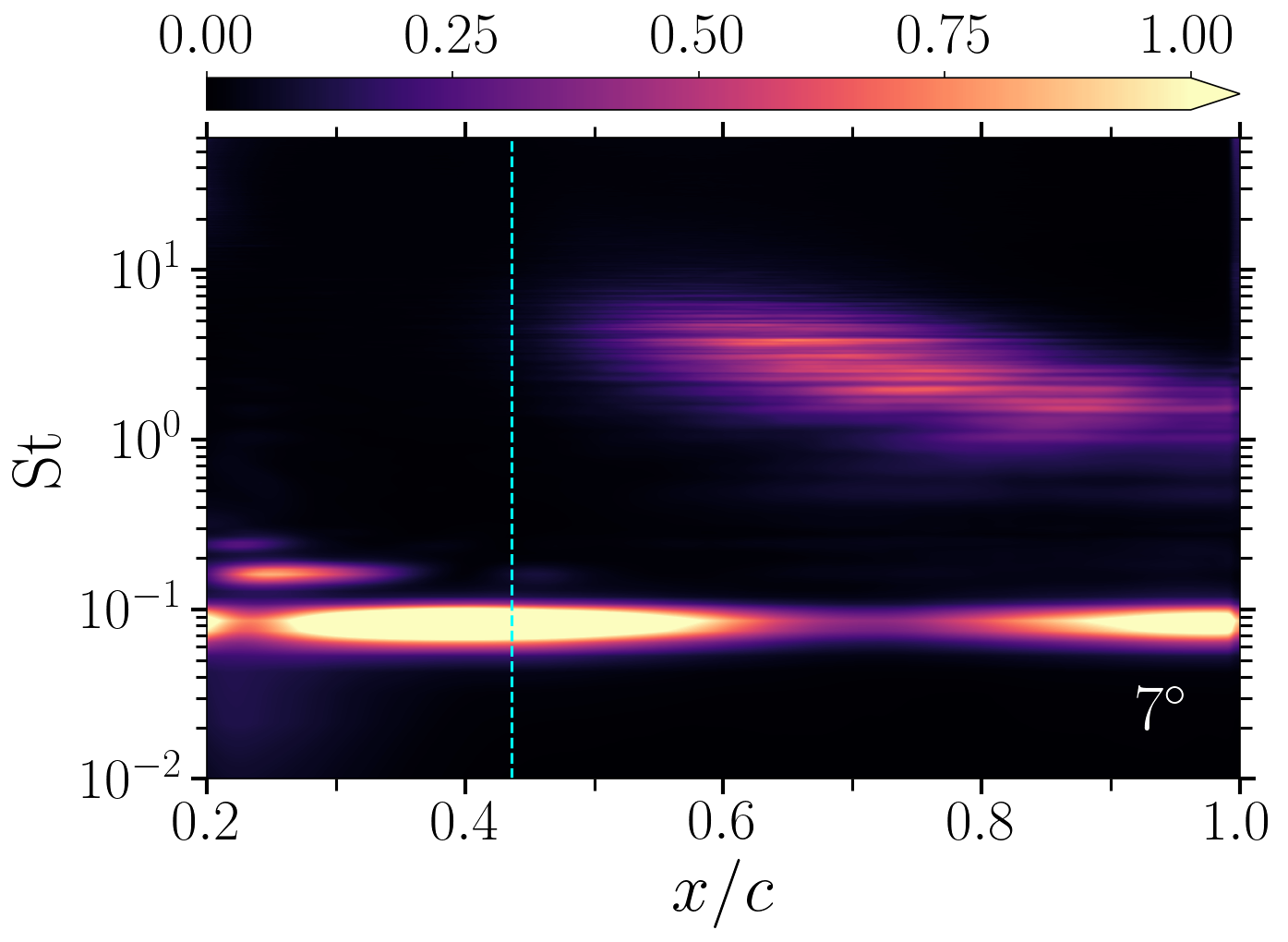}\\
(d)\includegraphics[width=.31\textwidth]{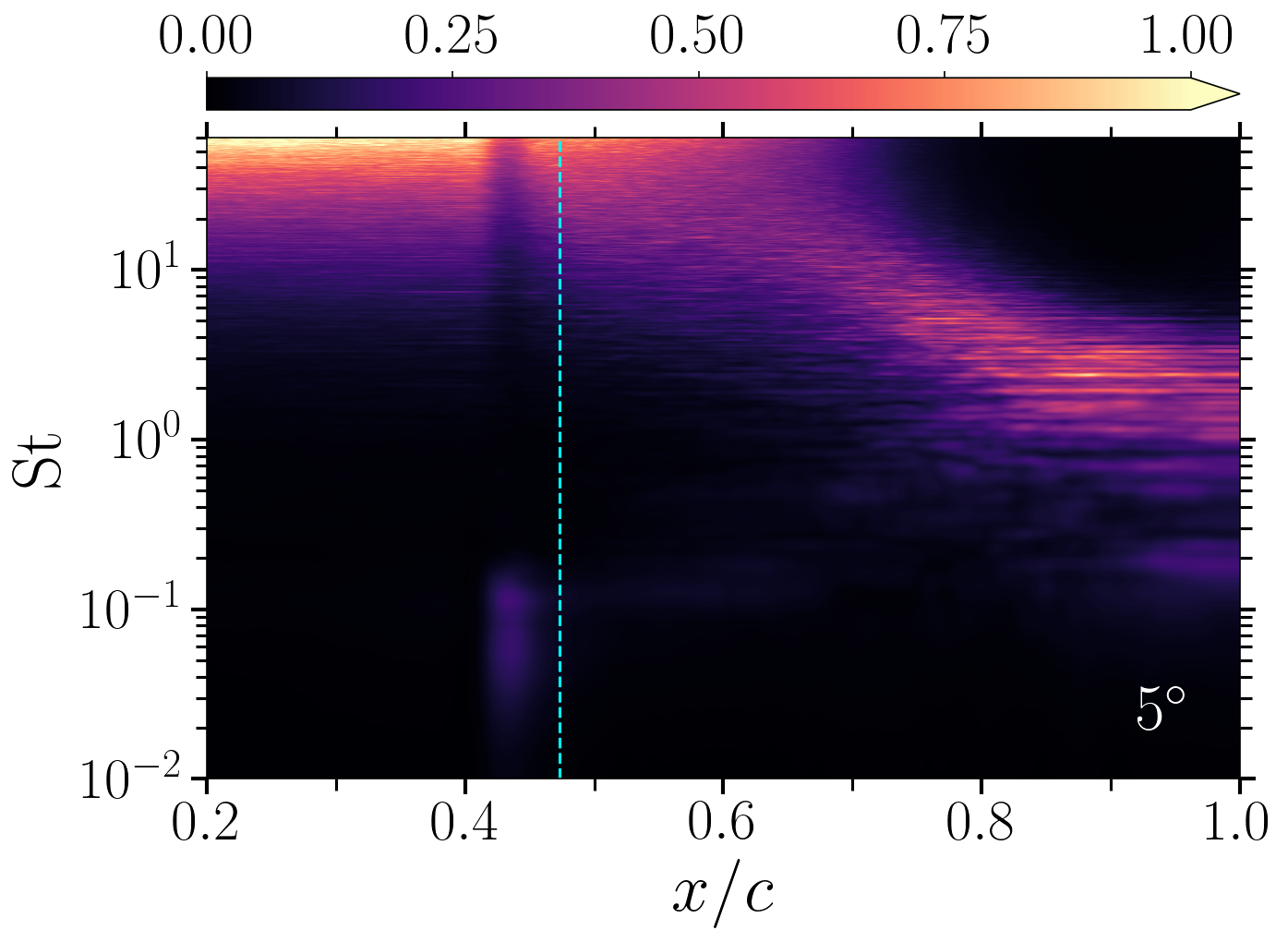}\hspace{-4mm}
(e)\includegraphics[width=.31\textwidth]{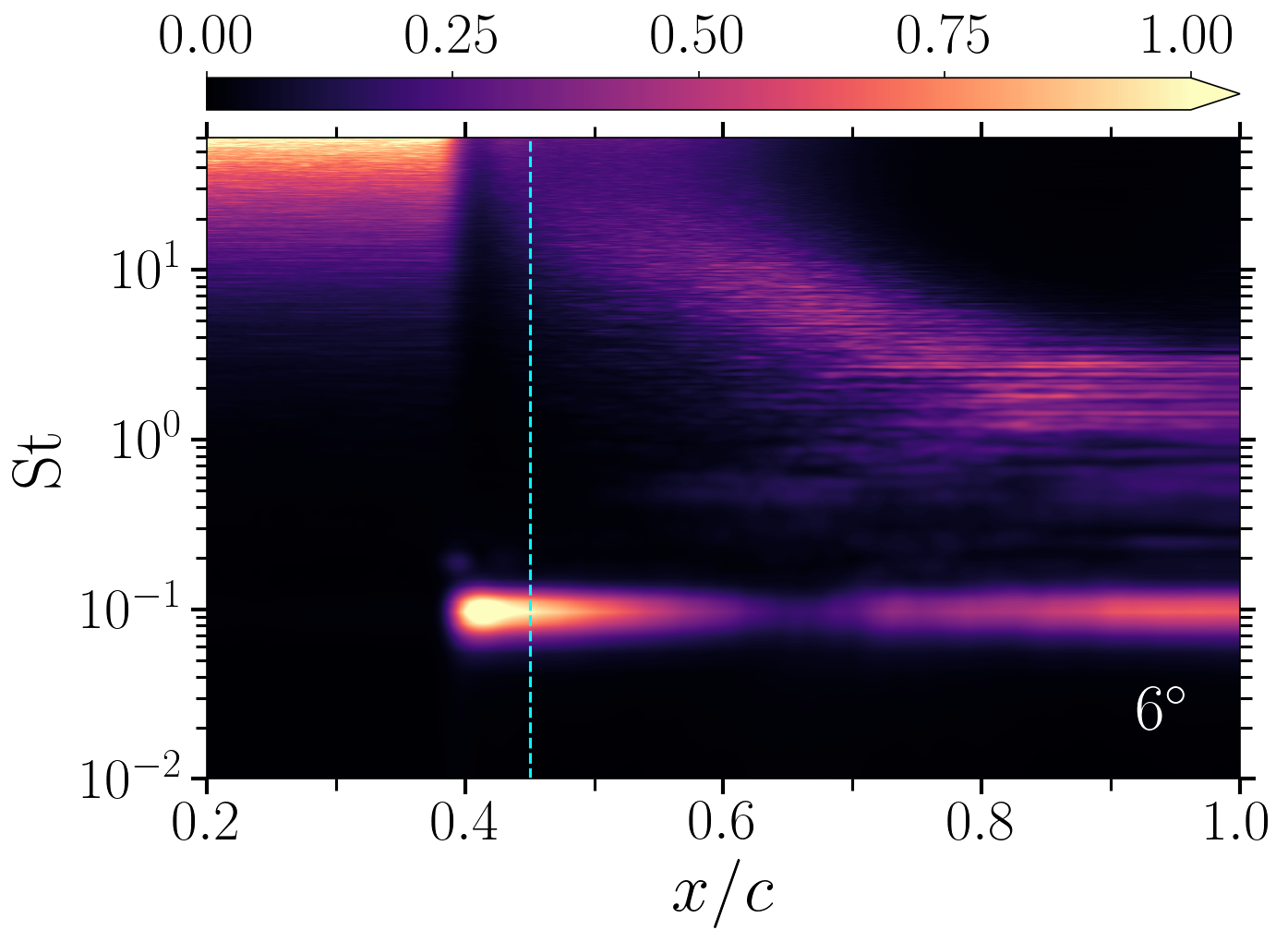}\hspace{-4mm}
(f)\includegraphics[width=.31\textwidth]{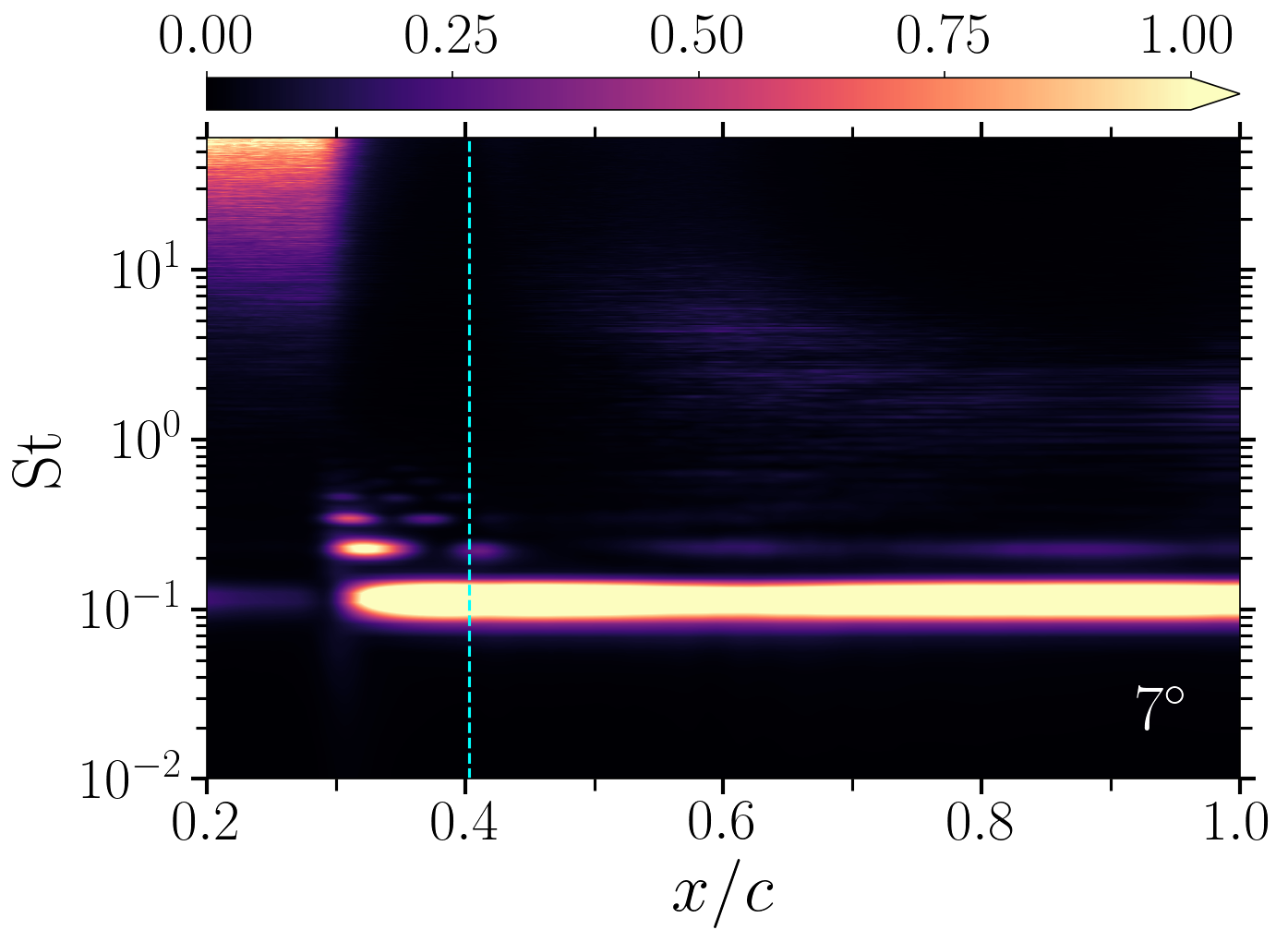}
\caption{
\blue{
Normalised and premultiplied PSD of the suction-side wall pressure 
as a function of the $x$-coordinate and Strouhal number for flow cases 
at $\Rey=3\times10^5$ (a, b, c) and $6\times10^5$ (d, e, f) at 
$\alpha=5^\circ$ (a, d), $6^\circ$ (b, e), $7^\circ$ (c, f). 
Cyan dashed lines denote the mean shock position. 
}
}
\label{fig:psdcp}
\end{figure}

%==============================================
The spectral characteristics of wall-pressure fluctuations on the
suction side are examined using pre-multiplied power spectral densities
(PSD). Normalising the spectra by the local wall-pressure variance 
facilitates a direct comparison of characteristic time scales across 
the flow field. Figure \ref{fig:psdcp} presents the PSD distributions 
for both the low-$\Rey$ cases (top panels) and the high-$\Rey$ cases 
(bottom panels). The inspected angles of attack correspond to 
pre-buffet ($\alpha=5^\circ$), incipient buffet ($\alpha=6^\circ$), 
and fully developed buffet ($\alpha=7^\circ$) cases. 

Under non-buffet flow conditions (figure \ref{fig:psdcp}a, d) the spectra 
are consistent with canonical SBLI behaviour. 
The signature of the incoming turbulent boundary layer is evident 
at high frequencies ($\St \approx 50$), approximately five times the
characteristic frequency of wall-shear stress fluctuations (not shown).
This difference reflects the fact that wall-pressure disturbances are
primarily associated with motions in the buffer layer, whereas
wall-shear fluctuations scale with the friction velocity.

Slightly upstream of the mean shock position, a spectral energy 
peak emerges at the buffet frequency ($\St \approx 0.1$), associated 
with the low-amplitude oscillations of the shock foot. 
\green{
The offset 
between this peak and the shock location stems from the upstream 
influence mechanism, whereby pressure perturbations propagate upstream 
through the subsonic near-wall layer and spread the pressure rise ahead 
of the shock. The rapid growth in displacement thickness deflects the 
streamlines, generating compression waves upstream of the shock 
\citep{babinsky2011shock}. 
}
While the low-frequency peak is narrow-band 
at $\Rey = 3\times10^5$ (figure~\ref{fig:psdcp}a), it extends toward 
lower frequencies at $\Rey = 6\times10^5$ (figure~\ref{fig:psdcp}d), 
reflecting a broader spectral footprint of the instability.

In the trailing-edge separation region, energy is concentrated 
at intermediate frequencies ($1 < \St < 4$), associated with the {\em wake 
mode} and its underlying vortex-shedding dynamics \citep{moise2022large}.
The incipient buffet case at $\alpha = 6^\circ$ (figure \ref{fig:psdcp}b, e) 
exhibits transitional spectral characteristics, featuring a sharp peak at the 
buffet frequency that is concentrated immediately upstream of the mean shock 
location and spatially spread around it. A peak at the same frequency emerges 
in the trailing-edge region (especially evident at $\Rey = 6\times10^5$),
indicating a spectral coupling between the onset of shock oscillations 
and the unsteadiness of the trailing-edge separation. 

\green{
As the buffet enters its fully developed regime (figure~\ref{fig:psdcp}c, f), 
this shock--trailing-edge spectral coupling 
intensifies and the dependence on the Reynolds number becomes more pronounced.
At $\Rey = 6\times10^5$ (figure~\ref{fig:psdcp}f), the buffet mode dominates 
the spectral energy to such an extent that the wake-mode signature is nearly 
suppressed, indicating a frequency lock-in of the global instability. 
Conversely, 
at $\Rey = 3\times10^5$ (figure~\ref{fig:psdcp}c), the higher-frequency 
signature persists 
at the trailing edge and modulates the shock dynamics, as evidenced above by 
the shock-position spectrum (see figure~\ref{fig:psdxs}a). Under these 
low-$\Rey$ conditions, the high-frequency broadband energy characteristic 
of fully developed turbulence is absent upstream of the shock. These features 
are consistent with the results of \citet{lusher2024effect}, who identified 
spectral peaks in the frequency range $1 < \St < 2$ only in 
transitional interactions, whereas these energy scales are suppressed in 
fully turbulent buffet regimes.
}

%==============================================
\subsection{Modal decomposition}
\label{sec:spod}
%==============================================

\begin{figure}
\centering
(a)\includegraphics[width=.47\textwidth]{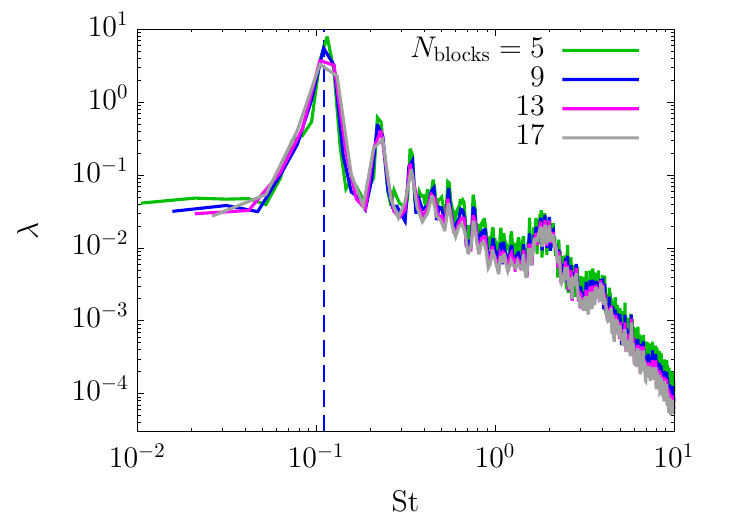}
(b)\includegraphics[width=.47\textwidth]{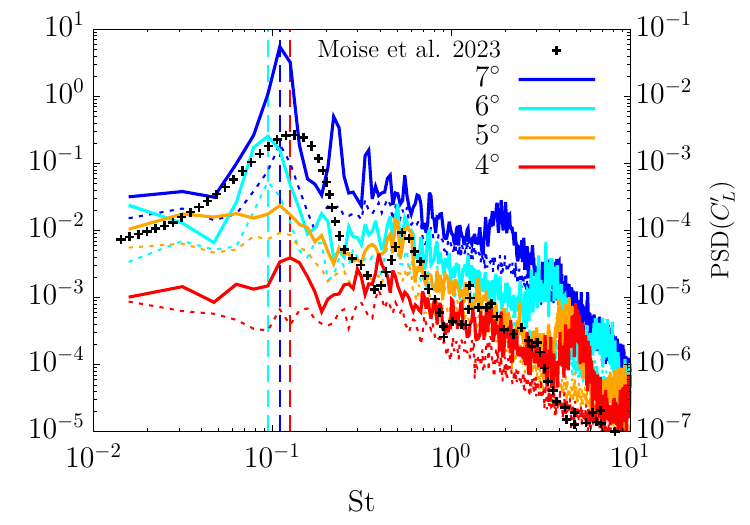}
\caption{
SPOD eigenvalue spectra for flow cases at $\Rey = 6 \times 10^5$, 
computed using a pressure-based norm \eqref{eq:pnorm}.
(a) Sensitivity of the leading eigenvalue $\lambda_1$ to the number of
realisations at $\alpha=7^\circ$. 
(b) Distribution of the first (solid lines) and second 
(dotted lines) eigenvalues using $N_\mathrm{blocks} = 9$.
Dashed lines mark the 
fundamental buffet frequencies ($\St \approx 0.09 \text{ -- } 0.11$). 
Results are compared with the lift spectrum of the V2C airfoil 
at $\alpha = 5^\circ$ and $M = 0.735$ \citep{moise2023transonic}. 
}
\label{fig:spod}
\end{figure}

Spectral proper orthogonal decomposition (SPOD) is employed  
to characterise the space--time coherence of the buffet instability. 
\green{
This section and the remainder of the paper (unless otherwise stated) 
consider the $\Rey = 6\times10^5$ cases only, 
thereby focusing on fully turbulent interactions and 
providing a consistent basis for comparison with other studies.
}
As established by \citet{towne2018spectral}, the SPOD provides 
a set of modes that are mutually orthogonal at each frequency and 
optimally represent the energy (variance) of the flow field.
The technique is rooted in the filtering theory of
\citet{lumley1970stochastic} and is mathematically equivalent to
performing a spatial POD on the cross-spectral density tensor.
\blue{
A key advantage of SPOD over traditional POD is its ability to 
decompose the flow field into frequency-specific modes, which 
enables coherent spatial structures associated with dominant 
frequencies to be isolated. 
}

\red{
SPOD is performed on the fluctuating pressure field, which is extracted from 
two-dimensional snapshots in the $(x,y)$ plane at $z = 0$. The sampling duration 
and frequency follow the procedures described in \S\ref{sec:method-time}.
The pressure-based inner product between two given spatial snapshots, 
represented as column vectors $\mathbf{p}_1$ and $\mathbf{p}_2$, is defined as 
}
\begin{equation}
\langle \mathbf{p}_1, \mathbf{p}_2 \rangle = \mathbf{p}_2^H \, W \, \mathbf{p}_1, 
\quad W = \int_A \mathrm{d}A, 
\label{eq:pnorm}
\end{equation}
\red{
where the superscript $H$ denotes the Hermitian, $W$ is the weight matrix, and 
$\mathrm{d}A$ is the local area of each grid cell. Following \citet{schmidt2020guide}, 
the weight matrix represents a weighted 2-norm. 
}

Given the sensitivity of spectral estimates to windowing choices, the
number of realisations is varied to ensure that the identified modes and
their associated energy levels are independent of the decomposition
parameters. Figure \ref{fig:spod}(a) shows the sensitivity of the SPOD 
eigenvalue spectrum to the number of segments ($N_\mathrm{blocks} = 5, 9, 13, 17$), 
for the fully developed buffet case. A satisfactory collapse of the spectral 
energy is observed across all configurations, indicating robustness with 
respect to the windowing. Accordingly, $N_\mathrm{blocks} = 9$ is selected as 
a compromise between spectral resolution and statistical convergence.

\red{
The SPOD eigenvalues at all angles of attack are 
shown in figure \ref{fig:spod}(b), where solid lines refer to 
the leading eigenvalues ($\lambda_1$) and dotted lines to 
the second eigenvalues ($\lambda_2$). 
For all angles of attack, $\lambda_1$ peaks at the fundamental buffet 
frequency ($\St \approx 0.1$). This peak grows by approximately one order 
of magnitude with each one-degree increase from $\alpha = 4^\circ$ to $7^\circ$, 
reflecting the transition from the pre-buffet to the fully developed state. 
In the latter regime, the second and third harmonics of the 
fundamental frequency also emerge, 
consistent with the shock-position spectrum (see figure~\ref{fig:psdxs}b). 
}

\red{
Wake modes associated with trailing-edge vortex shedding occur at 
higher frequencies ($2 < \St < 5$). 
Within this frequency band, the leading eigenvalue is an order of magnitude 
greater than the subdominant mode, indicating low-rank behaviour 
associated with the dominant shear-layer instability.
}

\red{
An additional broadband peak at intermediate frequencies 
($\St \approx 0.5$) is revealed by the eigenvalue spectra. 
Although this peak is quite evident at other angles of attack, 
at $\alpha=7^\circ$ it overlaps with the expected fifth harmonic 
of the buffet frequency, which makes it difficult to determine 
whether a distinct physical mechanism is at play.  
For cross-validation, figure \ref{fig:spod}(b) 
overlays the lift spectrum reported by \citet{moise2023transonic} for the V2C 
airfoil under buffet conditions. Their data 
exhibit a corresponding energy peak at $\St \approx 0.5$ in the absence of 
higher buffet harmonics, supporting the presence of an independent physical process.
}

\begin{figure}
\centering
(a)\includegraphics[width=.3\textwidth]{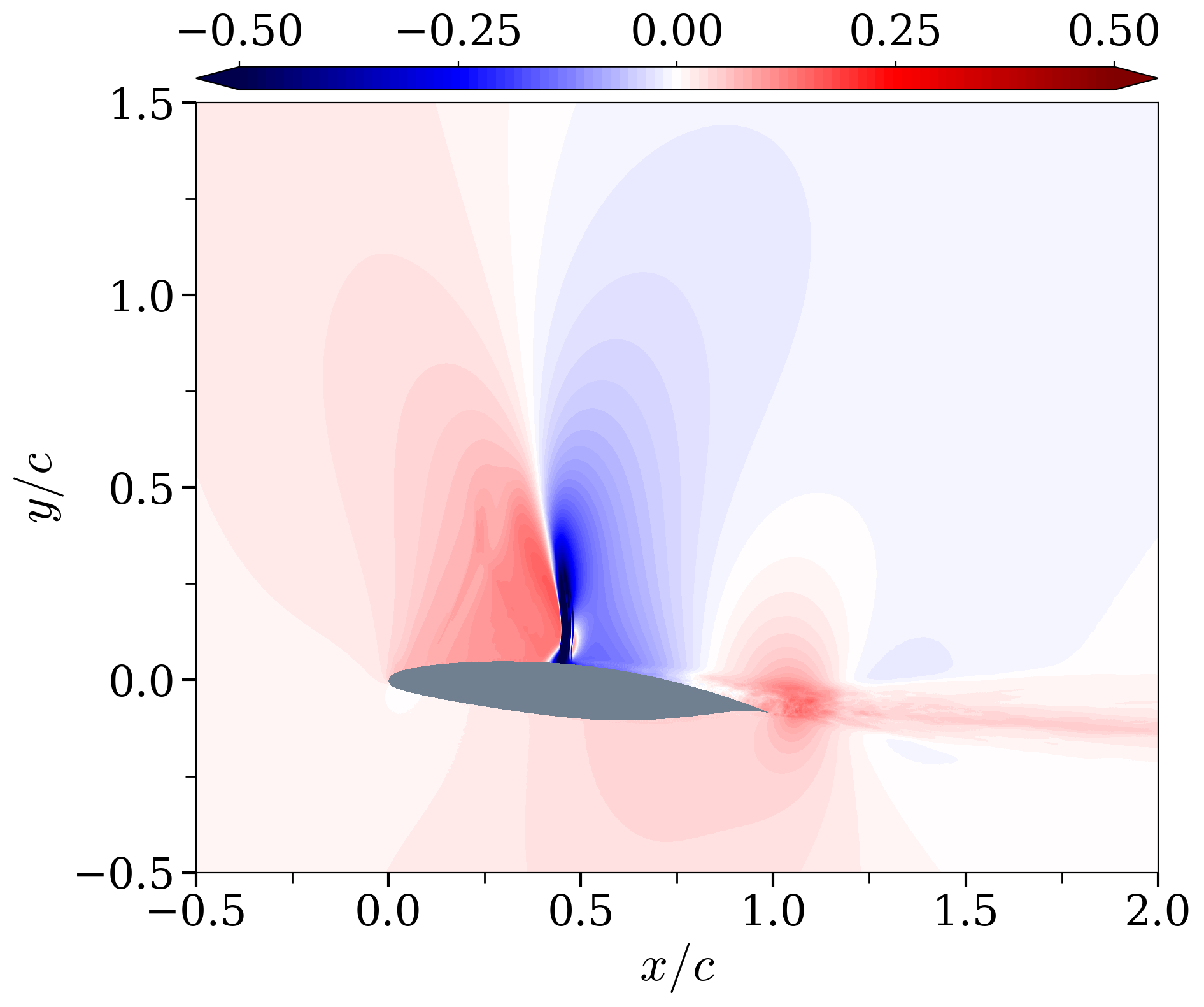}
(b)\includegraphics[width=.3\textwidth]{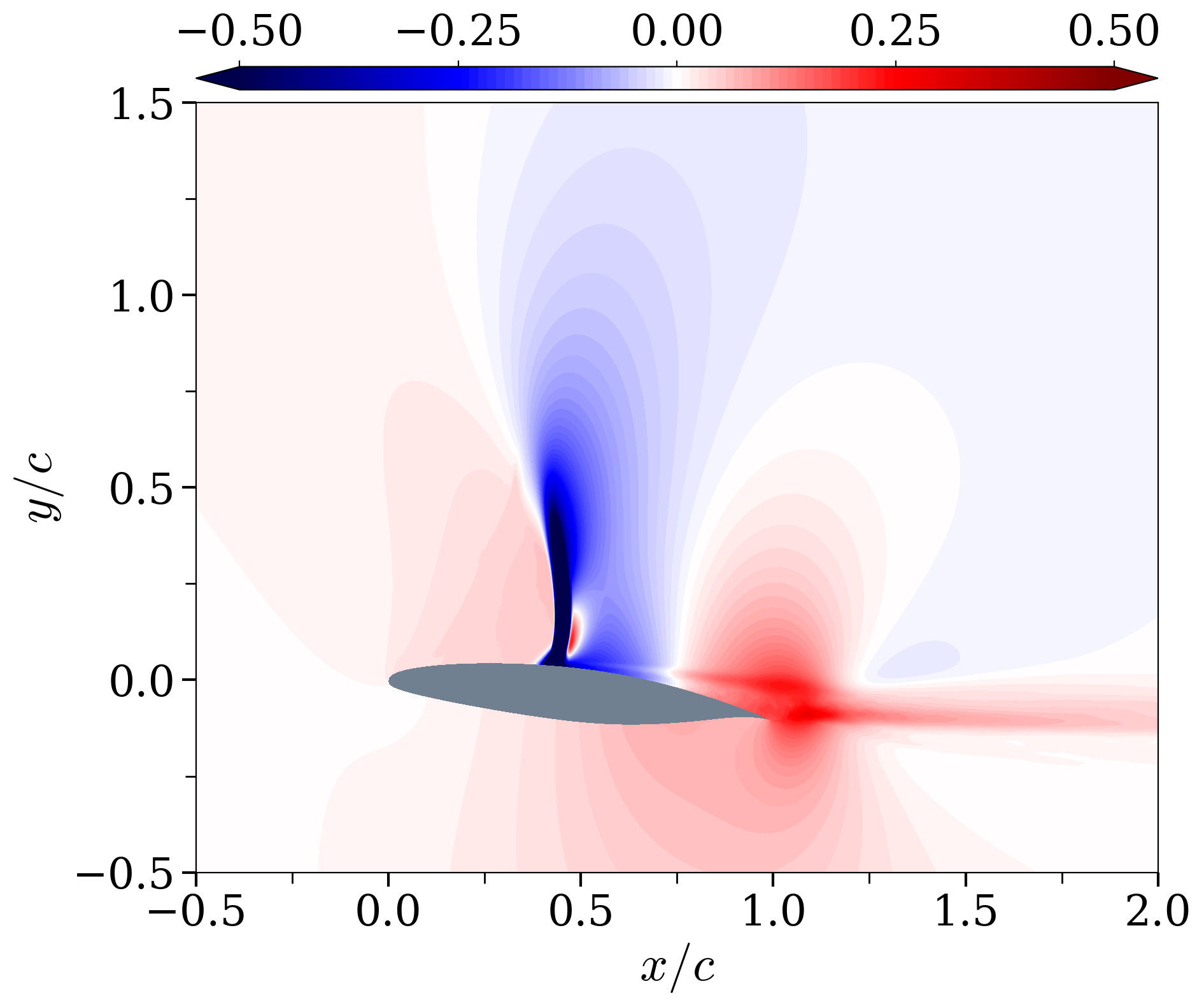}
(c)\includegraphics[width=.3\textwidth]{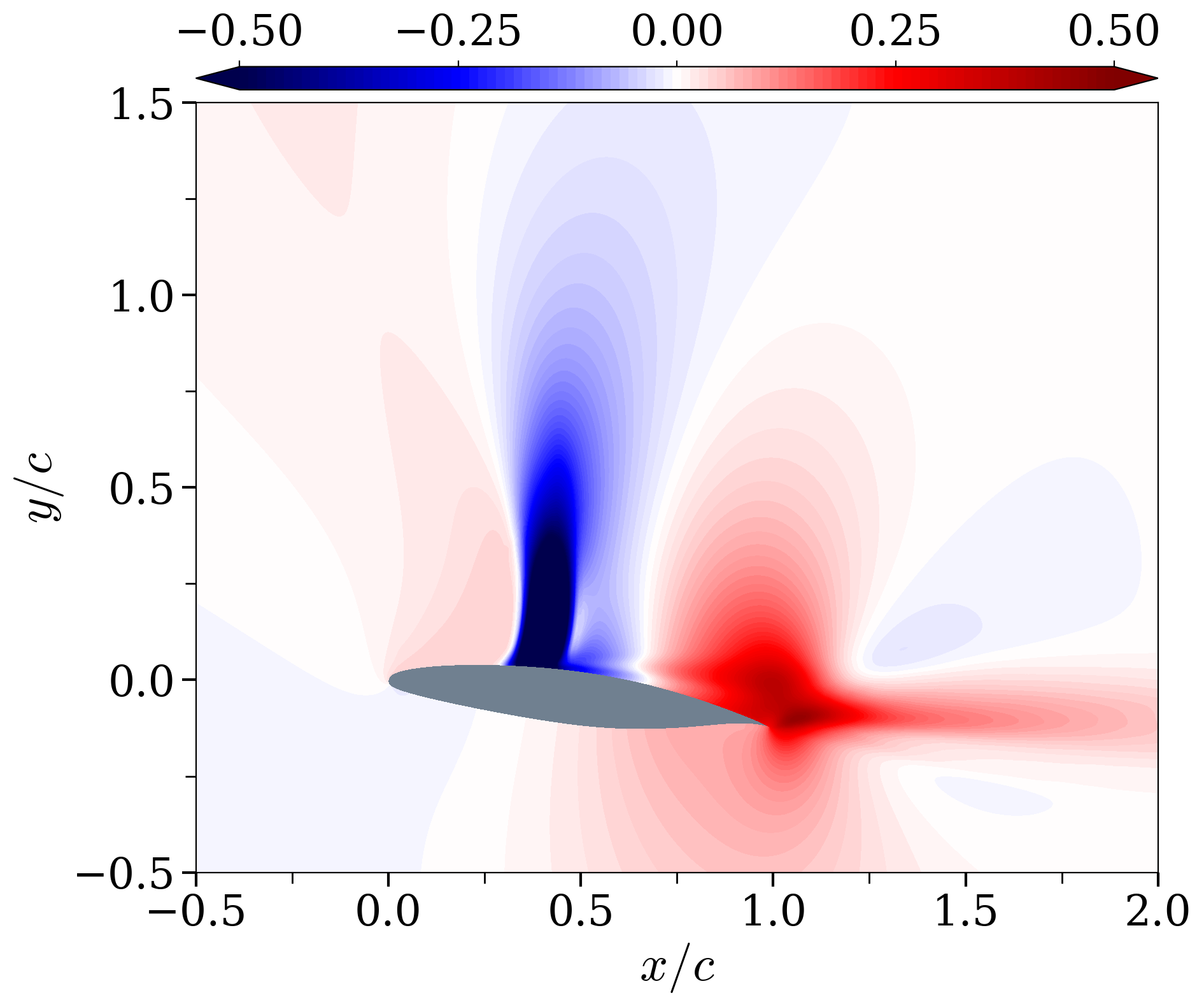}\\
(d)\includegraphics[width=.3\textwidth]{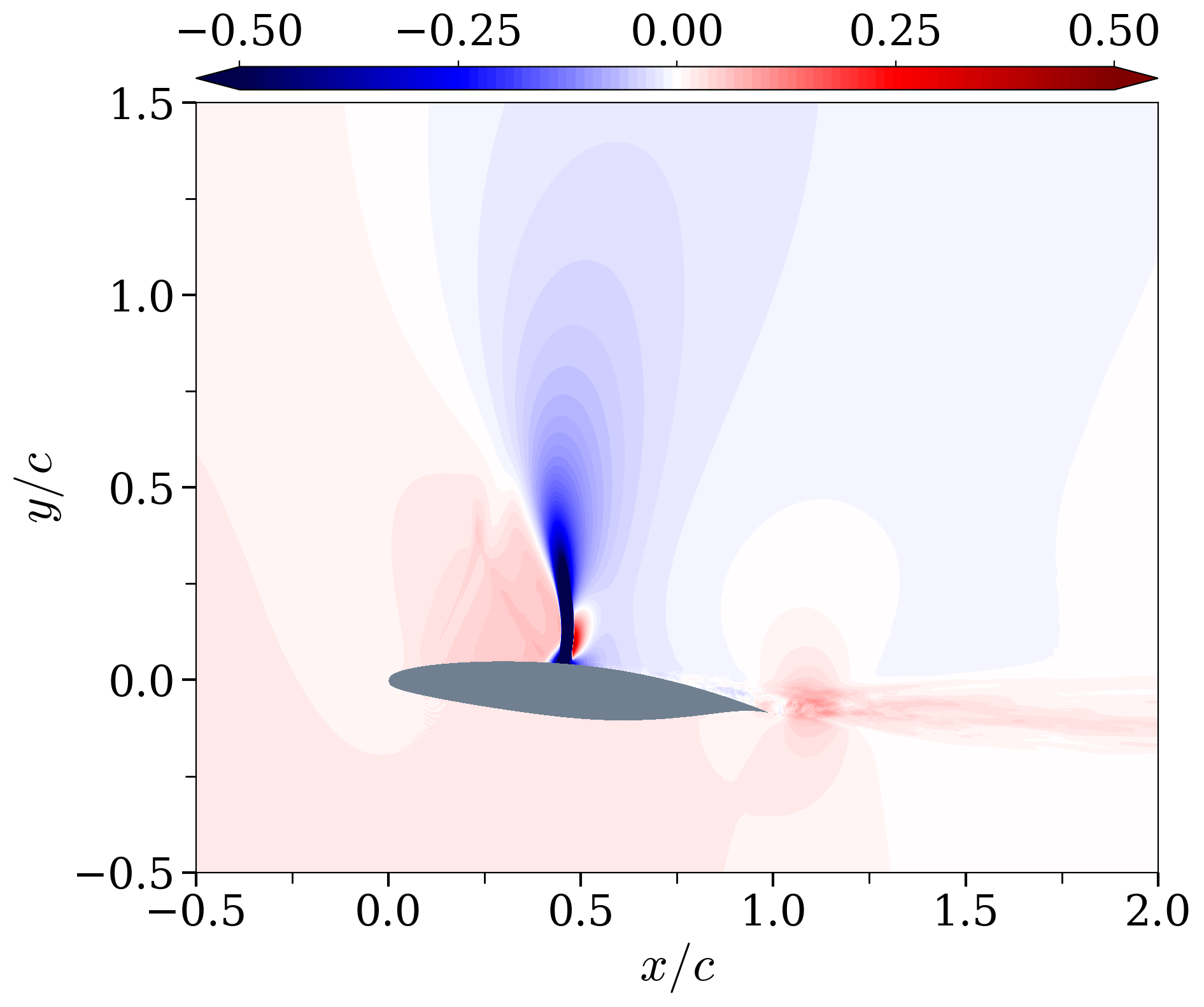}
(e)\includegraphics[width=.3\textwidth]{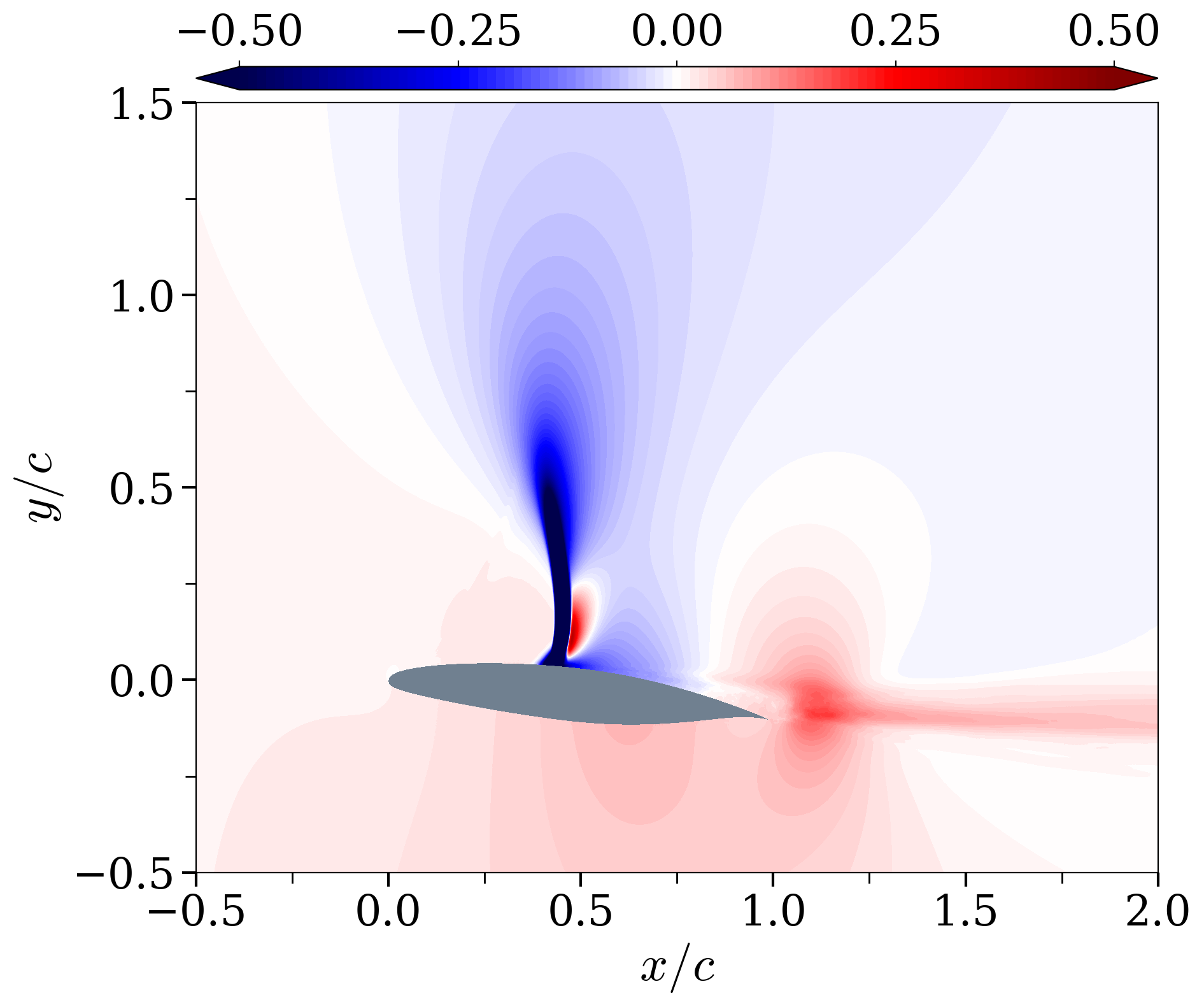}
(f)\includegraphics[width=.3\textwidth]{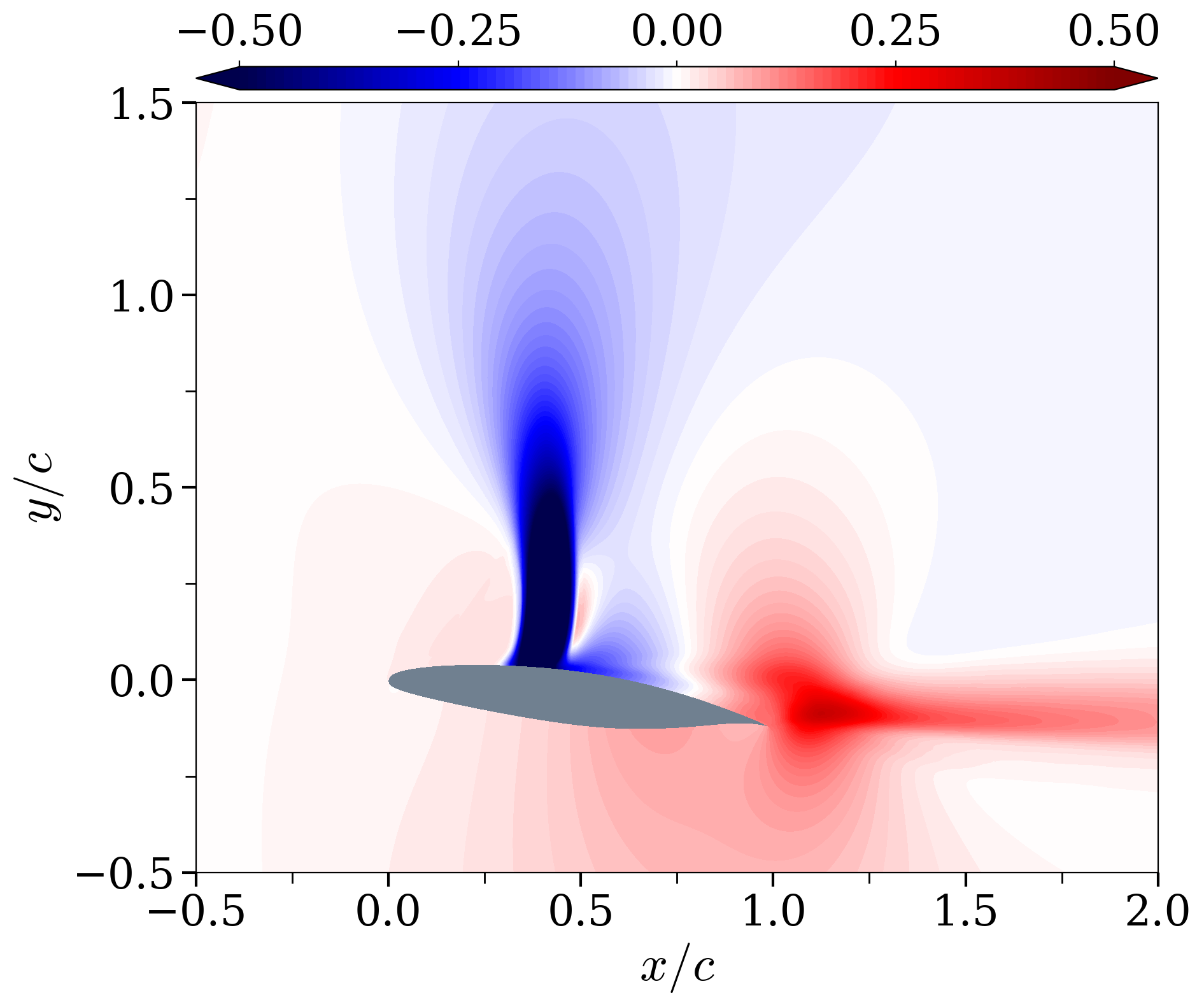}\\
(g)\includegraphics[width=.3\textwidth]{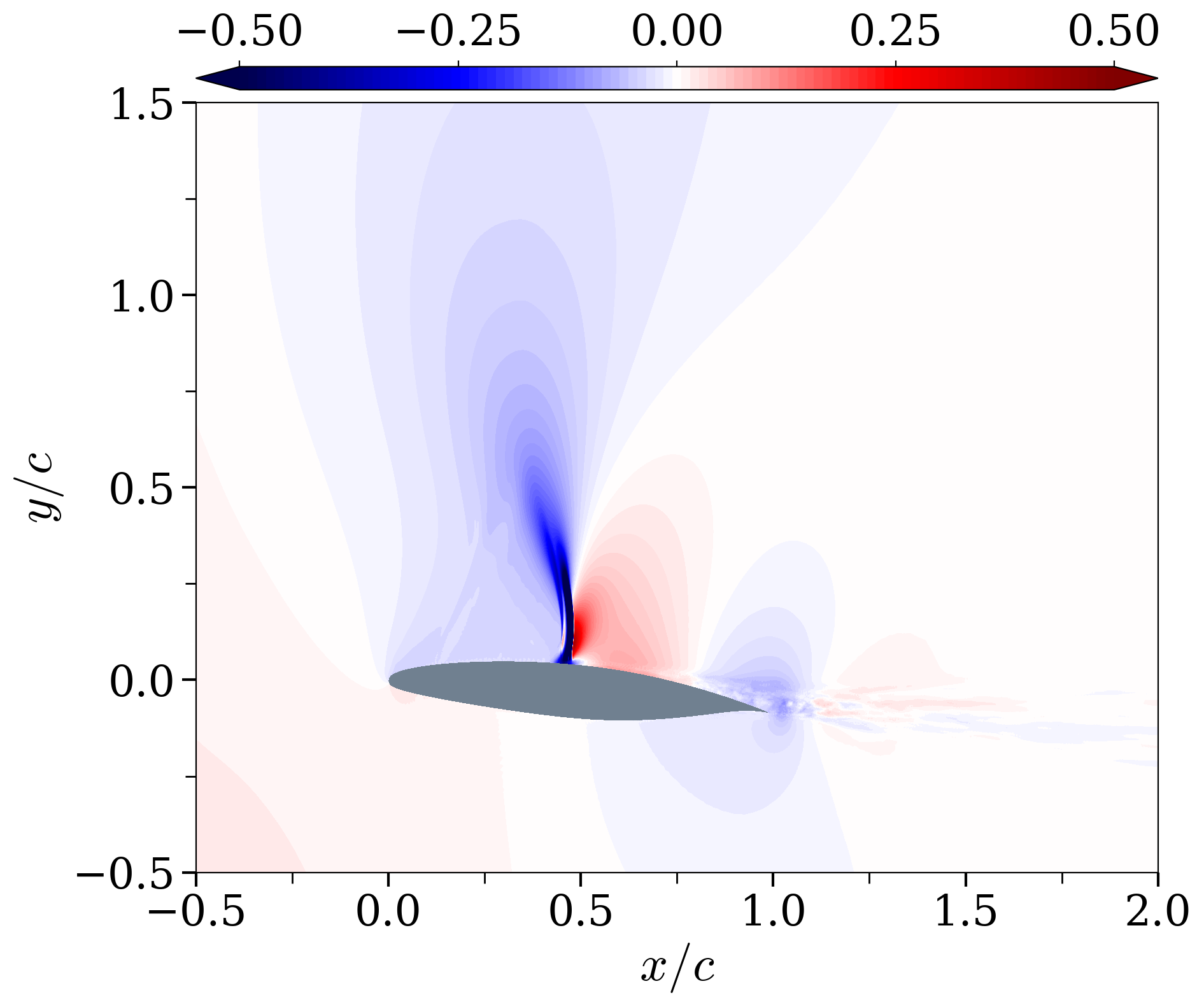}
(h)\includegraphics[width=.3\textwidth]{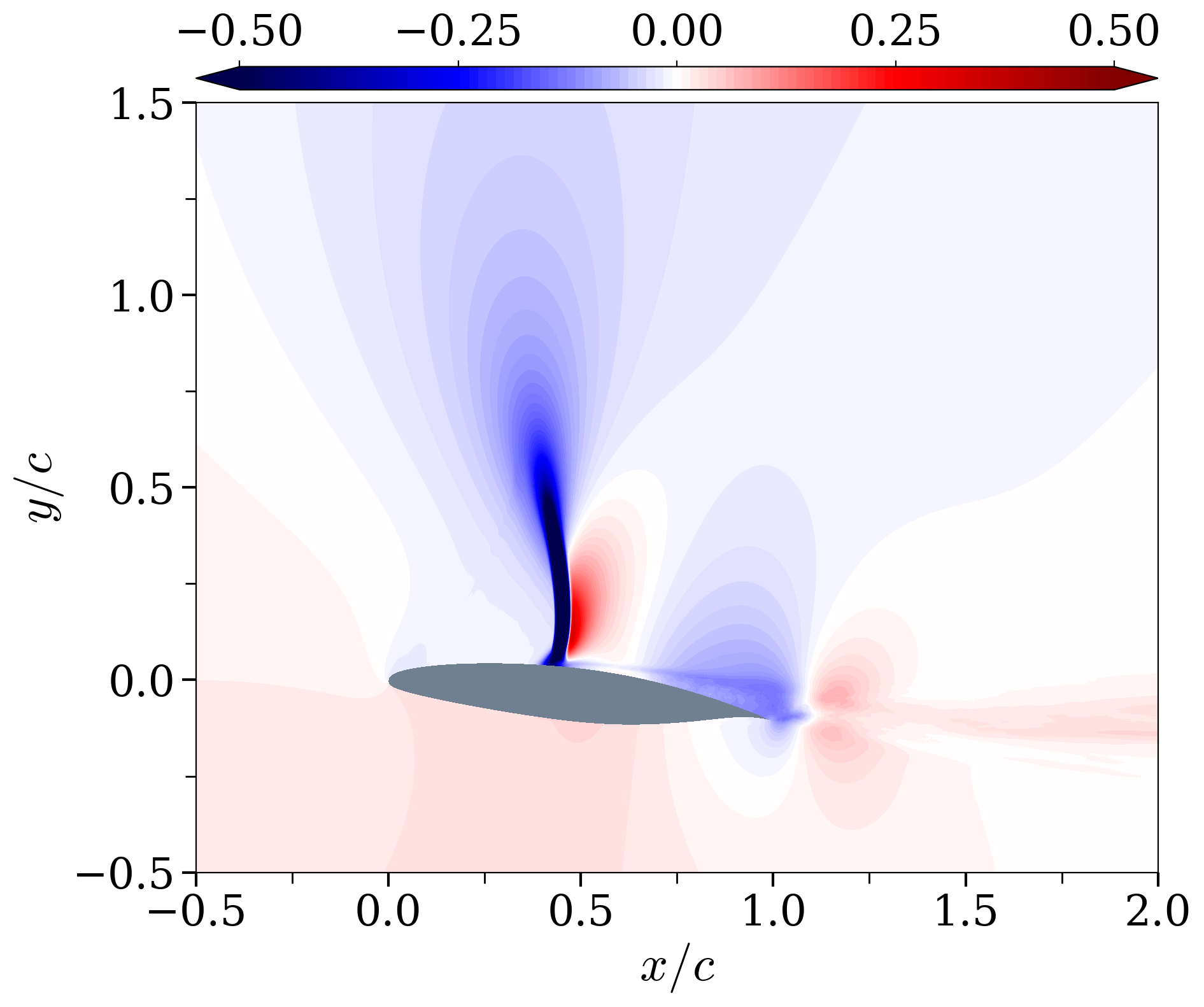}
(i)\includegraphics[width=.3\textwidth]{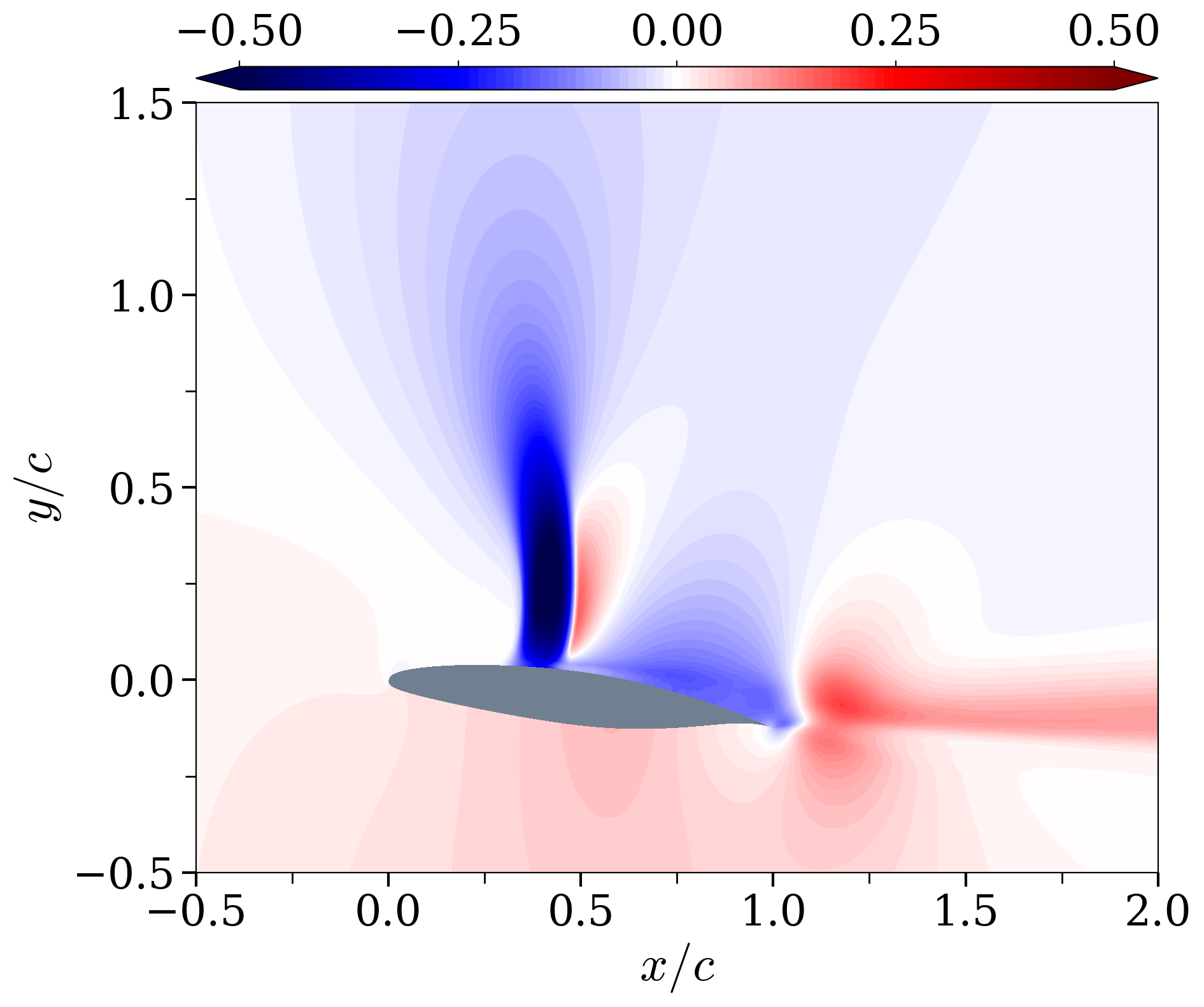}\\
(j)\includegraphics[width=.3\textwidth]{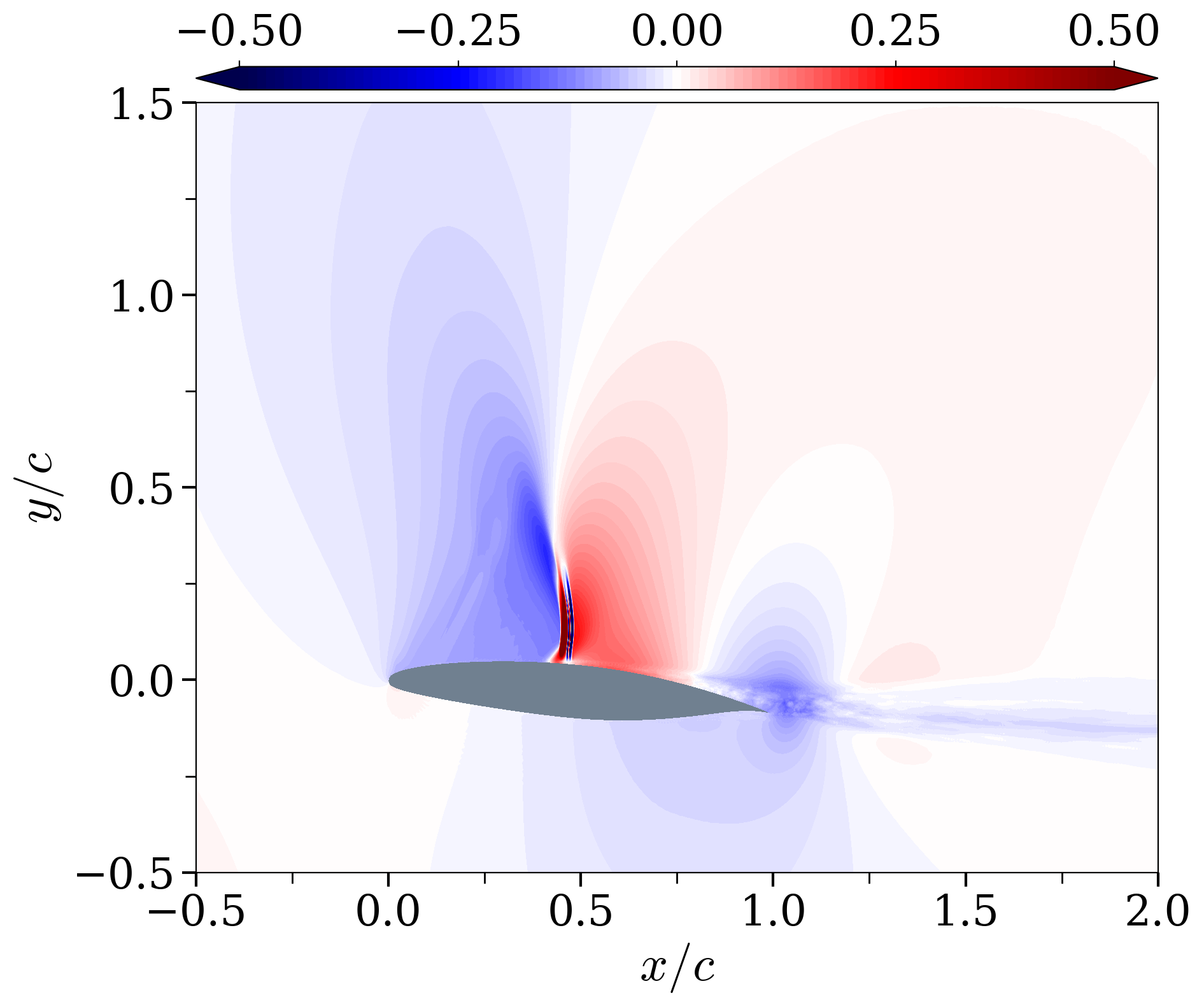}
(k)\includegraphics[width=.3\textwidth]{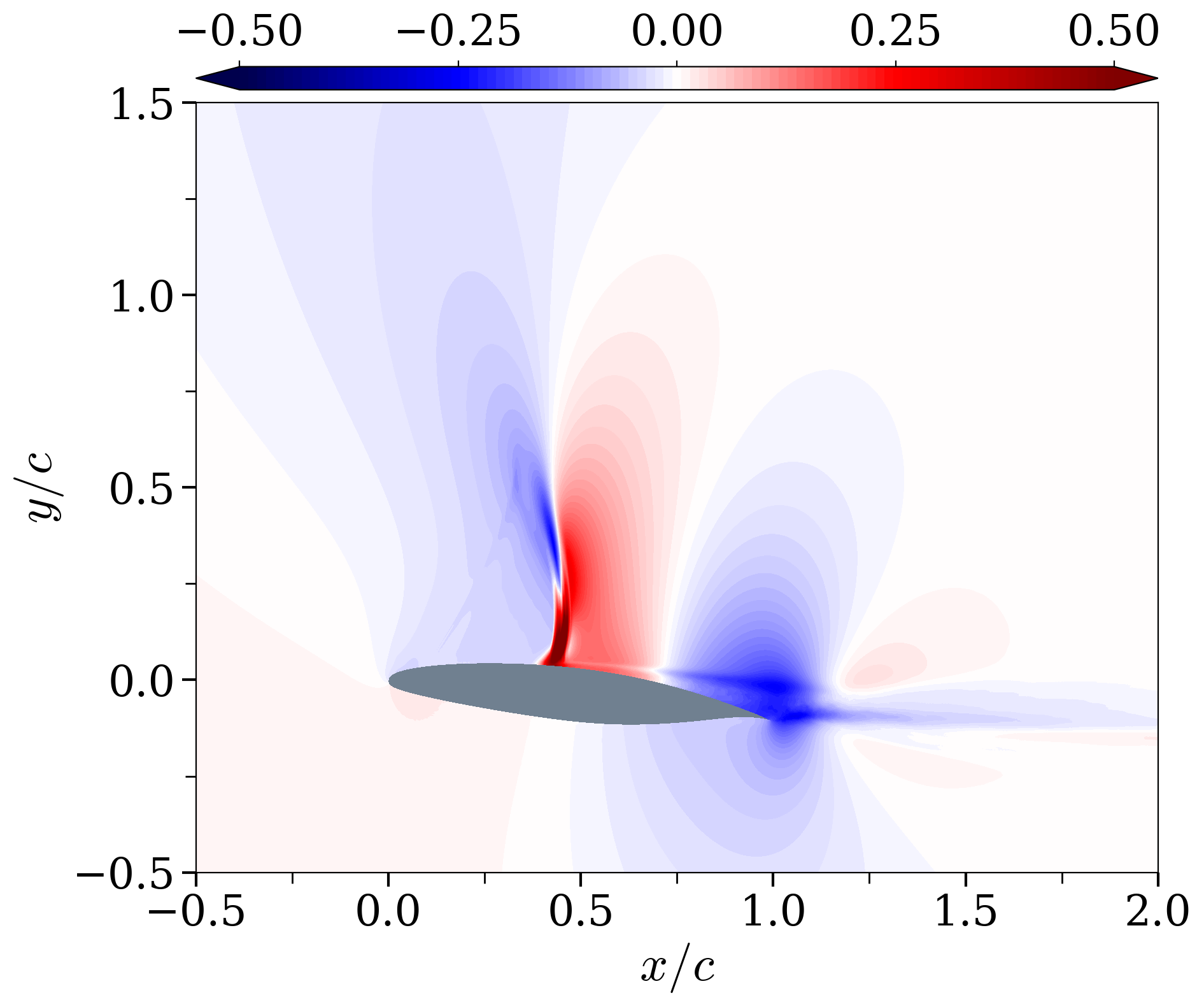}
(l)\includegraphics[width=.3\textwidth]{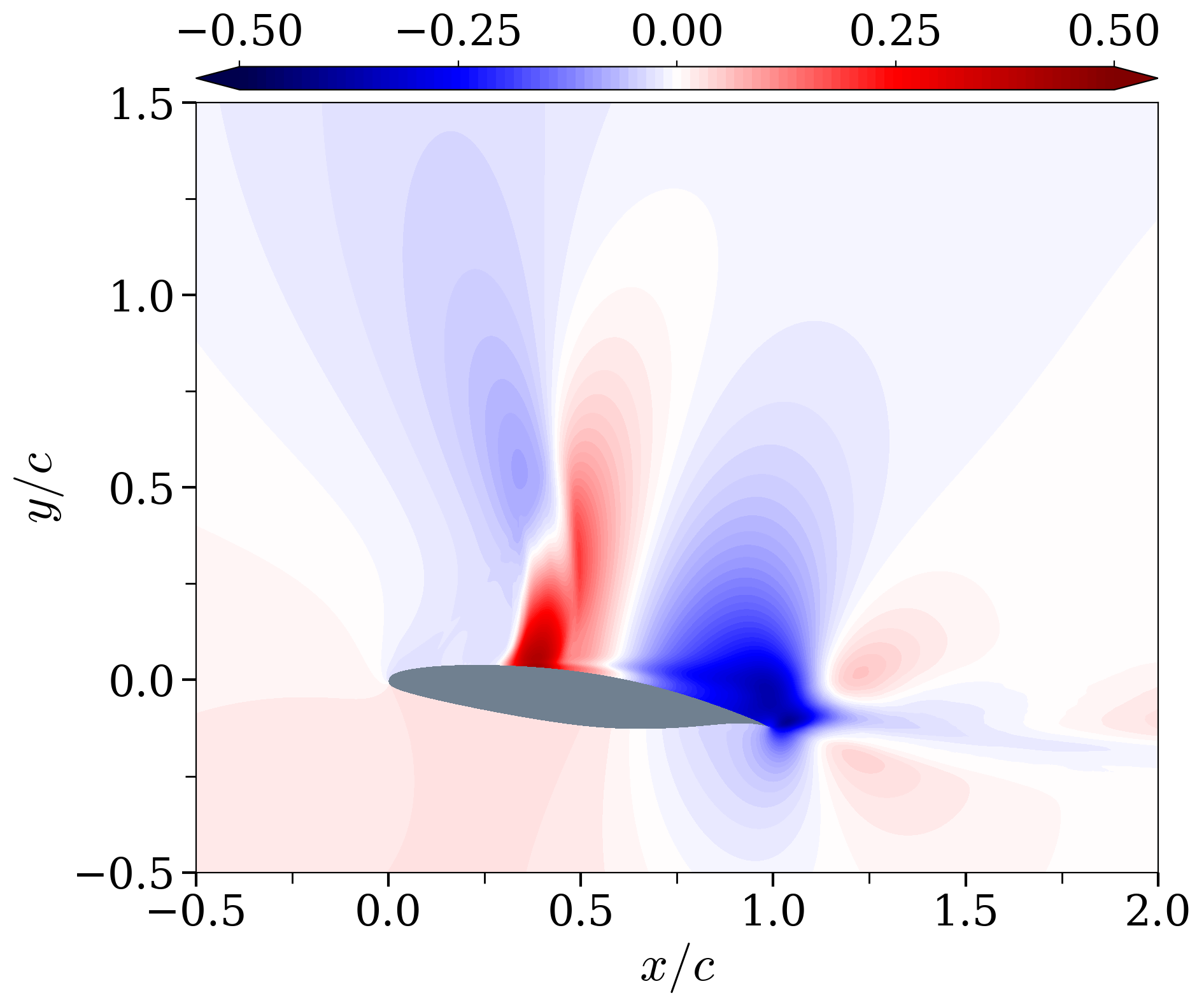}\\
(m)\includegraphics[width=.3\textwidth]{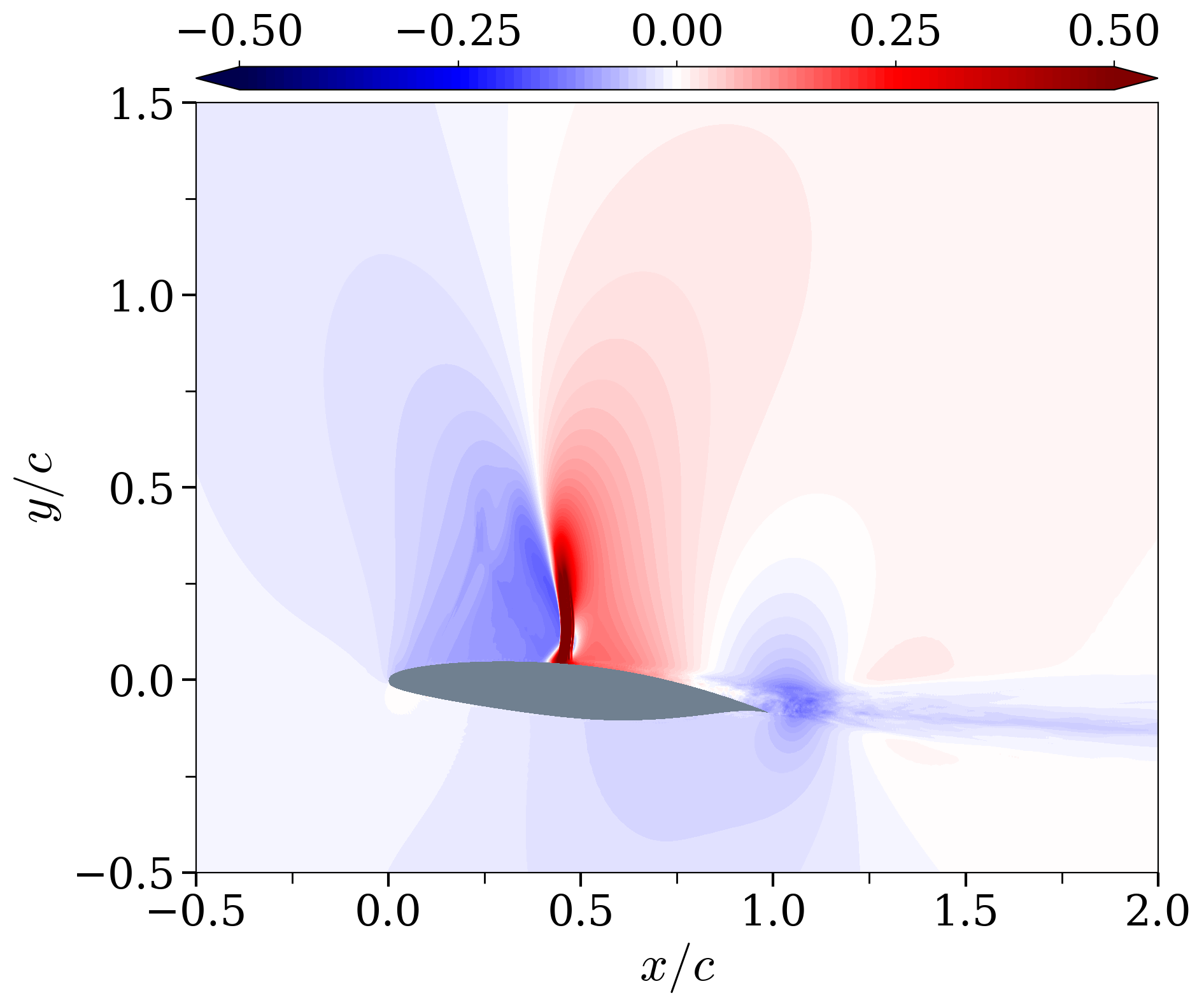}
(n)\includegraphics[width=.3\textwidth]{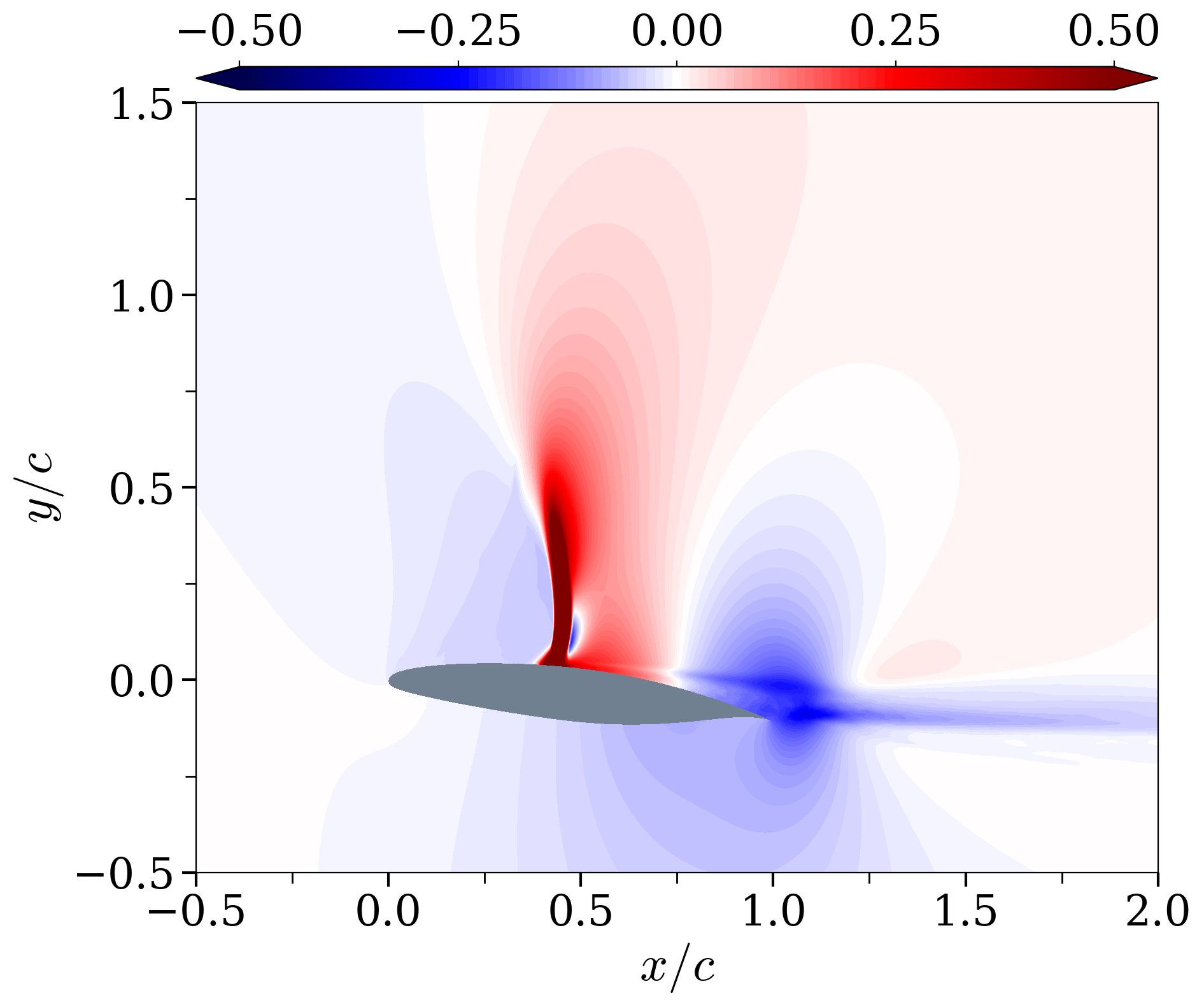}
(o)\includegraphics[width=.3\textwidth]{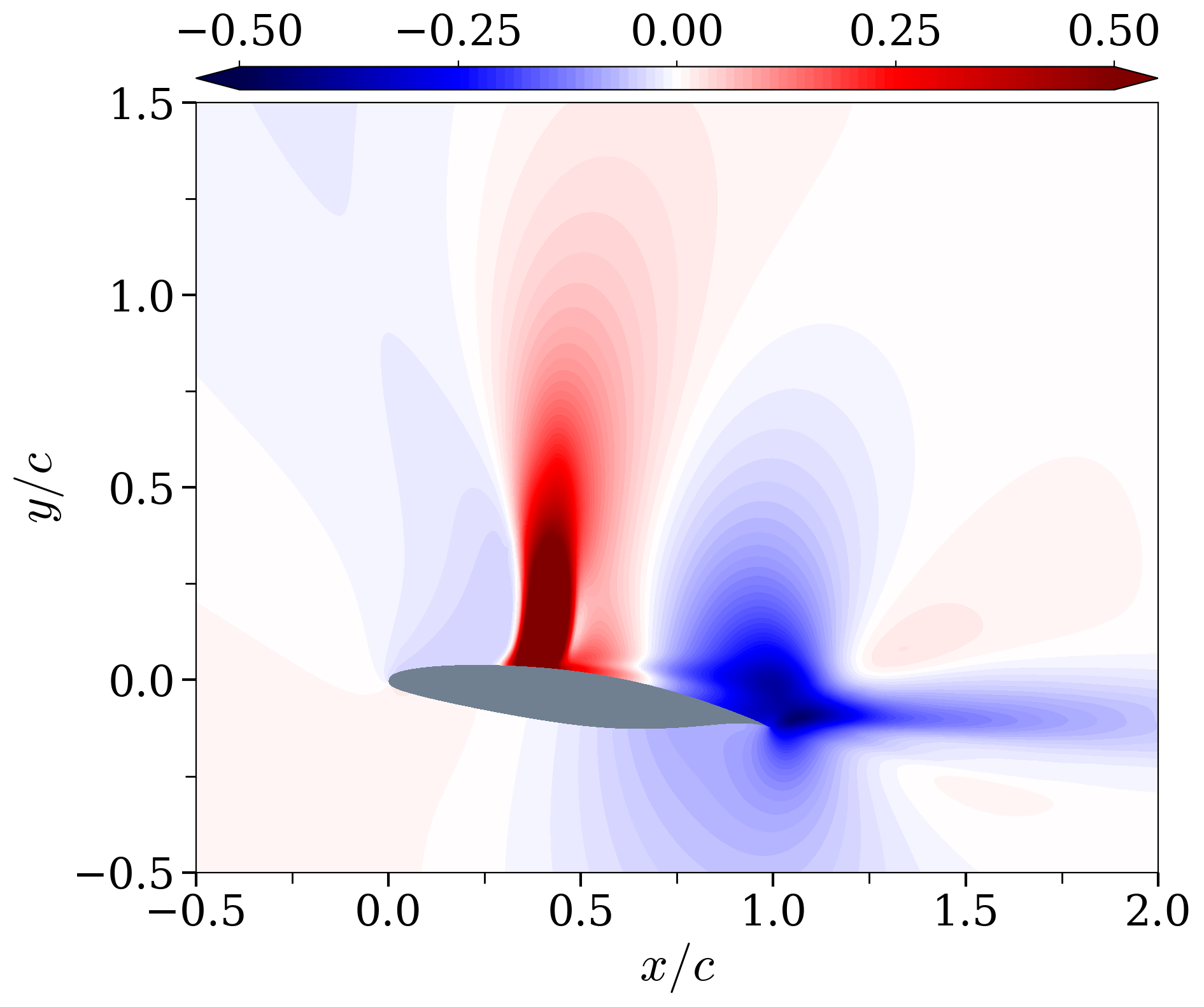}
\caption{
Phase-reconstructed pressure fluctuations of the leading SPOD
mode at the buffet frequency ($\St \approx 0.1$) 
for flow cases at $\Rey = 6 \times 10^5$. Left, middle, and right 
columns refer to $\alpha = 5^\circ$, $6^\circ$, and $7^\circ$, 
respectively. The real part of the phase-conditioned mode is shown 
at phase angles $\varphi = -\pi/2$ (a--c), $\varphi = -\pi/4$ (d--f), 
$\varphi = 0$ (g--i), $\varphi = \pi/4$ (j--l), and 
$\varphi = \pi/2$ (m--o).
}
\label{fig:mode} 
\end{figure}

The spatial shape of the leading SPOD mode at the buffet frequency 
is examined in figure \ref{fig:mode}, where each column refers to 
an angle of attack ($\alpha = 5^\circ$, $6^\circ$, $7^\circ$), 
and each row represents a phase angle ($\varphi = 2\pi \St \, t u_0 / c$) 
within the buffet cycle. The selected phase angles span from 
$\varphi = -\pi/2$ (maximum downstream excursion of the shock) 
to $\varphi = \pi/2$ (maximum upstream excursion). The zero-phase angle has been 
selected consistently with that used in figure \ref{fig:sch2}. 

\red{
By comparing the rows individually, it is evident that the spatial 
structure of the global buffet mode is essentially unchanged 
from the pre-buffet to the fully developed regime, with its 
frequency remaining close to the fundamental throughout the 
transition (see figure \ref{fig:spod}b). This is consistent 
with \citet{poplingher2019modal}, who identified buffet-like 
modes in the pre-buffet regime with frequencies close to the 
fundamental and damping rates approaching zero at the onset point. 
}

\red{
The persistence of both mode shape and fundamental frequency across 
the onset indicates that a single eigenmode, with an amplitude that 
varies smoothly with $\alpha$, captures the dominant dynamics throughout 
the transition. 
Below onset, \citet{nitzsche2009numerical} showed that small perturbations
excite a damped aerodynamic resonance associated with the dominant
complex eigenvalue. As $\alpha$ increases, the real part of this
eigenvalue approaches zero, reducing the damping until the mode becomes
unstable at onset. Beyond onset, nonlinear effects limit the modal
growth and lead to a self-sustained limit cycle, consistent with the
weakly nonlinear formulation of \citet{crouch2024weakly} and the
reduced-order model of \citet{sansica2022system}. 
}

\red{
The spatiotemporal evolution of the leading SPOD mode during the 
semi-cycle is consistent with the pressure eigenmode associated with 
the global instability reported by \citet{crouch2009global, crouch2009origin}, 
as well as with the periodic low-frequency pressure fluctuations described by 
\citet{fukushima2019self} and \citet{iwatani2022pod}. 
At $\varphi = -\pi/2$ (figure \ref{fig:mode}a--c), the shock is at its 
maximum downstream position, and the shock region is characterised by a 
coherent negative pressure fluctuation. During its upstream motion 
(figure \ref{fig:mode}d--f), the shock strength increases and a positive 
disturbance originates near the base of the shock. At the same time, 
a negative disturbance associated with shock-induced separation is 
generated at the shock foot, closer to the wall, and extends further downstream. 
}

\red{
At $\varphi = 0$, the positive 
disturbance propagates upwards along the shock towards the upper 
boundary of the sonic line. The negative disturbance exhibits 
flow-dependent features: in the pre-buffet condition 
(figure \ref{fig:mode}g), it dissipates past the shock, 
where the mean flow is attached, and re-emerges at the trailing-edge 
separated zone. In the fully developed buffet state 
(figure \ref{fig:mode}i), it spans the entire post-shock 
region, consistent with a fully separated boundary layer. 
The incipient buffet case (figure \ref{fig:mode}h) represents 
an intermediate state. 
}

\red{
As the shock continues to move upstream (figure \ref{fig:mode}j--l), 
this coherent negative disturbance convects downstream, amplifying 
along the shear layer toward the trailing edge.
At $\varphi = \pi/2$ (figure \ref{fig:mode}m--o), the shock reaches its 
maximum upstream position, reversing the pattern observed at $\varphi = -\pi/2$. 
A coherent positive disturbance now appears in the shock region, while the 
negative disturbance peaks in intensity and coherence in the near wake. 
}
The anti-phase relationship between the shock and trailing-edge 
regions reflects the coupled evolution of shock motion and separation dynamics: 
as the shock moves upstream, the separated region thickens, limiting pressure 
recovery over the suction side.

\begin{figure}
\centering
(a)\includegraphics[width=.3\textwidth]{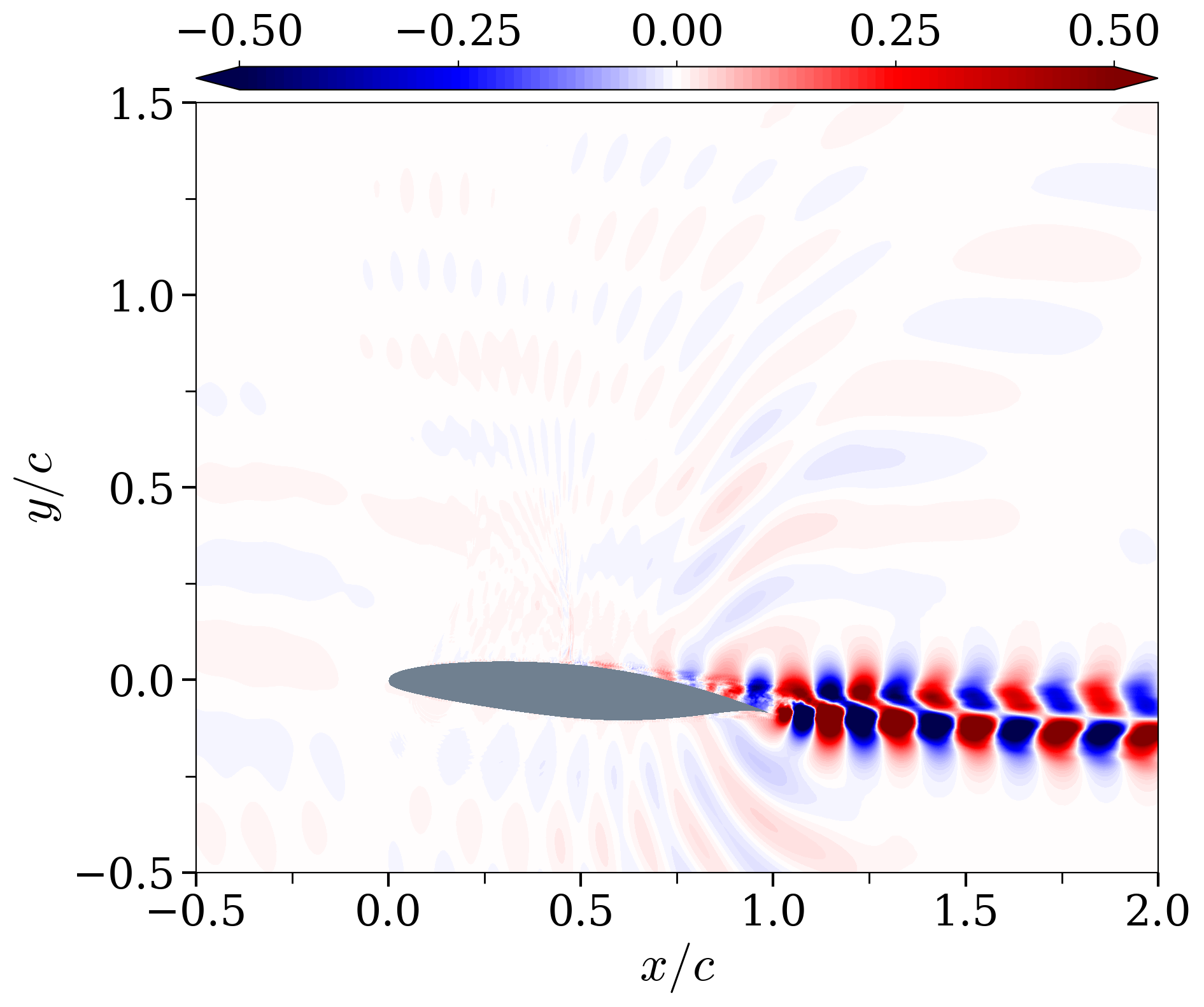}
(b)\includegraphics[width=.3\textwidth]{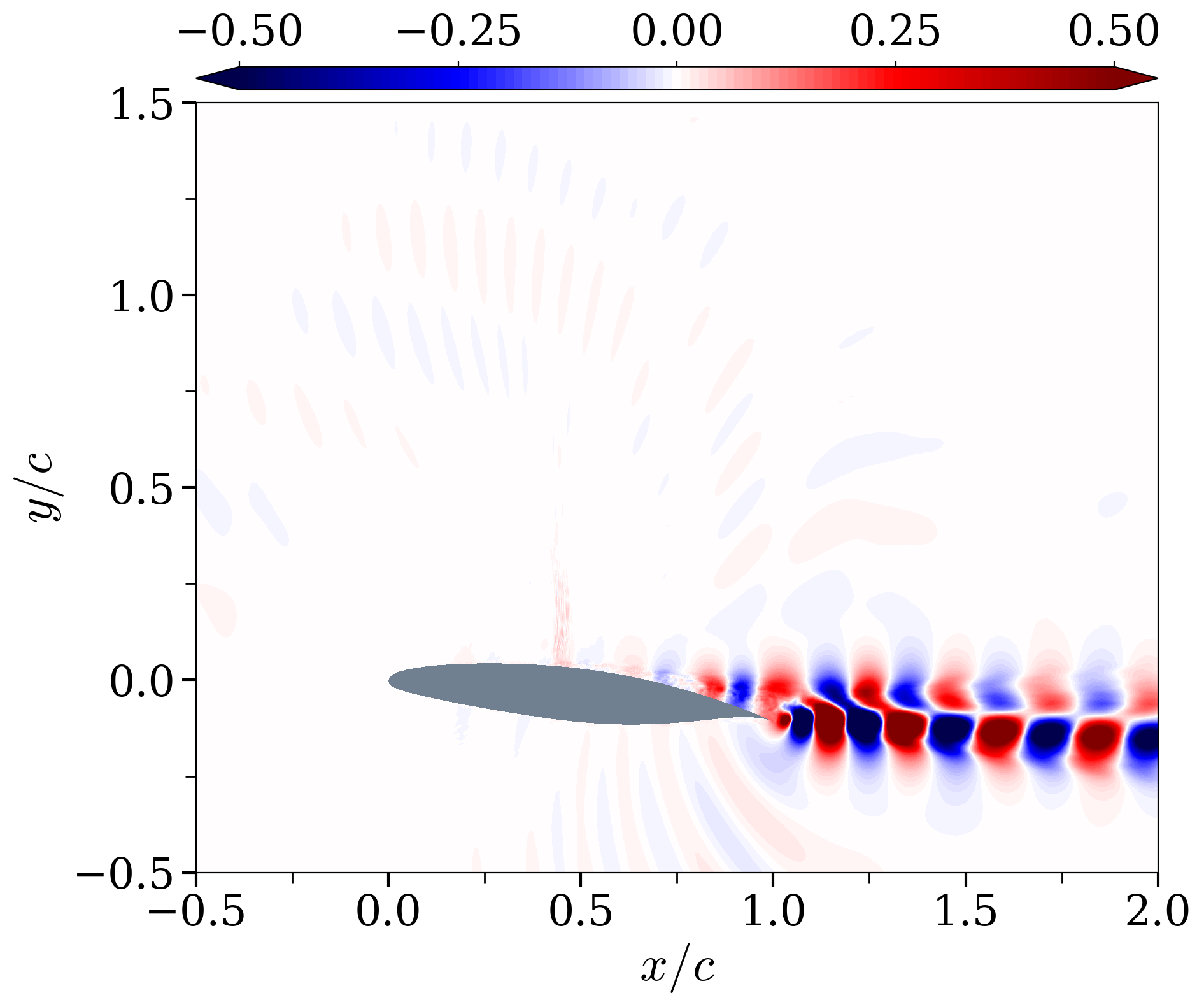}
(c)\includegraphics[width=.3\textwidth]{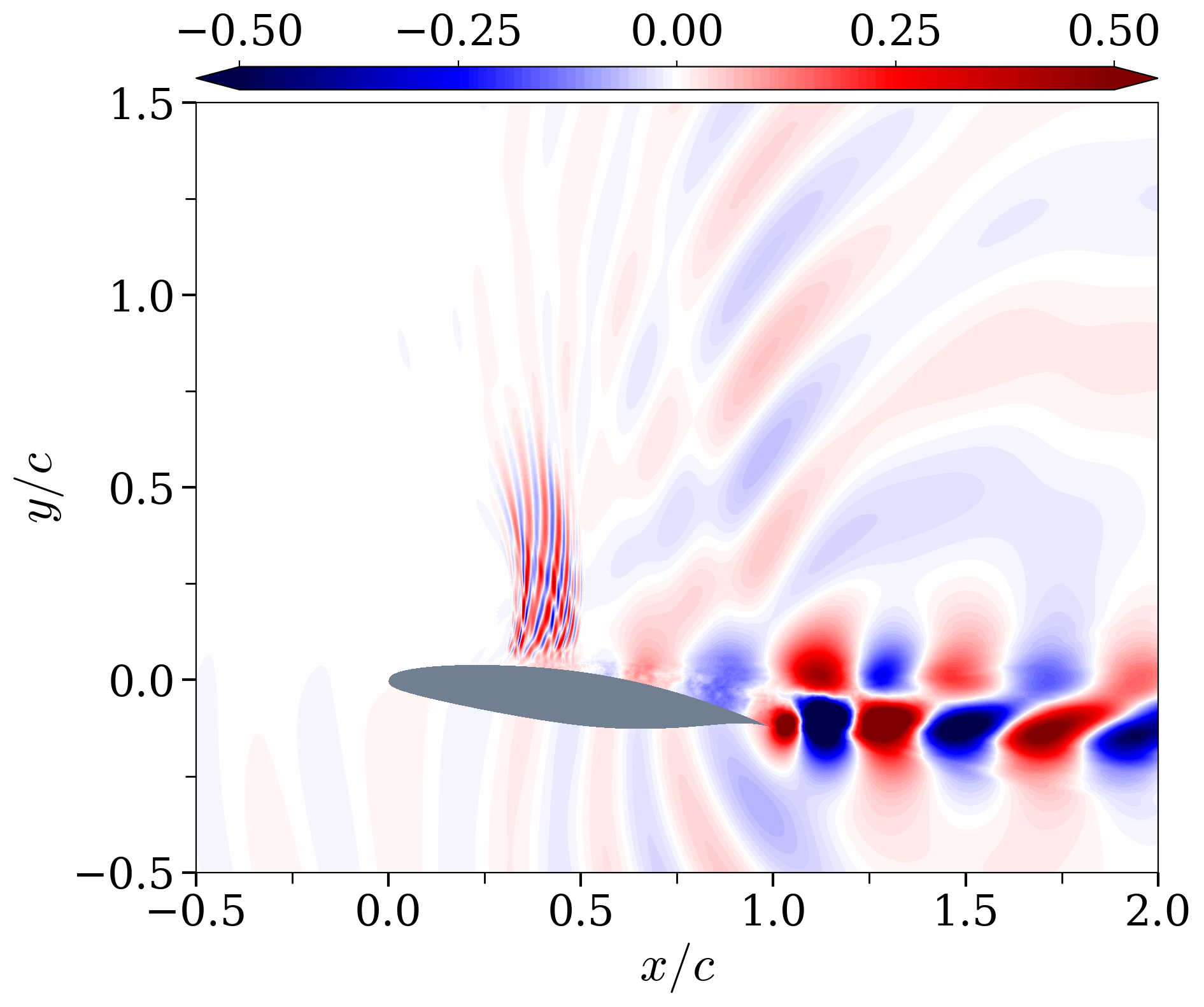}\\
(d)\includegraphics[width=.3\textwidth]{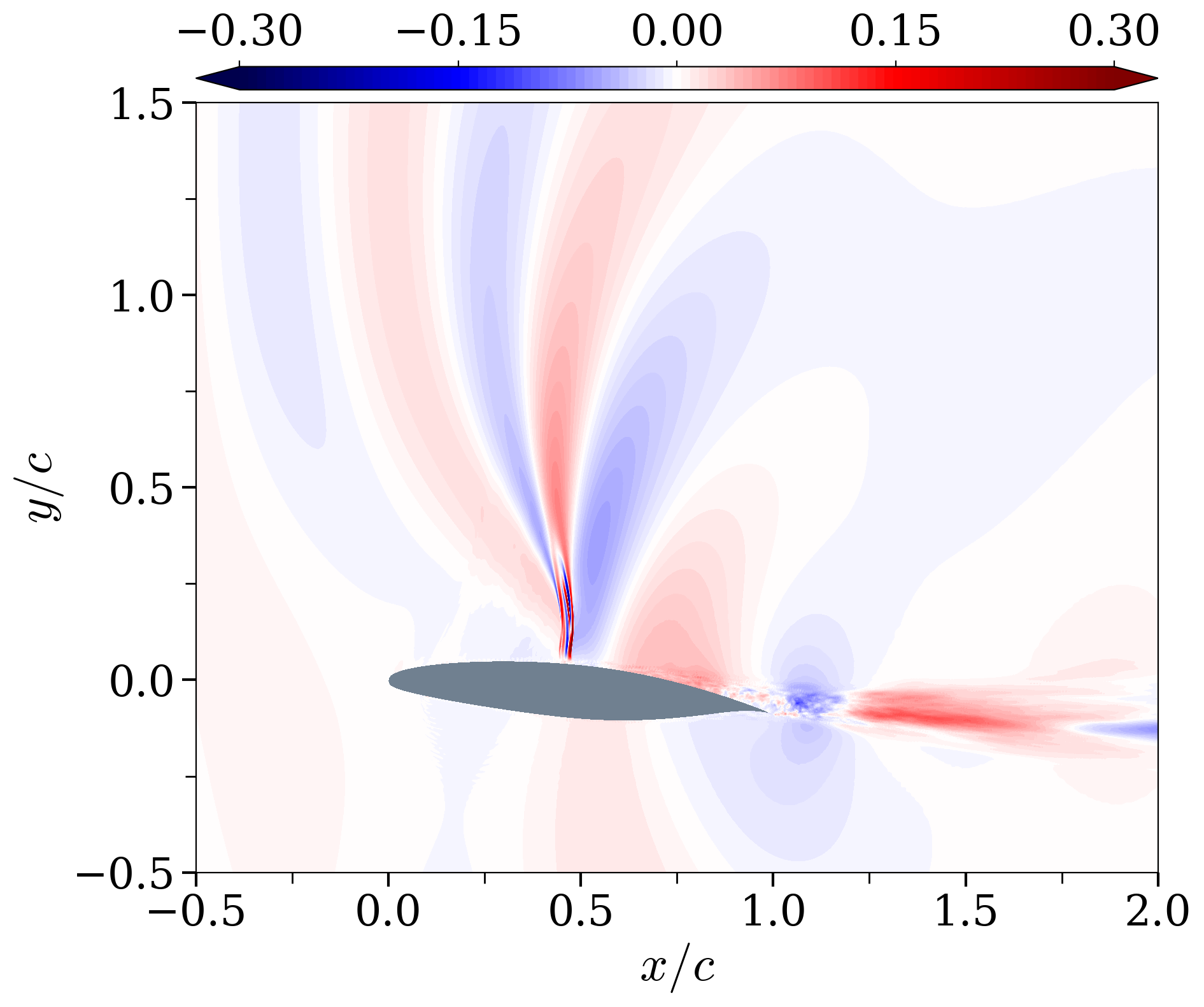}
(e)\includegraphics[width=.3\textwidth]{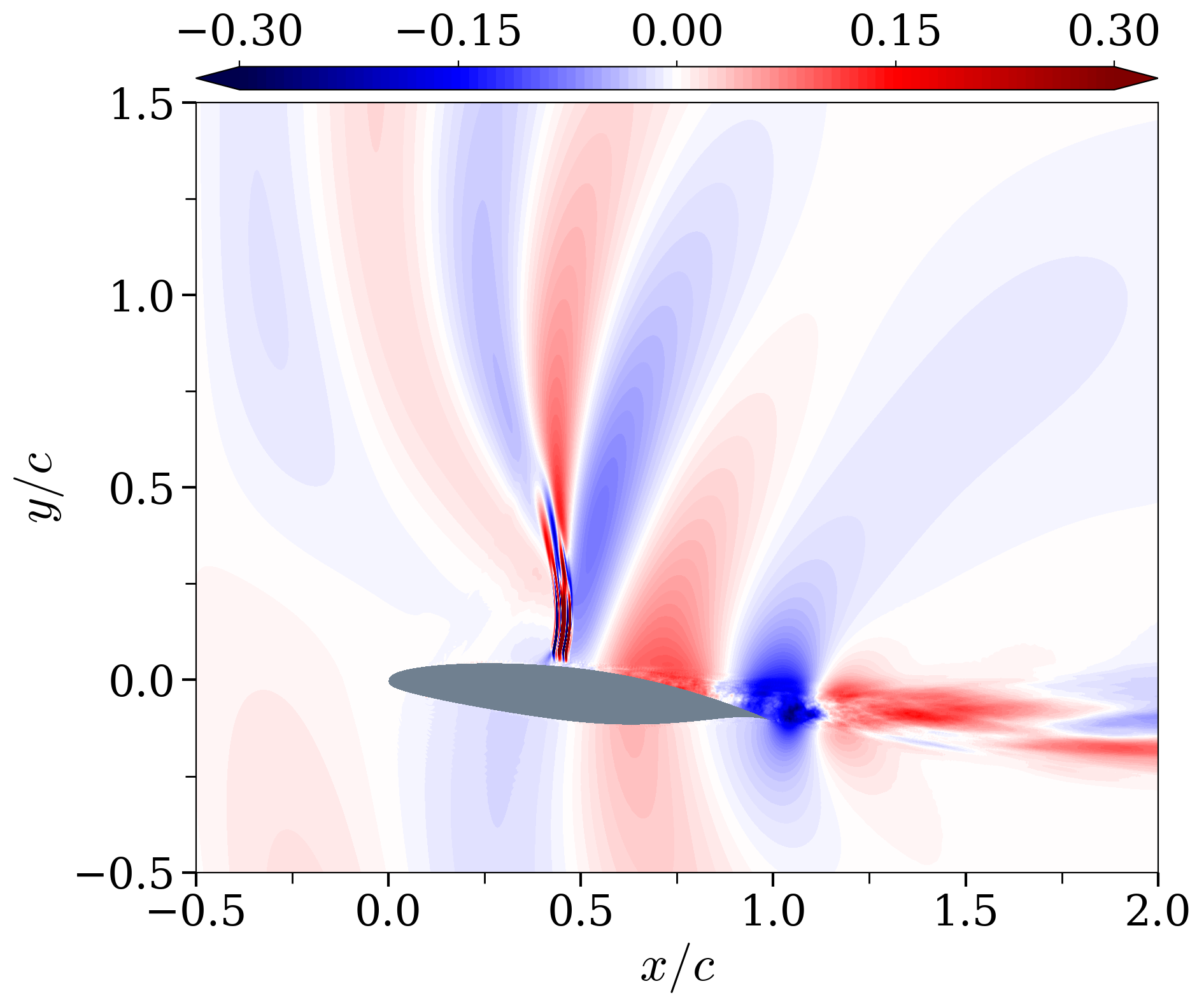}
(f)\includegraphics[width=.3\textwidth]{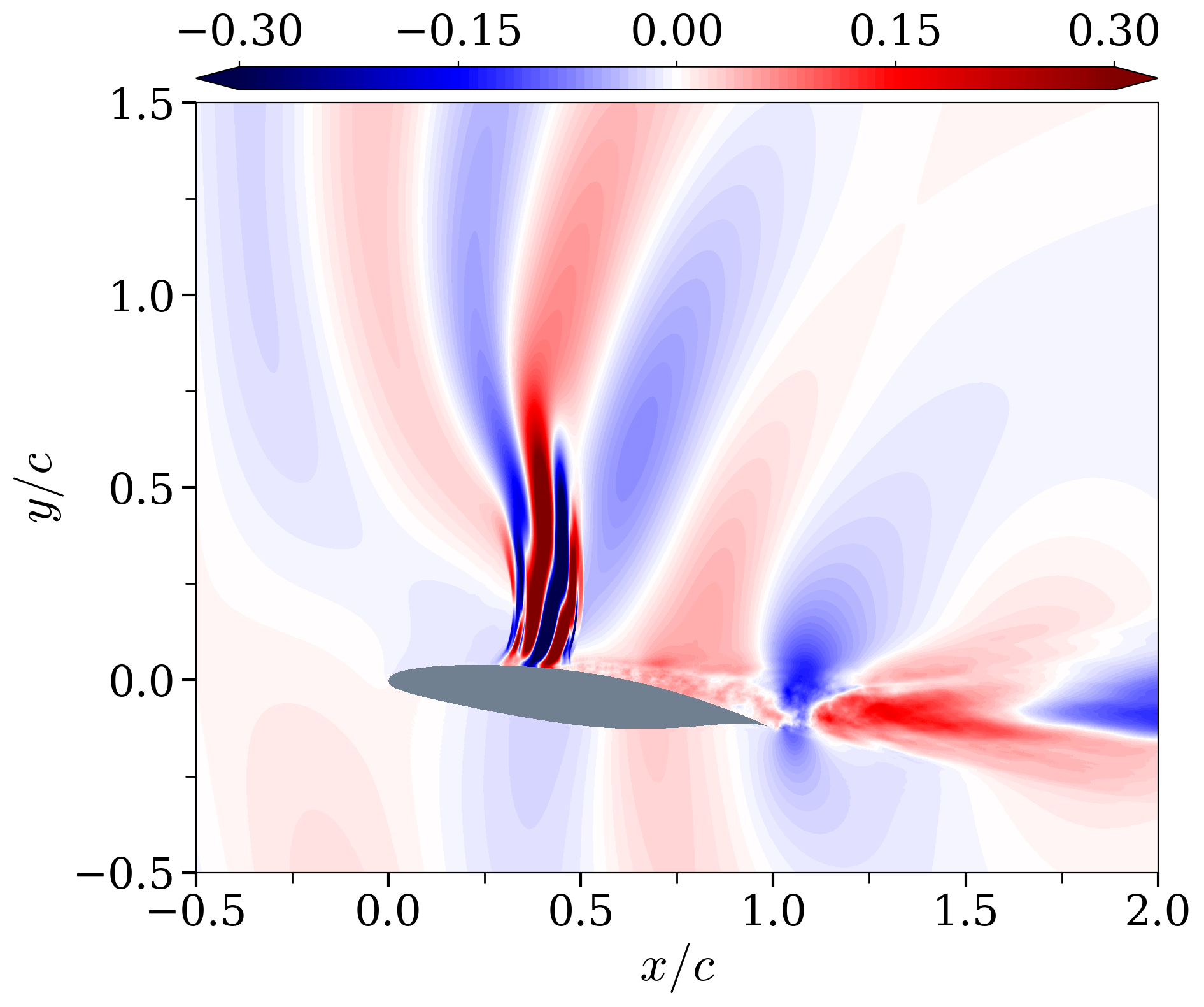}\\
\caption{
Phase-reconstructed pressure fluctuations of the leading SPOD mode 
at $\St \approx 4.0$ (a), $3.4$ (b), $1.9$ (c), and 
at $\St \approx 0.5$ (d--f),
for flow cases at $\Rey = 6 \times 10^5$. Left, middle, and right 
columns refer to $\alpha = 5^\circ$, $6^\circ$, and $7^\circ$, 
respectively. 
}
\label{fig:mode2} 
\end{figure}

\red{
Figure \ref{fig:mode2}(a--c) shows the near-wake vortical structures 
captured by the higher-frequency eigenmodes ($\St \approx 4.0$, $3.4$, $1.9$ 
for $\alpha = 5^\circ$, $6^\circ$, $7^\circ$, respectively), extracted 
from the local peaks of the eigenvalue spectrum in figure \ref{fig:spod}(b). 
A well-organised von Kármán vortex street forms at all angles of attack,
similar to that reported in \citet{moise2023transonic}. 
The roll-up initiates along the suction-side shear layer, immediately 
downstream of the shock, followed by the shedding of a stronger 
pressure-side vortex at the trailing edge. 
The interaction between these structures produces the alternating wake 
pattern. As the incidence increases, the detached shear layer 
is displaced further from the wall and the wake thickens. This yields 
a longer vortex wavelength, which is consistent with the observed drop 
in shedding frequency.
}

\red{
The intermediate-frequency modes at $\St\approx0.5$ are visualised in 
figure \ref{fig:mode2}(d--f). For the three flow cases, the 
eigenmodes display features characteristic of upstream-travelling 
acoustic waves, as will be substantiated below. An alternating pattern of 
elongated pressure fluctuations originates in the near wake and 
propagates symmetrically along the two sides of the airfoil. 
Under pre-buffet conditions (figure \ref{fig:mode2}d), pressure 
fluctuations are relatively weak and tend to decay near the shock. Conversely, 
the incipient and fully developed buffet regimes (figure \ref{fig:mode2}e--f) 
exhibit stronger disturbances, which propagate upstream across 
the shock and refract into the supersonic region. These modal structures 
align with the acoustic signature discussed in \S\ref{sec:visua}. 
}

%==============================================
\section{Buffet mechanics}\label{sec:buffet}
%==============================================

%==============================================
\subsection{Acoustic propagation}\label{sec:acoustic}
%==============================================

\red{
The acoustic propagation discussed in \S\ref{sec:spod} is assessed 
against the
predictions of geometrical acoustics \citep{candel1977numerical,
pierce1981acoustics}.
Assuming that wave fronts propagate according to Huygens' principle, the
trajectory of an acoustic ray $x_i$ is governed by
}
\begin{equation}
\frac{\mathrm{d} x_i}{\mathrm{d} t}=\frac{a s_i}{|\mathbf{s}|}+v_i, \qquad
\frac{\mathrm{d} s_i}{\mathrm{d} t}=-|\mathbf{s}| \frac{\partial a}{\partial x_i} -
\sum_{j=1}^2 s_j \frac{\partial v_j}{\partial x_i}, \qquad i=1,2,
\label{eq:rays}
\end{equation}
\red{
where $\mathbf{s}=\mathbf{n}/(a+\mathbf{v}\cdot\mathbf{n})$ is the
wave-slowness vector, $\mathbf{n}$ is the normal to the wave front,
$\mathbf{v}$ is the mean velocity vector in the $(x,y)$ plane,
$a$ is the local speed of sound. Integrating equations \eqref{eq:rays}
from a localised source yields the acoustic wave fronts and the
local phase speed $v_\varphi=1/|\mathbf{s}|$.
}

\red{
\citet{lee1994role} employed the same ray-tracing approach 
to study the propagation of {\em Kutta waves}. These waves are generated 
by the scattering of shear-layer disturbances at the trailing edge, 
through a mechanism analogous to that
driving self-excited oscillations in transonic cavity flows
\citep{rossiter1964wind}.
In Lee's model, the acoustic source is located at the trailing edge. 
This choice 
is based on the acoustic analogy, according to which solid boundaries act as 
dipole sources \citep{curle1955the} whose radiation is amplified by sharp 
edges \citep{williams1970aerodynamic}. Hence, an attached boundary layer at 
the trailing edge can be treated as an effective dipole source \citep{amiet1976noise, howe1978review}. 
However, this compact-source assumption may be inaccurate in the present 
configuration due to the massively separated mean flow. 
This motivates inspecting 
the turbulence field to estimate the source location.
}

\begin{figure}
\centering
(a)\includegraphics[width=.45\textwidth]{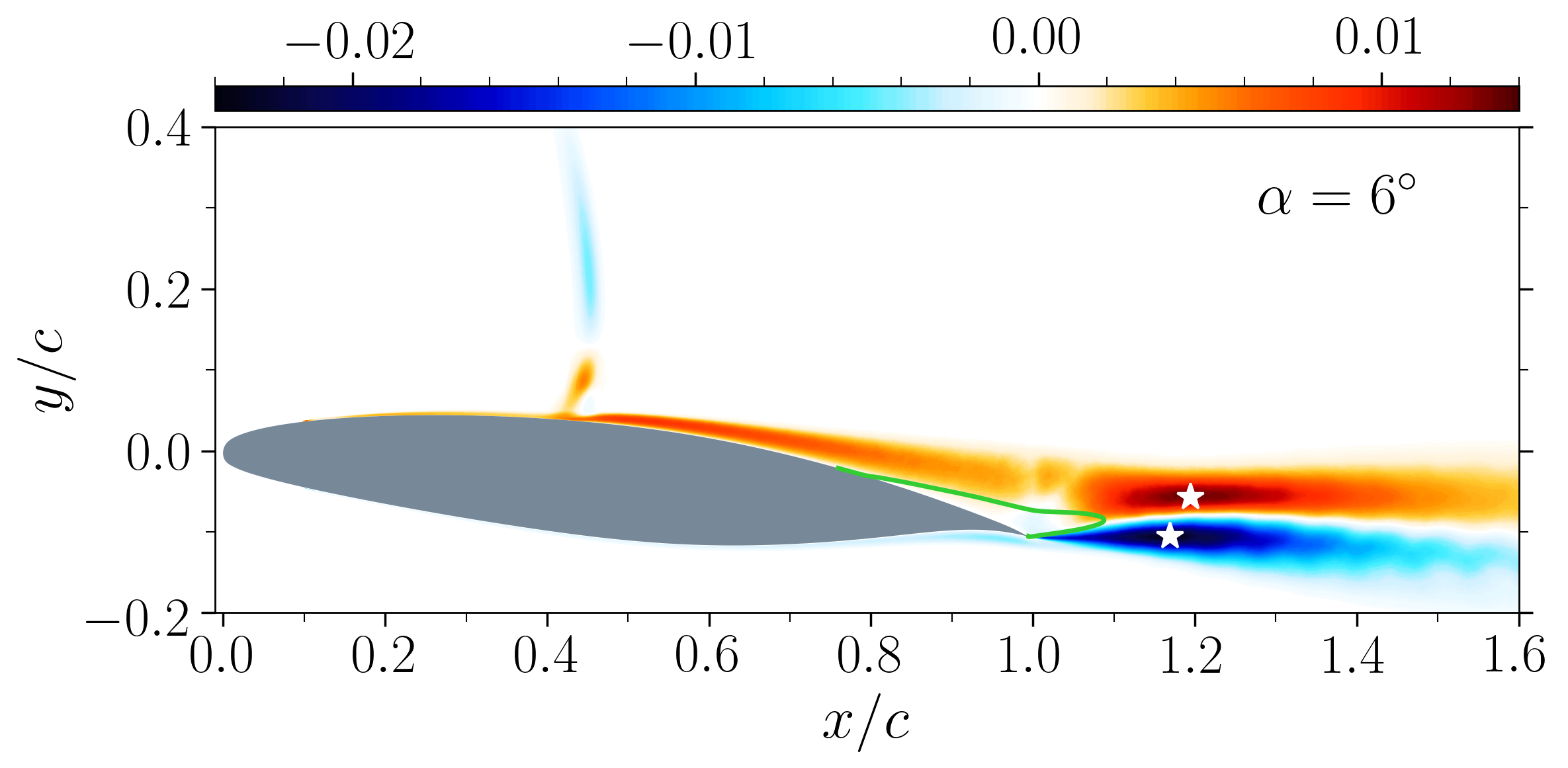}
(b)\includegraphics[width=.45\textwidth]{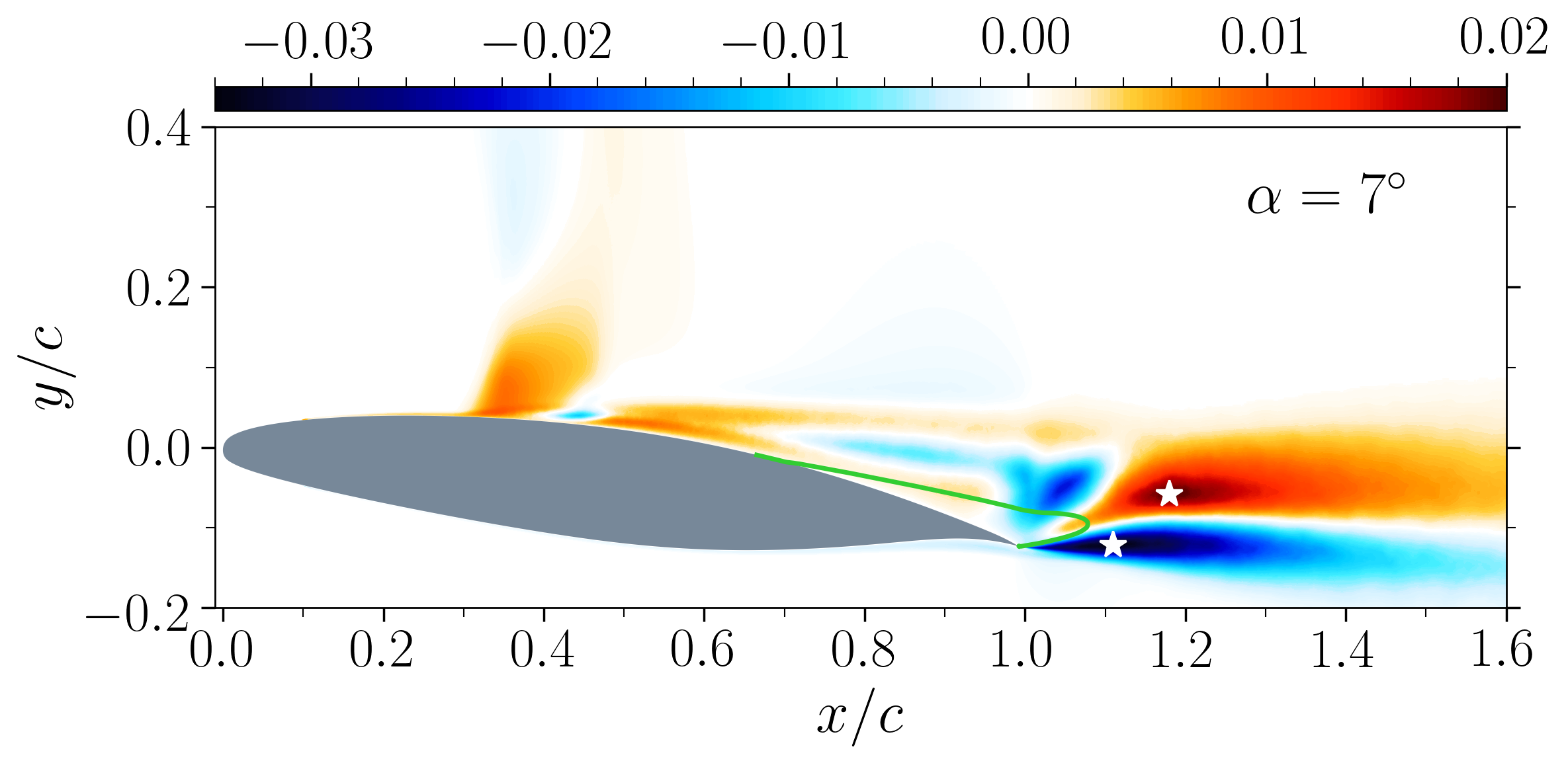}
\caption{
Contour plots of the mean turbulent shear stress
$\smash{-\overline{\rho\,u\pp v\pp }/(\rho_0 u_0^2)}$ 
for flow cases at $\Rey = 6 \times 10^5$ and 
$\alpha=6^\circ$ (a), $\alpha=7^\circ$ (b). 
Green lines denote mean flow reversal ($\tilde{u}=0$) and 
white stars mark the absolute maximum and minimum of 
$\smash{-\overline{\rho\,u\pp v\pp}}$. 
}
\label{fig:uv} 
\end{figure}

\red{
Figure \ref{fig:uv} shows the mean turbulent shear stress,
$\smash{-\overline{\rho\,u\pp v\pp}/(\rho_0 u_0^2)}$. 
%where $u\pp = u-\tilde u$, and
%\tilde u\equiv\overline{\rho u}/\bar\rho$ denotes the Favre average.
Turbulence activity is concentrated in the suction-side shear layer, 
which develops immediately downstream of the shock. This region of 
positive Reynolds shear stress intensifies as it convects downstream, 
extending into the near wake where it merges with the shear layer 
detaching from the trailing edge. Compared to the $6^\circ$ case 
(figure \ref{fig:uv}a), the $7^\circ$ case (figure \ref{fig:uv}b) 
exhibits more complex dynamics. A narrow region of negative 
$\smash{-\overline{\rho\,u\pp v\pp }}$ 
emerges at the shock foot, associated with incipient separation, 
and a larger negative peak dominates the trailing-edge separated 
area (the green solid line denotes mean flow reversal). 
}

\red{
Despite these differences, the near-wake features are similar 
in both cases. The turbulent shear stress is concentrated within 
two detached shear layers, with the peak locations occurring 
near their merging point, as marked by 
the white stars above and below the wake centreline. 
This flow organisation limits the applicability of 
the trailing-edge dipole assumption, which requires the source to 
be coupled to the solid body through an attached boundary layer. 
}

\red{
According to the aerodynamic noise theory by \citet{lighthill1952on, 
lighthill1954sound}, sound is radiated by the double divergence of 
the Lighthill stress tensor,
}
\begin{equation}
T_{i j}=\rho u_i u_j+p_{i j}-\delta_{i j} c_0^2 \rho
\end{equation}
\red{
where $p_{ij}$ is the stress tensor including pressure and viscous
stress contributions. Since the fluctuating Reynolds stresses provide 
the dominant contribution to $T_{ij}$ in turbulent flows 
\citep{crighton1975basic}, we use the turbulent shear stress 
as a proxy for the acoustic source distribution. Specifically, 
the maxima shown in figure \ref{fig:uv} are used as a practical 
indicator of the dominant source regions.
}

\red{
This approximation is intended only to localise the source, 
not to predict the amplitude or directivity of the radiated sound. 
A quantitative description would require evaluating the double 
divergence of the full Lighthill stress tensor, 
together with the associated propagation and retarded-time effects. 
The far-field directivity associated with the Lighthill analogy 
is also not applicable here, since the shock lies within the 
acoustic near field at a distance from the source that is a 
fraction of the acoustic wavelength. Using the near wake as 
the source region is consistent with previous experimental 
and DNS studies of trailing-edge noise 
\citep{tam1974discrete, arbey1983noise, desquesnes2007numerical}, 
as well as with experimental observations of coupled near-wake 
and shock oscillations \citep{roos1977measurements, mohan1991periodic}.
}

\begin{figure}
\centering
(a)\includegraphics[width=.46\textwidth]{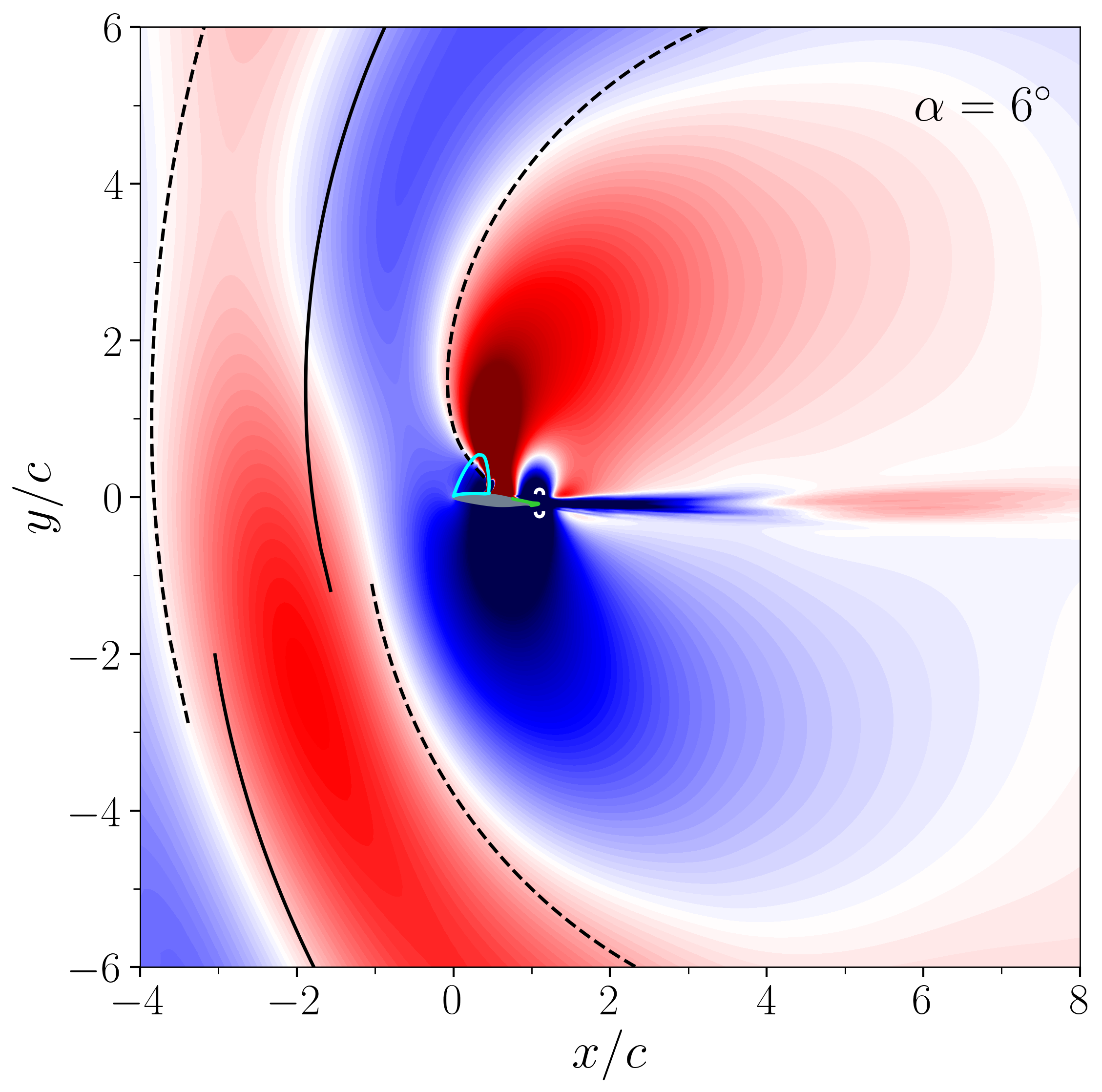}
(b)\includegraphics[width=.46\textwidth]{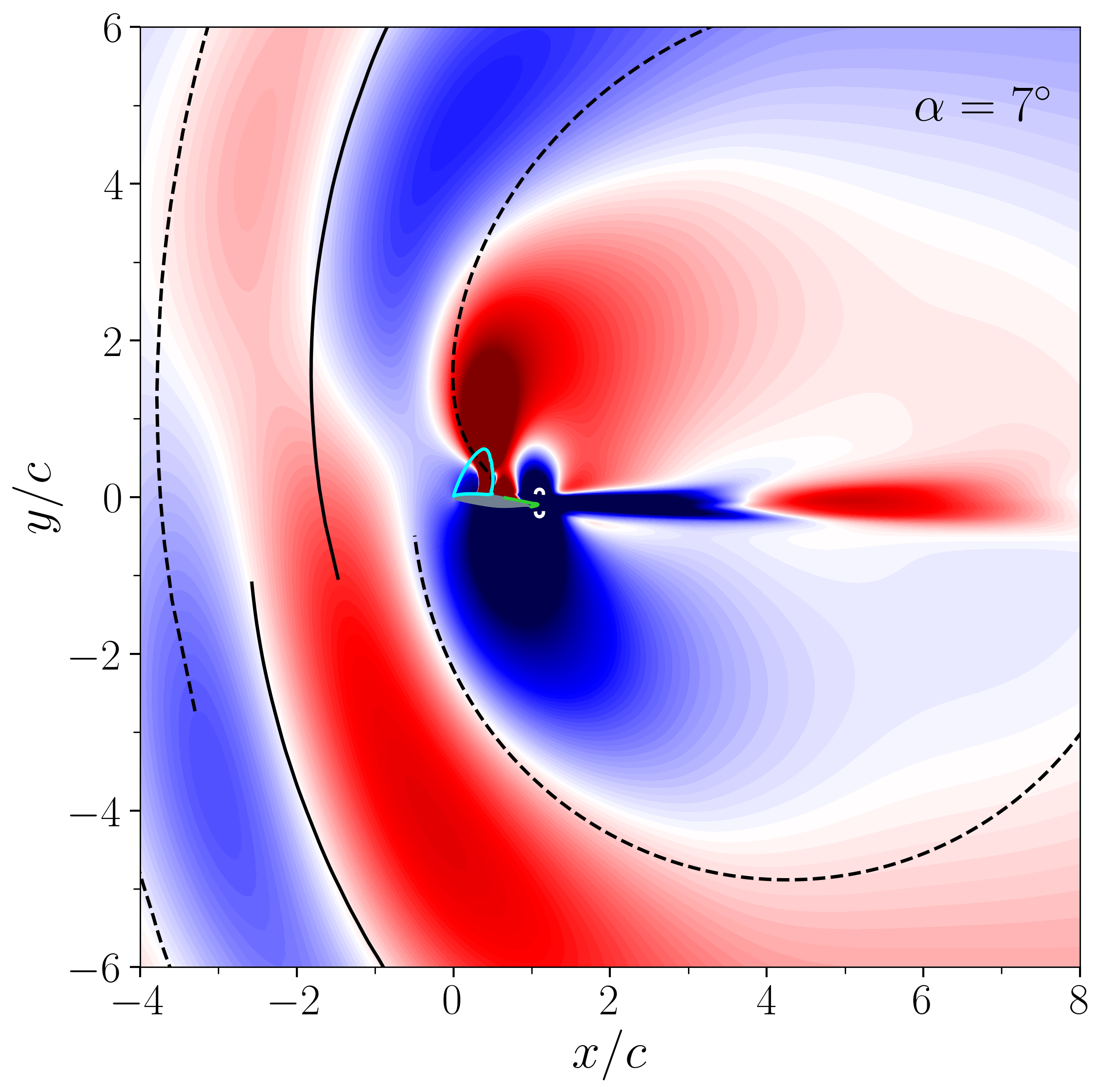}
(c)\includegraphics[width=.46\textwidth]{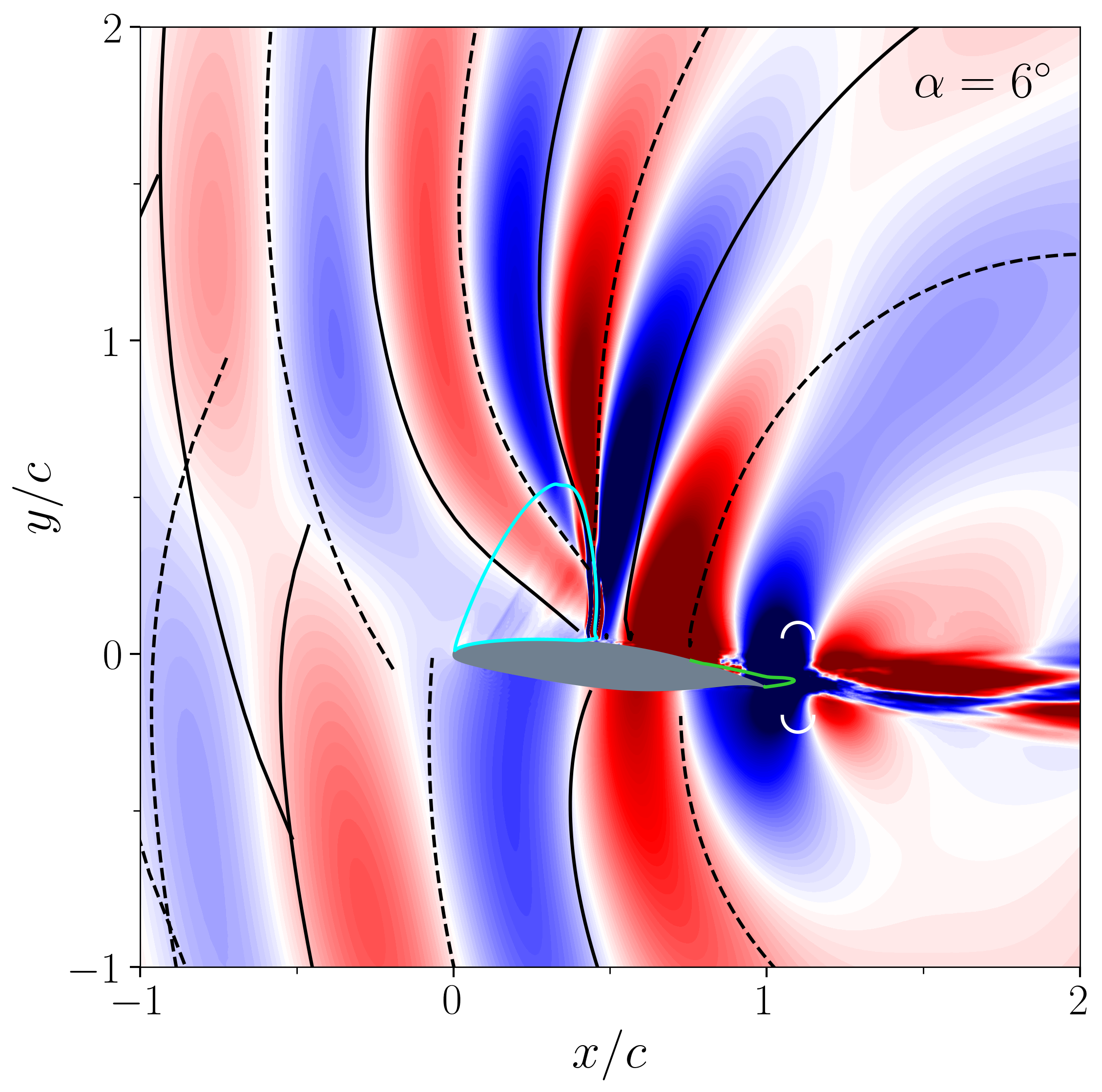}
(d)\includegraphics[width=.46\textwidth]{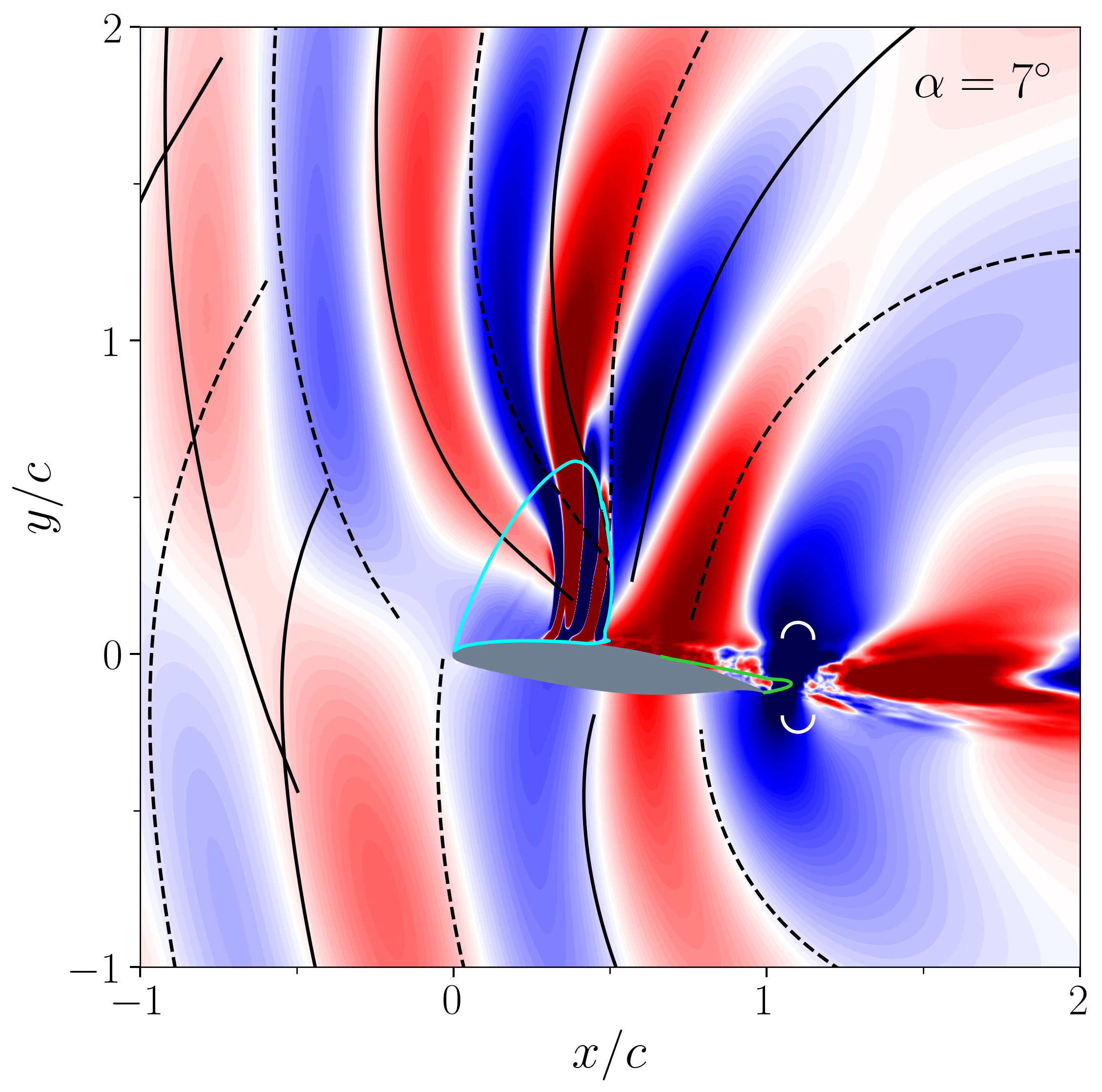}
\caption{
Phase-reconstructed pressure fluctuations of the leading SPOD mode 
at $\St\approx0.1$ (a, b) and $\St\approx0.5$ (c, d) 
for $\alpha=6^\circ$ (a, c) and $\alpha=7^\circ$ (b, d) at $\Rey=6\times10^5$. 
Black solid lines mark wave fronts computed from geometrical acoustics 
spaced by one wavelength; dashed lines mark half-wavelength fronts. 
Rays are launched from two semicircular arcs (white lines) 
centred at $(x/c, y/c) = (1.1,\,0.05)$ and $(1.1,\,-0.2)$ 
with radius of $0.05\,c$. 
The mean sonic line is marked in cyan and the flow reversal line in green.   
}
\label{fig:wave} 
\end{figure}

\red{
Having identified the near wake as the dominant acoustic emission region, 
ray trajectories and wavefront distributions are computed by integrating 
equations \eqref{eq:rays}. Rays are launched from two semicircular arcs 
centred near the locations of the Reynolds shear stress peaks. 
No rays are seeded within the wake, where the slowly-varying 
medium assumption underlying geometrical acoustics is not expected to hold,
due to the presence of strong shear and hydrodynamic fluctuations. 
The origin arcs are thus displaced from the exact peak locations 
to $(x/c, y/c) = (1.1,\,0.05)$ and $(1.1,\,-0.2)$. 
The arc radius is set to $0.05c$, which is comparable to the 
separation between the two peak locations. The computed 
trajectories are found to be nearly insensitive to slight variations 
in the source position and arc radius, which are both much smaller than 
the acoustic wavelength. 
}

\red{
Figure \ref{fig:wave} overlays the computed wave fronts onto the 
SPOD modes at the buffet frequency (figure \ref{fig:wave}a--b) and at the 
intermediate frequency (figure \ref{fig:wave}c--d). 
Wave fronts are extracted at discrete time 
intervals of $0.5/\St$, corresponding to half a period or, equivalently, 
to a phase increment of $\pi$. These increments 
correspond locally to a half-wavelength spatial separation. Solid lines 
denote wave fronts separated by a $2\pi$ phase increment, while dashed lines 
mark the intermediate half-period wave fronts.
}

\red{
A signature of predominant acoustic propagation is a spatial 
alignment between the predicted wave fronts and the zero
contours of the SPOD modes, which delimit regions of 
alternating positive and negative disturbances. 
At buffet frequency (figure \ref{fig:wave}a--b), 
the predicted wave fronts reproduce the curvature of the SPOD 
structures reasonably well, most notably near the upper boundary 
of the supersonic pocket, where local velocity gradients refract 
the acoustic field. The agreement is even closer at the 
intermediate frequency (figure \ref{fig:wave}c--d), 
at which the shorter acoustic wavelength more closely satisfies 
the scale separation required by the slowly-varying-medium 
approximation.
}

\red{
At both frequencies, the quantitative correspondence improves 
at $\alpha=6^\circ$ (figure \ref{fig:wave}a, c), 
whereas a larger discrepancy is observed at $\alpha=7^\circ$ 
(figure \ref{fig:wave}b, d). 
This difference can be attributed to the increasing flow 
unsteadiness, and highlights a limitation of applying 
geometrical acoustics to the mean flow field.
Under fully developed buffet conditions, shock oscillations smear 
the velocity gradients in the mean field, providing a 
coarser representation of the instantaneous acoustic medium. 
Despite these limitations, geometrical acoustics captures 
the qualitative wave pattern at both frequencies, providing a 
first-order estimate of the acoustic propagation time from 
the source to the shock wave. 
}

\begin{figure}
\centering
(a)\includegraphics[width=.47\textwidth]{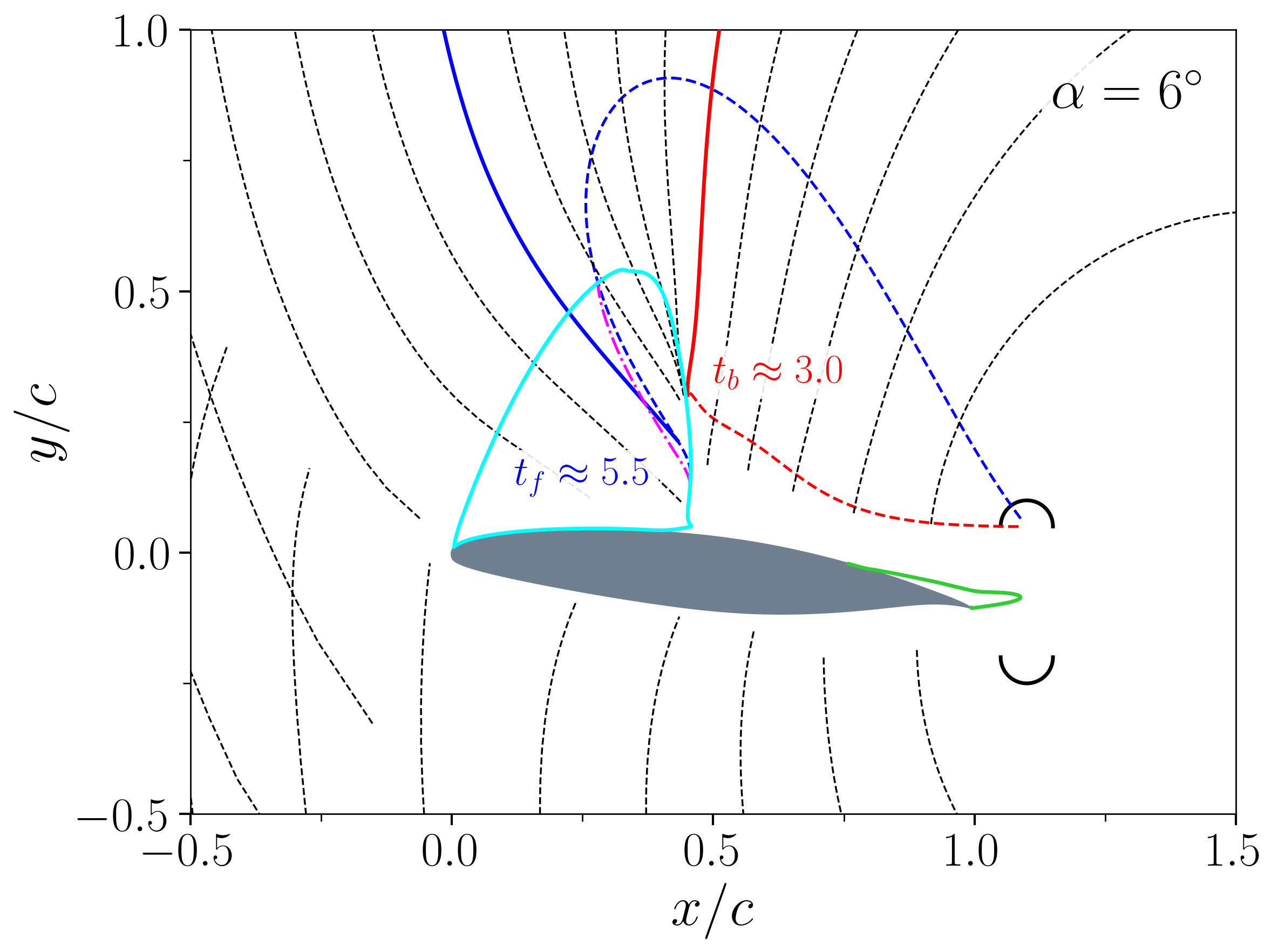}
(b)\includegraphics[width=.47\textwidth]{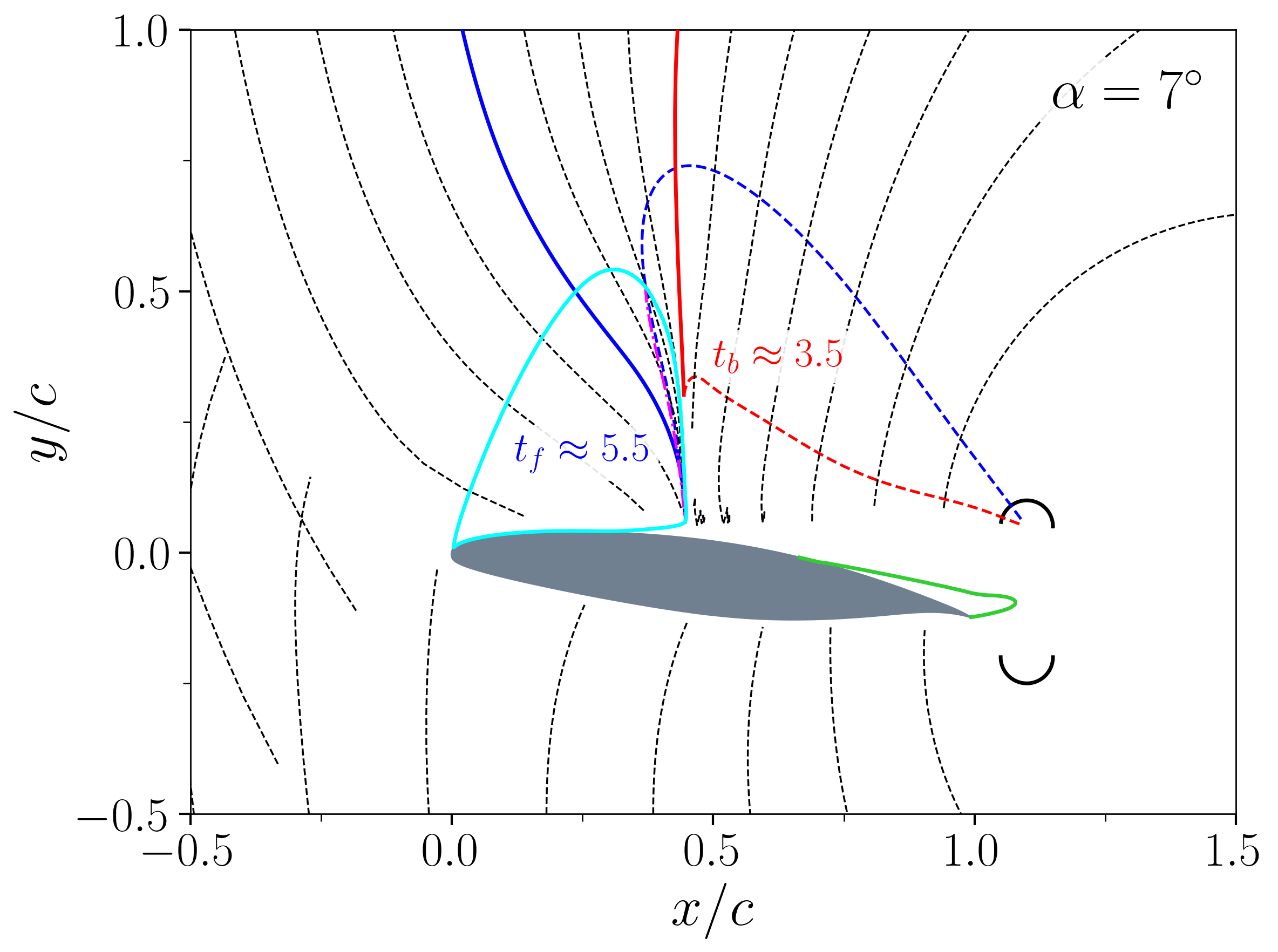}
\caption{
Wave fronts computed via geometrical acoustics for
$\alpha=6^\circ$ (a) and $\alpha=7^\circ$ (b) 
at $\Rey=6\times10^5$. The red and blue fronts
correspond to back- and front-impinging waves at the shock, with
associated propagation times $t_b$ and $t_f$. 
Dashed lines denote the corresponding acoustic rays; 
the magenta dash-dotted line indicates the characteristic 
integrated from the intersection of the front-impinging 
ray with the sonic line.
}
\label{fig:GA} 
\end{figure}

\red{
Figure \ref{fig:GA} shows the predicted wave fronts at finer time intervals
of $0.05/\St_B$ to resolve their temporal evolution. Two specific wave fronts 
are highlighted. 
The first (red solid line) represents the earliest wave front to reach the shock, 
impinging on it from the back as proposed in Lee's model. 
The associated travel time is $t_{{b}}\approx3.0\,c/u_0$ at 
$\alpha=6^\circ$ (figure \ref{fig:GA}a) and $3.5\,c/u_0$
at $7^\circ$ (figure \ref{fig:GA}b). 
The second wave front (blue solid line) 
refracts around the supersonic pocket and reaches the shock foot from the 
front at $t_{{f}}\approx5.5\,c/u_0$. The corresponding acoustic ray (blue dashed line) follows 
a bypass mechanism analogous to that 
identified by \citet{memmolo2018scrutiny}.
}

\blue{
Our focus on this front-impinging ray is motivated by prior stability and 
resolvent studies. Although transonic buffet acts as a global instability 
\citep{crouch2009origin, iorio2014direct}, the specific spatial region 
that drives the instability can be isolated using 
the \textit{wavemaker} concept \citep{monkewitz1990local}. 
The wavemaker is identified through structural sensitivity, 
defined by the spatial overlap of the direct and adjoint global modes 
\citep{giannetti2007structural}. 
\citet{paladini2019transonic} showed that the buffet wavemaker is highly 
localised at the shock foot, a finding subsequently corroborated by 
\citet{kojima2020resolvent} using resolvent analysis. 
Furthermore, both the adjoint 
global mode \citep{iorio2014direct,sartor2015stability,paladini2019various} 
and the optimal resolvent forcing mode 
\citep{houtman2022resolvent,iwatani2023identifying,gao2025optimal} have been 
shown to concentrate within the supersonic region, clustering specifically 
along the rightward-propagating characteristic that impinges on the shock foot.
}

\blue{
These observations suggest that acoustic waves reaching 
the shock foot along this trajectory play a key role 
in the buffet dynamics. To verify that our acoustic model 
captures this sensitive region, we assess the alignment of 
the selected front-impinging ray with the right characteristic. 
The characteristics are inclined at the local Mach angle 
$\mu=\arcsin(1/M)$ 
relative to the flow direction $\vartheta$, yielding 
$\mathrm{d}y/\mathrm{d}x = \tan(\vartheta-\mu)$. 
Integrating this relation from 
the point where the selected ray crosses the sonic line 
gives the characteristic curves depicted 
by the magenta dash-dotted lines in figure \ref{fig:GA}. 
Within the supersonic region, the local tangent of the 
acoustic ray deviates from the characteristic line direction 
by approximately $4^\circ$ and $2^\circ$ on average for 
$\alpha=6^\circ$ and $7^\circ$. This close alignment confirms 
that the acoustic ray traverses the highly sensitive 
region identified by linear theory.
}
\blue{
The relative impact of the back- and 
front-impinging acoustic waves on the shock displacement is evaluated below.
}

%==============================================
\subsection{Shock--acoustics interaction}\label{sec:lin}
%==============================================

We analyse the linear response of a normal shock wave to impinging
acoustic disturbances, grounded in the classical theory of 
shock--disturbance interactions. In their seminal works, 
\citet{moore1953unsteady} and \citet{ribner1954convection} 
introduced the modal decomposition 
of impinging disturbances into acoustic, entropy, and vorticity 
modes to describe their transmission and generation across the shock. 
This framework has been generalised 
by \citet{mckenzie1968interaction} to oblique shocks, 
and the accuracy of the linear approximation has been validated against 
high-fidelity numerical data \citep{mahesh1995interaction}. 
Subsequently, \citet{robinet2001critical} applied this linear theory 
to self-sustained shock oscillations on a transonic airfoil, 
reporting theoretical predictions qualitatively consistent with experiments.

Building on this well-established framework, we employ the linear model 
to interpret the present DNS results. 
The primary objectives are: 1) to elucidate the frequency 
dependence of the shock response to broadband acoustic forcing generated by
trailing-edge scattering, and to assess whether the response favours
low-frequency components, thereby supporting frequency lock-in between
the acoustics and the shock motion; and 2) to quantify the shock
receptivity to perturbations incident from the upstream direction
(\emph{front impact}) versus the downstream direction (\emph{back
impact}).

\begin{figure}
\centering
(a)\includegraphics[width=.49\textwidth]{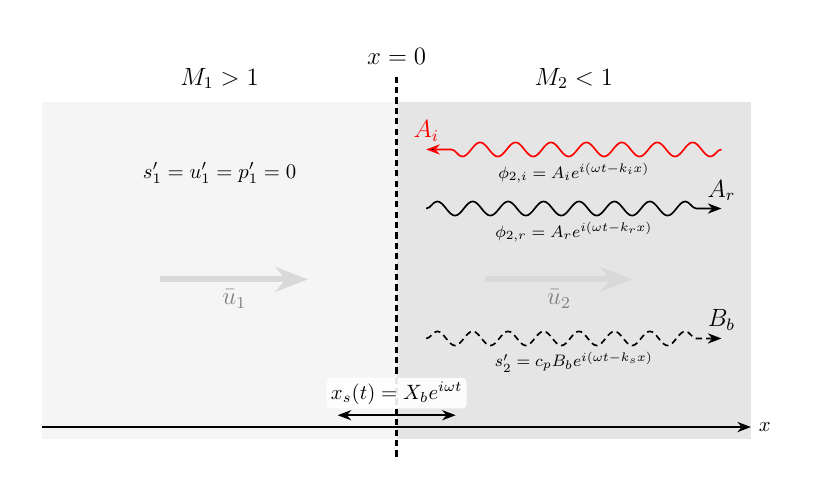}\hspace{-.07\textwidth}
(b)\includegraphics[width=.49\textwidth]{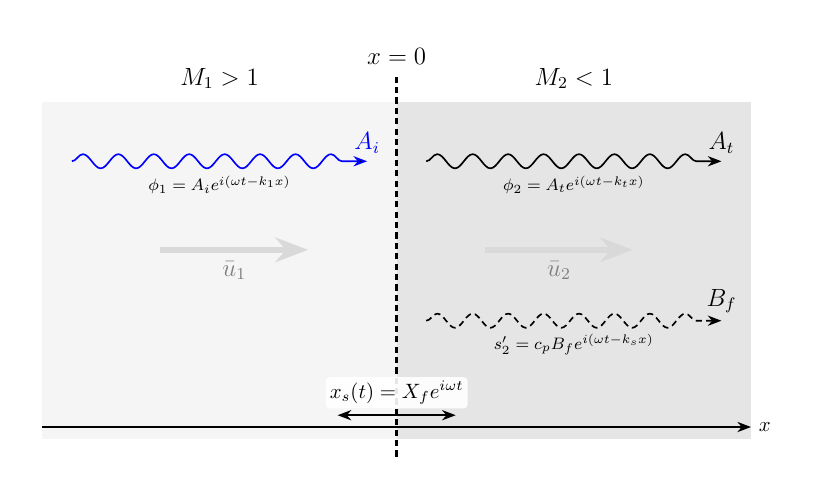}
\caption{
Schematic representation of the linear shock--acoustics interaction 
problem for the back impact scenario (a) and the front impact scenario (b).
}
\label{fig:sketch}
\end{figure}

The problem is modelled as one-dimensional flow of an inviscid ideal gas, 
\blue{
as sketched in figure \ref{fig:sketch}. 
}
We adopt a linear perturbation approach, decomposing the instantaneous
flow variables into a steady, uniform mean state and small unsteady
fluctuations,
$q(x,t) = \bar{q}_i + q'_i(x,t)$,
where the subscript $i$ denotes the spatial domain ($i=1$ upstream,
$i=2$ downstream).
The overline denotes mean quantities, while the prime denotes
perturbations of order $O(\varepsilon)$ with $\varepsilon \ll 1$.
Substituting this decomposition into the Euler equations yields the
linearised system (see appendix \ref{app:euler}).
We introduce a velocity potential $\phi_i$ such that
$u'_i = \partial \phi_i/\partial x$.
Integrating the linearised momentum equation \eqref{eq:momentum} yields
the unsteady Bernoulli relation
\begin{equation}
p'_i = -\bar{\rho}_i \left(
\frac{\partial\phi_i}{\partial t}
+ \bar{u}_i\frac{\partial\phi_i}{\partial x}
\right).
\label{eq:bernoulli}
\end{equation}
Substituting \eqref{eq:bernoulli} into the mass conservation law \eqref{eq:mass},
under the isentropic assumption $p'_i = \bar{c}_i^2\rho'_i$ (where
$\bar{c}_i = \sqrt{\gamma \bar{p}_i/\bar{\rho}_i}$ is the local speed of sound),
we obtain the convected wave equation 
\begin{equation}
\left(\frac{\partial}{\partial t} + \bar{u}_i\frac{\partial}{\partial x}\right)^2 \phi_i
- \bar{c}_i^2\frac{\partial^2\phi_i}{\partial x^2} = 0.
\label{eq:convected}
\end{equation}
The entropy perturbations ($s'_i$) satisfy the transport equation \eqref{eq:entropy}.

In the \emph{back impact} scenario, an acoustic disturbance originates in
the subsonic downstream region ($i=2$) and impinges on the shock.
The velocity potential in region 2 is expressed as the superposition of
an incident wave of amplitude $A_i$ and a reflected wave of amplitude
$A_r$,
\begin{equation}
\phi_2(x,t)=\left(A_i e^{-ik_i x} + A_r e^{-ik_r x}\right)e^{i\omega t}.
\label{eq:phi2}
\end{equation}
The wavenumbers correspond to the acoustic roots of the dispersion
relation \eqref{eq:convected}, representing the upstream-propagating
incident mode,
$k_i={\omega}/[{\bar{c}_2(M_2-1)}]$, and the downstream-propagating
reflected mode,
$k_r={\omega}/[{\bar{c}_2(M_2+1)}]$, where $M_2=\bar{u}_2/\bar{c}_2<1$.
The shock interaction generates entropy perturbations that convect
downstream as
$s'_2 = c_p B_b e^{i(\omega t - k_s x)}$, where the subscript $b$ denotes
the back-impact response and $k_s=\omega/\bar{u}_2$.
The shock responds by oscillating about its mean position,
$x_s(t)=X_b e^{i\omega t}$.

The unknown amplitudes ($A_r$, $B_b$) and the shock displacement ($X_b$)
are determined by enforcing conservation laws across the moving
discontinuity.
The linearised unsteady Rankine--Hugoniot relations (see
appendix \ref{app:RH}) are applied as boundary conditions at $x=0$.
Since the upstream flow is supersonic, the domain $x<0$ remains
unperturbed ($u'_1=p'_1=s'_1=0$).
Solving the resulting algebraic system for the shock motion and the
acoustic reflection coefficient $\mathcal{R}=A_r/A_i$ yields
\begin{subequations}
\label{eq:back_solution}
\begin{align}
\frac{X_b}{A_i} &=
-\,\frac{\bar{u}_1}{(\bar{u}_2-\bar{u}_1)\bar{u}_2}\,(1+\mathcal{R}),
\label{eq:X_back} \\[10pt]
\mathcal{R} &=
\frac{M_2\left[2 \gamma M_1^2-(\gamma-1)\right]
-\left[2 M_1^2+(\gamma-1)\right]}
{M_2\left[2 \gamma M_1^2-(\gamma-1)\right]
+\left[2 M_1^2+(\gamma-1)\right]},
\label{eq:R_back}
\end{align}
\end{subequations}
where $M_1=\bar{u}_1/\bar{c}_1>1$.
The expression for the entropy amplitude $B_b$ is given in
appendix \ref{app:back}.

We now consider an acoustic disturbance incident from the supersonic
upstream region (\emph{front impact} scenario).
The velocity potential in region 1 describes the incident wave
travelling downstream with amplitude $A_i$,
\begin{equation}
\phi_1(x,t) = A_i e^{i(\omega t - k_1 x)},
\label{eq:phi1}
\end{equation}
with $k_1 = {\omega}/[{\bar{c}_1(M_1+1)}]$.
No reflected waves can propagate upstream against the supersonic mean
flow, hence the upstream region remains isentropic ($s'_1=0$) and the
perturbations $u'_1$ and $p'_1$ associated with the incident wave act as
the forcing.

The interaction with the shock generates a transmitted acoustic wave
($A_t$) and an entropy wave ($B_f$) in the subsonic downstream region
($i=2$).
The downstream potential is therefore
$\phi_2 = A_t e^{i(\omega t - k_t x)}$, where
$k_t = \omega/[\bar{c}_2(M_2+1)]$ corresponds to the
downstream-propagating acoustic mode.
Entropy perturbations convect as
$s'_2 = c_p B_f e^{i(\omega t - k_s x)}$, where the subscript $f$ denotes
the front-impact response.
As in the previous case, the shock oscillates according to
$x_s(t) = X_f e^{i\omega t}$.
Substituting these wave definitions into the linearised
Rankine--Hugoniot conditions \eqref{eq:RHlin} yields the algebraic
solution for the shock displacement amplitude and the acoustic
transmission coefficient $\mathcal{T}=A_t/A_i$,
\begin{subequations}
\label{eq:front_solution}
\begin{align}
\frac{X_f}{A_i} &=
\frac{1}{\bar{u}_1-\bar{u}_2}
\left[ \frac{\bar{u}_1}{\bar{u}_2}\mathcal{T}
- \left( 1 + M_1 - M_2 C\right) \right],
\label{eq:X_front} \\[10pt]
\mathcal{T} &=
\frac{M_2 C}{M_1}
\left[
\frac{C(M_1+1) - M_1 M_2 (\gamma-1)(M_1 - M_2 C)}
{C(M_2+1) - M_2 (\gamma-1)(M_1 - M_2 C)}
\right],
\label{eq:T_front}
\end{align}
\end{subequations}
where $C=\bar{c}_2/\bar{c}_1$ is the sound-speed ratio across the shock.
The expression for the entropy amplitude $B_f$ is provided in
appendix \ref{app:front}.

\begin{figure}
\centering
(a)\includegraphics[width=.47\textwidth]{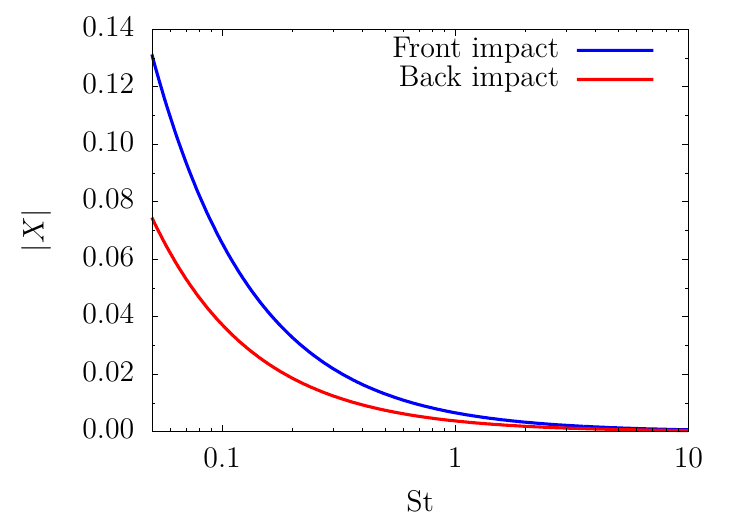}
(b)\includegraphics[width=.47\textwidth]{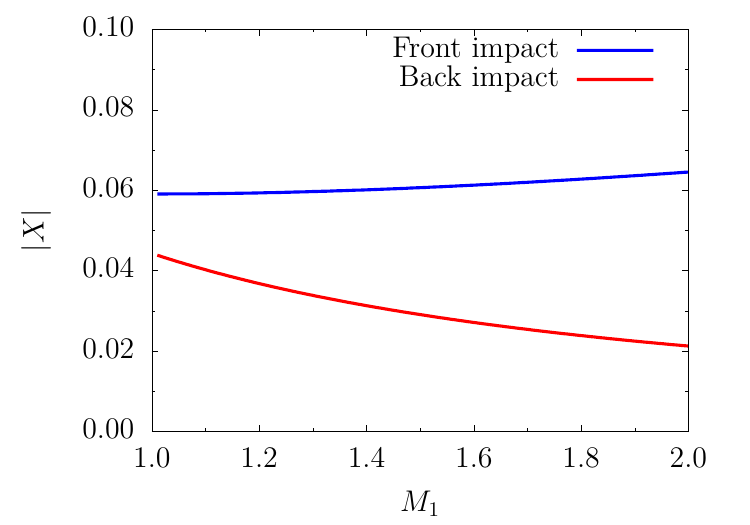}
\caption{Linear response of the shock displacement amplitude $|X|$ to
acoustic forcing.
(a) Strouhal-number dependence at fixed upstream Mach number $M_1=1.3$.
(b) Upstream Mach-number dependence at fixed frequency $\St=0.11$.
Results are normalised by imposing a constant incident pressure
amplitude $|p'_i| = 0.05\,p_0$.}
\label{fig:xs_lin}
\end{figure}

The solutions \eqref{eq:back_solution} and \eqref{eq:front_solution}
permit evaluation of the sensitivity of the shock position to the
frequency of the incoming disturbance.
The analytical forcing is normalised by prescribing a constant incident
pressure amplitude $|p'_i| = 0.05\,p_0$, consistent with the acoustic
disturbance levels observed in the DNS.
Since $|p'_i| \sim \omega |A_i|$ from \eqref{eq:bernoulli}, a fixed
incident pressure load implies $\smash{|A_i| \sim \omega^{-1}}$.
The reference chord length $c$ is introduced through the Strouhal number
$\St = f c / u_0$, and the resulting displacement amplitude $|X|$ is
non-dimensionalised by $c$.

Figure \ref{fig:xs_lin}(a) reports the shock displacement $|X|$ as a
function of Strouhal number for $M_1=1.3$ (inferred from DNS).
The response exhibits a hyperbolic decay
$|X| \sim \St^{-1}$, revealing stronger sensitivity to low-frequency 
acoustic forcing. The shock therefore acts as a low-pass filter, 
providing a plausible basis for the lock-in between large-wavelength 
acoustic waves and low-frequency shock motion.

The predictive value of the linear analysis is assessed against DNS
results at the buffet frequency ($\St \approx 0.11$).
For upstream incidence (front impact), the model yields a normalised
displacement $|X_f| \approx 0.06\,c$, which is compatible with 
the amplitude of the harmonic component estimated from the DNS signal
as $\sqrt{2}\,\sigma_s \approx 0.07\,c$. 
At the same frequency, downstream incidence predicts 
a weaker response of $|X_b| \approx 0.03\,c$, suggesting that the shock 
is more receptive to acoustic disturbances incident from the upstream side.

Figure \ref{fig:xs_lin}(b) shows the dependence of shock displacement on
upstream Mach number at fixed frequency ($\St = 0.11$).
While the response amplitudes are comparable in the weak-shock limit
($M_1 \to 1$), increasing shock strength yields diverging trends in which 
the response to upstream incidence increases and that to downstream 
incidence decreases.
This difference stems from the jump in acoustic impedance 
$Z_i = \bar{\rho}_i \bar{c}_i$ across the shock. For a fixed acoustic pressure 
amplitude $|p'_i|$, the associated velocity fluctuation scales inversely with the 
local impedance, $|u'_i| \sim |p'_i|/Z_i$. Because $Z_1 < Z_2$, a front-impinging 
wave carries larger velocity fluctuations than a back-impinging wave of equal 
pressure intensity, resulting in a stronger kinematic forcing.
For flow conditions representative of transonic buffet, the
analysis indicates that front-impact incidence can provide 
a more effective coupling with the shock motion
than back-impact incidence.

%==============================================
\subsection{Downstream convection}\label{sec:conv}
%==============================================
As discussed in \S\ref{sec:acoustic}, acoustic waves 
are emitted in the near wake 
due to the scattering of pressure disturbances 
that are generated at the shock 
foot and convected downstream along the separated shear layer. 
Convective speed and travel time along the downstream path are 
quantified via the two-point space--time 
cross-correlation coefficient of pressure fluctuations,
\begin{equation}
R_{pp}(\Delta s, \Delta t) =
\frac{\overline{p'(s_{\mathrm{ref}}, t)\,
p'(s_{\mathrm{ref}}+\Delta s, t+\Delta t)}}
{\sqrt{\;\overline{p'(s_{\mathrm{ref}}, t)^2}\;\;
\overline{p'(s_{\mathrm{ref}}+\Delta s, t+\Delta t)^2}}}\;,
\label{eq:rpp}
\end{equation}
where $p'(s,t) = \langle p(s,t)\rangle_z - \bar{p}(s)$ is the 
spanwise-averaged fluctuating pressure, $\Delta s$ is the 
separation along the path, $\Delta t$ is the time lag, 
and the reference location is fixed at $s_{\mathrm{ref}}=0.6\,c$. 
\red{
The signal is sampled along a line offset from the suction 
surface by a fixed wall-normal distance of $n\approx0.02\,c$, 
corresponding to the peak of the mean shear at $x=0.6\,c$, 
which is nearly identical for the two angles of attack. 
The sampling path extends from the leading edge,  
past the trailing edge, to $x/c=1.1$, allowing the downstream 
propagation to be tracked 
up to the source region identified in \S\ref{sec:acoustic}. 
Only the transverse distance between the end of the sampling path 
and the launch arcs remains unaccounted for.
}

\begin{figure}
\centering
(a)\includegraphics[width=.45\textwidth]{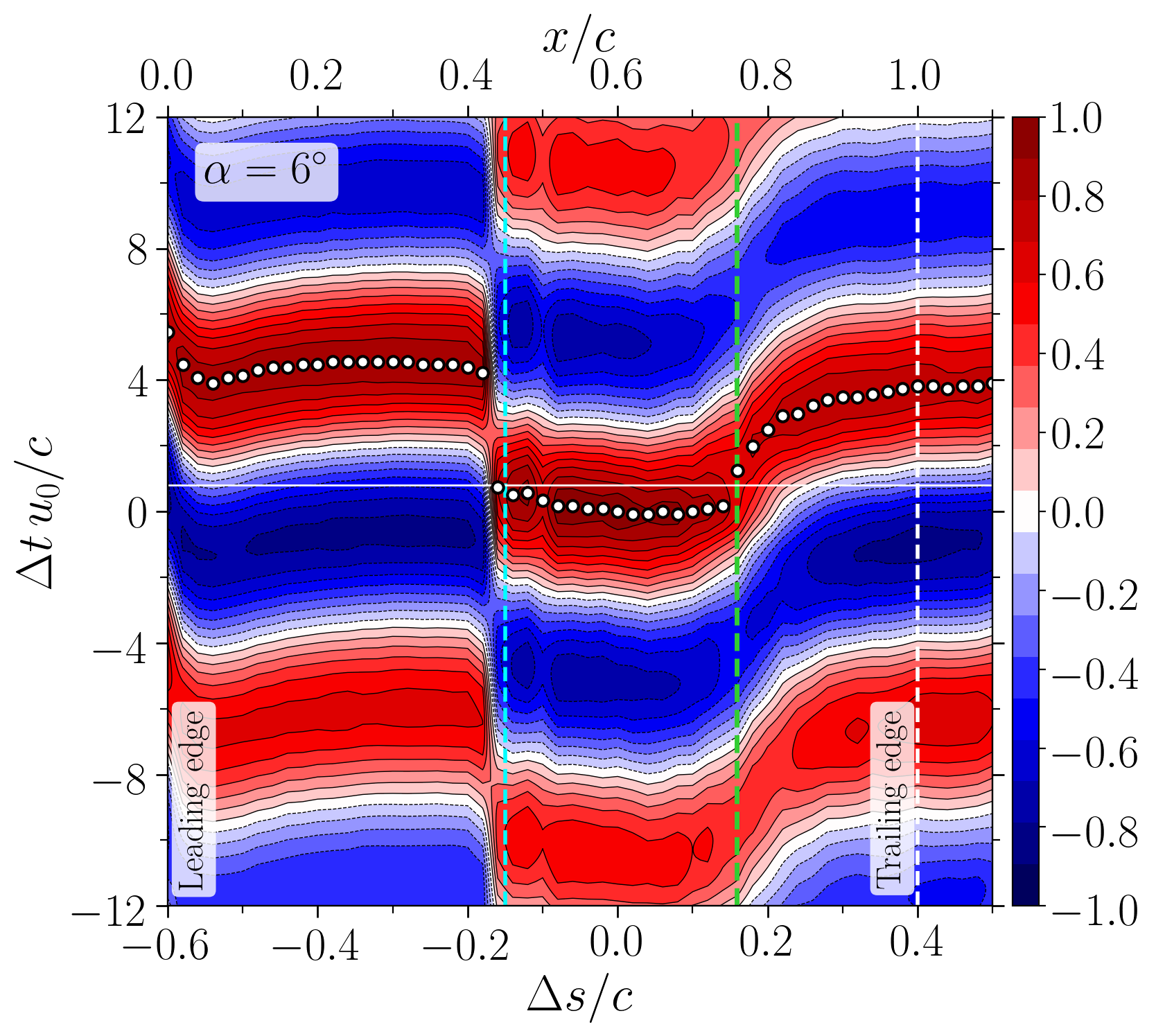}
(b)\includegraphics[width=.45\textwidth]{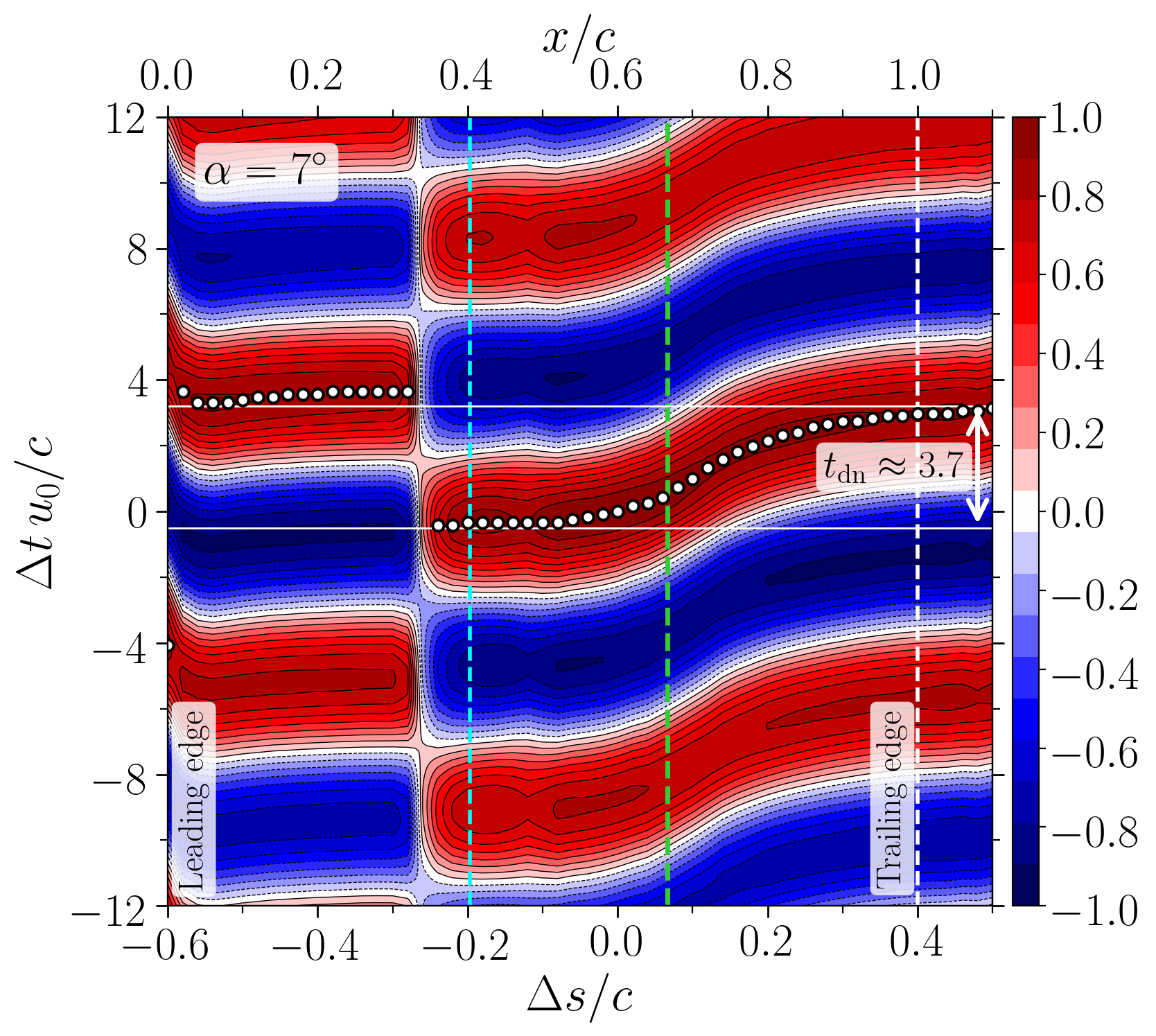}
\caption{
Space--time cross-correlation coefficient of the fluctuating 
pressure \eqref{eq:rpp} for $\alpha=6^\circ$ (a) and $7^\circ$ (b) 
at $\Rey=6\times10^5$. 
White circles trace the locus of $\max_{\Delta t}(R_{pp})$; 
the cyan and green dashed lines mark the mean shock position 
and separation point. 
}
\label{fig:rpp}
\end{figure}

\red{
Figure \ref{fig:rpp} shows the space--time correlation contours.  
White circles denote the local peaks $\max_{\Delta t}(R_{pp})$, 
obtained by maximising the correlation with respect to the time lag 
at each spatial separation, which trace the space--time trajectory 
of the dominant pressure disturbance.
}

\begin{figure}
\centering
(a)\includegraphics[width=.45\textwidth]{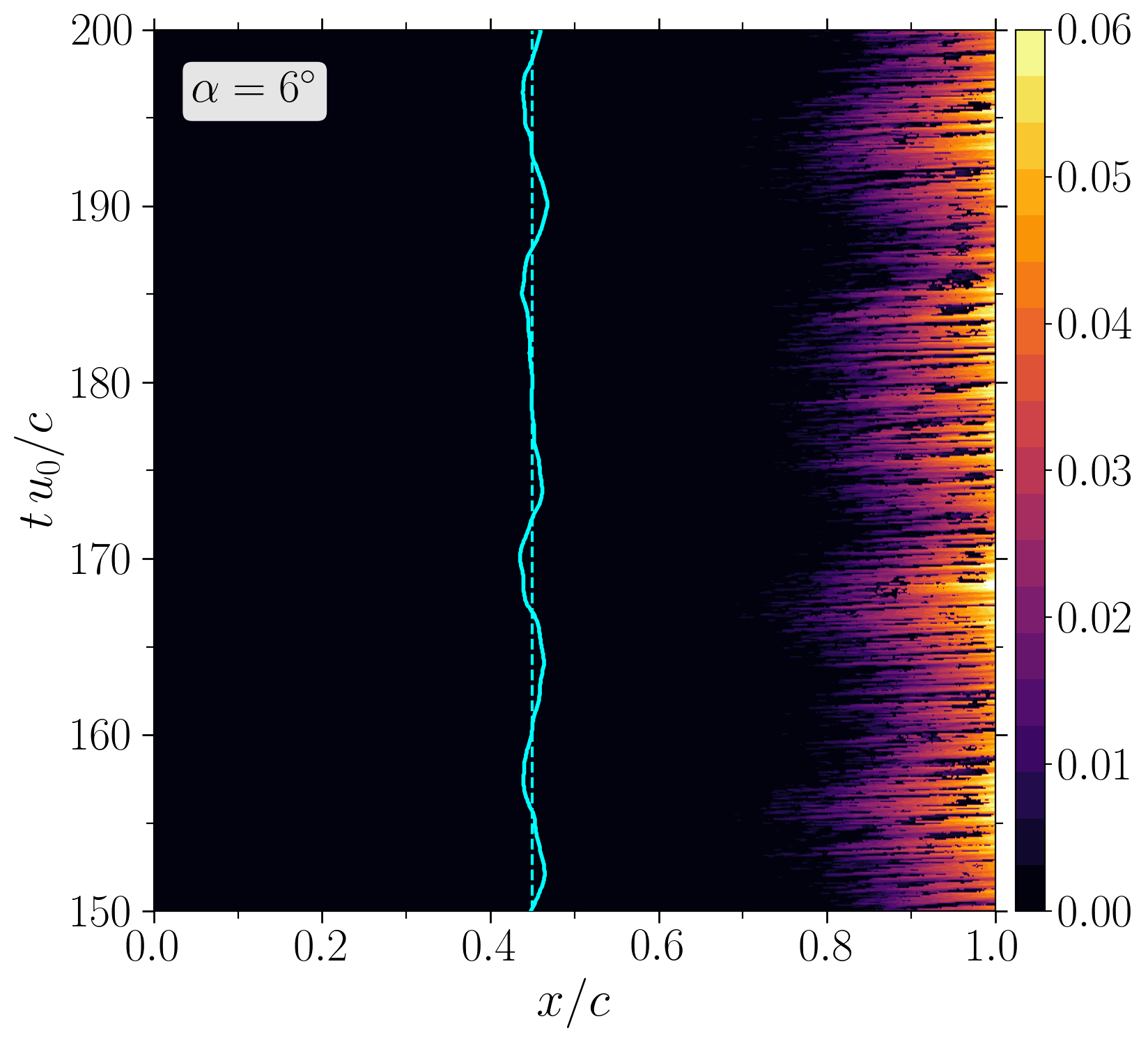}
(b)\includegraphics[width=.45\textwidth]{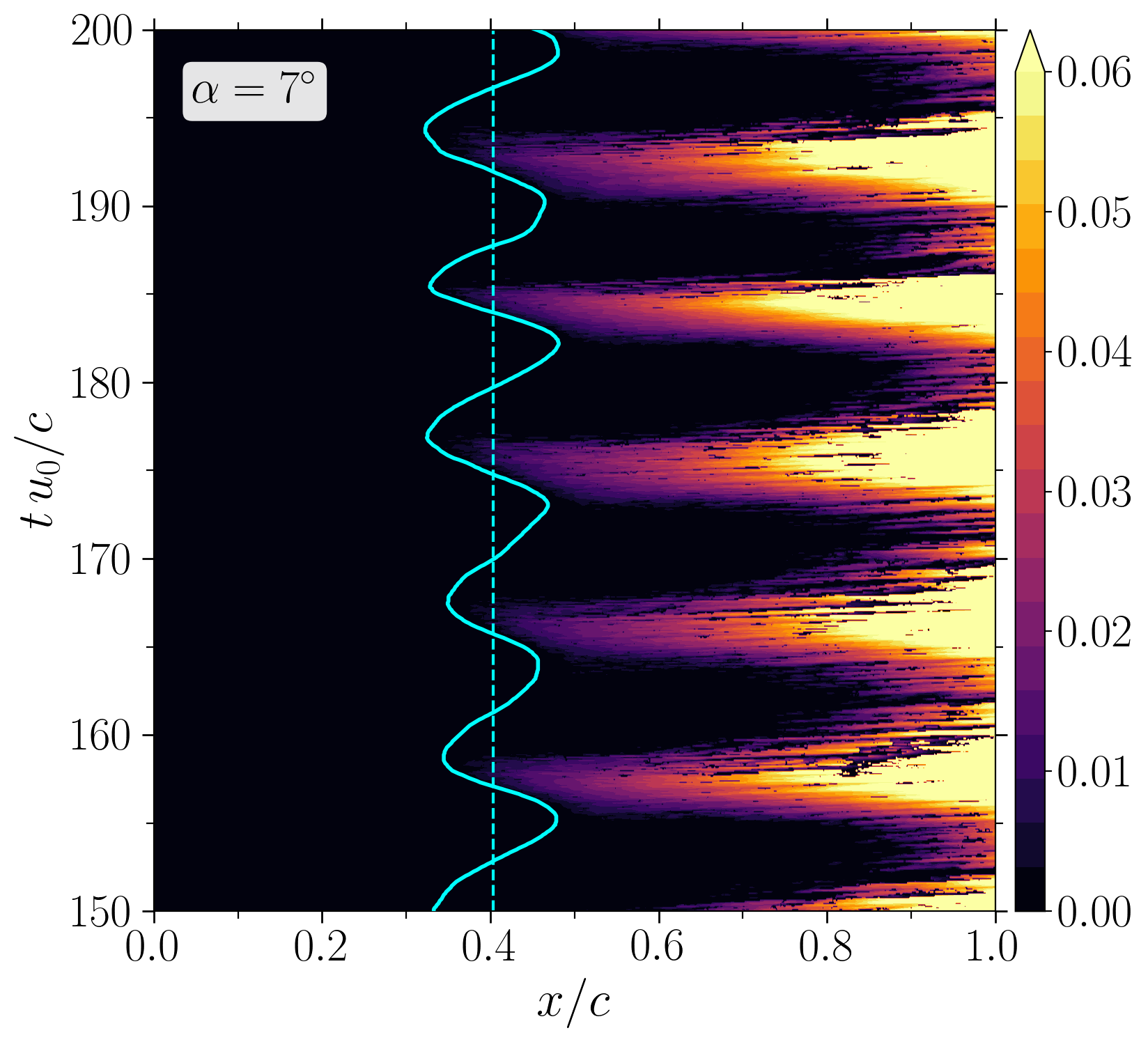}
\caption{
Space--time diagram of the spanwise-averaged separation 
height $n_{\psi=0}$ along the suction surface 
for $\alpha=6^\circ$ (a) and $7^\circ$ (b). 
Black contours denote attached flow; 
cyan solid lines track the instantaneous shock position 
and cyan dashed lines mark its time average.   
}
\label{fig:sep}
\end{figure}

\red{
A positive correlation is observed in the upstream 
supersonic region, associated with a coherent disturbance 
that undergoes a $180^\circ$
phase reversal upstream of the shock (cyan dashed line). 
The wall region downstream of 
the shock exhibits strong dependence on the flow regime. 
At $\alpha=6^\circ$ (figure \ref{fig:rpp}a), the trajectory 
is split at the mean separation point (green dashed line), 
from which information travels in two opposite directions: 
upstream towards the shock, and downstream towards the 
trailing edge. Since there is no continuous path 
connecting the shock to the near wake, 
a single downstream propagation time cannot be extracted.
Conversely, 
at $\alpha=7^\circ$ (figure \ref{fig:rpp}b), the dominant 
disturbance travels in the downstream direction only. 
The time lag accumulated 
along the correlation ridge between the mean shock position 
and the end of the sampling path provides an estimate 
of the downstream propagation time, $t_{\mathrm{dn}}\approx3.7\,c/u_0$, 
which also includes the convection from the trailing edge 
into the near wake.
}

\red{
The difference in the propagation trajectories of pressure 
disturbances between the incipient and fully developed 
buffet regimes is rooted in the dynamics of separation. 
Figure 
\ref{fig:sep} shows the spatiotemporal evolution of the separation
height $n_{\psi=0}$, computed as the wall-normal distance 
to the dividing 
streamline ($\psi=0$) after spanwise averaging. 
The cyan solid line tracks the instantaneous 
shock position and the 
cyan dashed line marks its time average. 
At $\alpha=6^\circ$ (figure \ref{fig:sep}a), a 
region of attached mean flow persists between the shock and 
the trailing-edge separation at all times, as shown by the 
black contours corresponding to $n_{\psi=0}=0$. 
Under fully developed buffet conditions (figure \ref{fig:sep}b), 
the separation region expands during the upstream shock motion,
as anticipated in \S\ref{sec:visua} and in agreement with the 
experimental findings of \citet{schauerte2023experimental}. 
The boundary layer remains fully separated for nearly half buffet 
period with peaks of $n_{\psi=0}>0.06\,c$ 
near the trailing edge. During the remaining portion, the flow 
is mostly attached. This intermittency explains the statistical 
backflow probability in figure \ref{fig:bl1}(d).
}

\red{
The shift in propagation pathways shown in figure \ref{fig:rpp} is 
associated with the periodic establishment of complete flow separation, 
which is a distinctive feature of fully developed buffet. 
An attached flow allows disturbances to propagate upstream 
through the boundary layer. Upon full separation, 
downstream-travelling pressure waves are dominant, 
in agreement with experimental observations \citep{roos1980some, lee2001self}. 
}
In addition, a fully detached flow promotes the rapid 
growth of shear-layer instabilities and their roll-up into coherent 
vortical structures \citep{pearcey1962simple, jacquin2009experimental, 
d2021experimental}. The enhanced shear associated with the upstream-moving 
shock coincides with stronger large-scale disturbances, which are convected 
towards the trailing edge and accompanied by increased acoustic scattering 
activity in the near wake \citep{hartmann2013interaction}. 

%==============================================
\subsection{Aeroacoustic feedback assessment}\label{sec:loop}
%==============================================
Interpreting the buffet instability as a self-sustained aeroacoustic 
feedback implies that the buffet period, $T_B$, is governed by the 
cumulative travel time of the downstream- and upstream-propagating 
disturbances,
$T_B = t_{\mathrm{dn}} + t_{\mathrm{up}}$,
where $t_{\mathrm{dn}}$ and $t_{\mathrm{up}}$ denote the downstream and
upstream propagation times, respectively. 
\red{
The propagation times identified in \S\ref{sec:acoustic} bracket the 
plausible range of $t_{\mathrm{up}}$, from $t_b\approx3.5\,c/u_0$ 
to $t_f\approx5.5\,c/u_0$. In \S\ref{sec:lin}, we showed that a normal 
shock is more receptive to front-impinging waves, exhibiting a response
approximately twice that of back-impinging disturbances. 
The front-impinging path is therefore expected to dominate the 
upstream feedback, placing $t_{\mathrm{up}}$ near the upper bound 
of the range.
Combining this range with the downstream convective time, 
$t_{\mathrm{dn}}\approx3.7\,c/u_0$ obtained in \S\ref{sec:conv}, 
gives a predicted period in the range 
$7.2$ -- $9.2\,c/u_0$, 
compared with the DNS value of $T_B\approx9.1\,c/u_0$. 
The agreement with the upper bound supports the front-impinging 
path as the relevant upstream leg.
These estimates should be considered characteristic time scales 
rather than an exact prediction, as they are based on the 
geometrical acoustics approximation and do not take into 
account the time lag between a shear-layer disturbance reaching 
the near wake and the subsequent acoustic emission.
}

\begin{figure}
\centering
(a)\includegraphics[width=.49\textwidth]{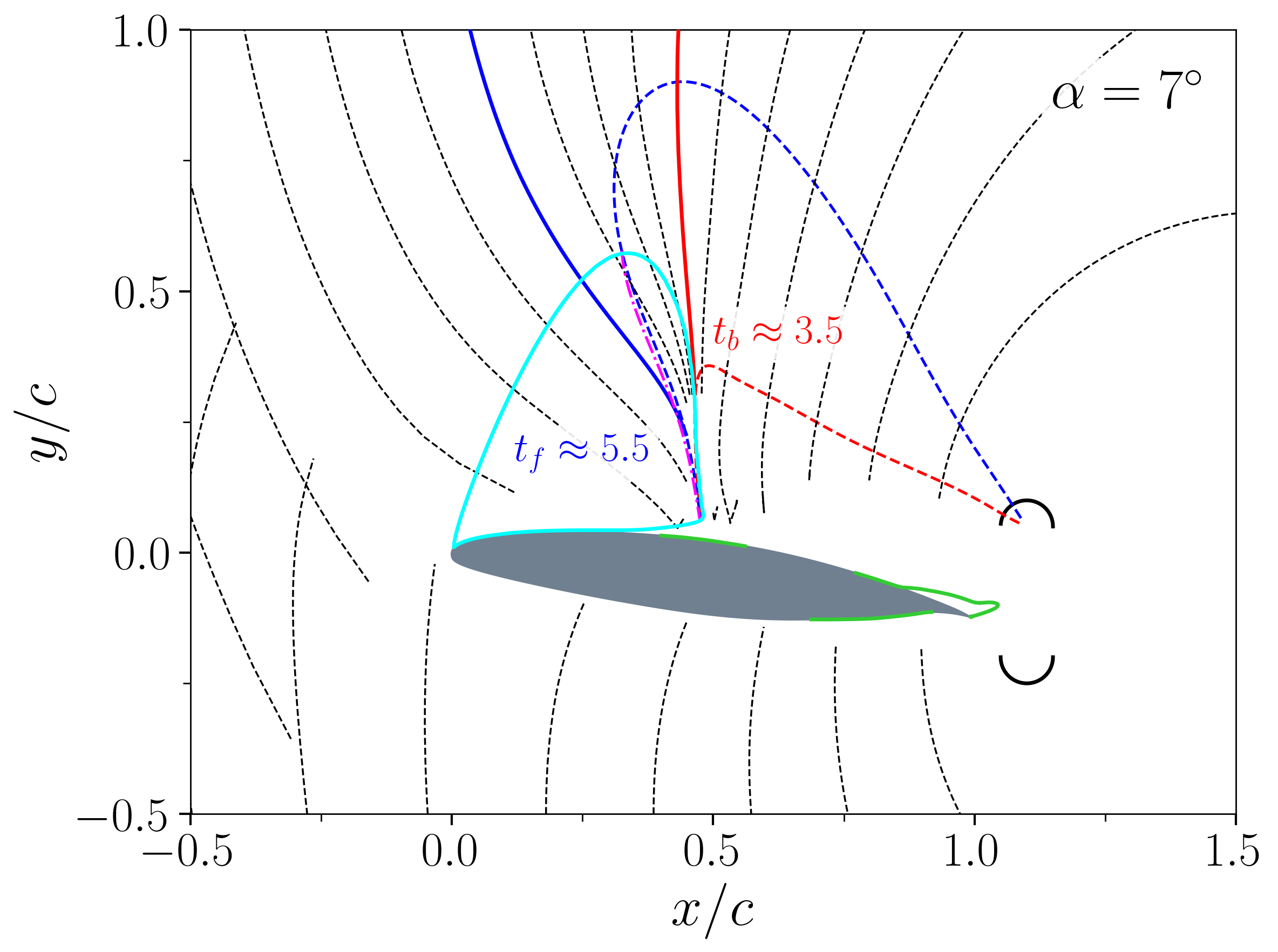}
(b)\includegraphics[width=.442\textwidth]{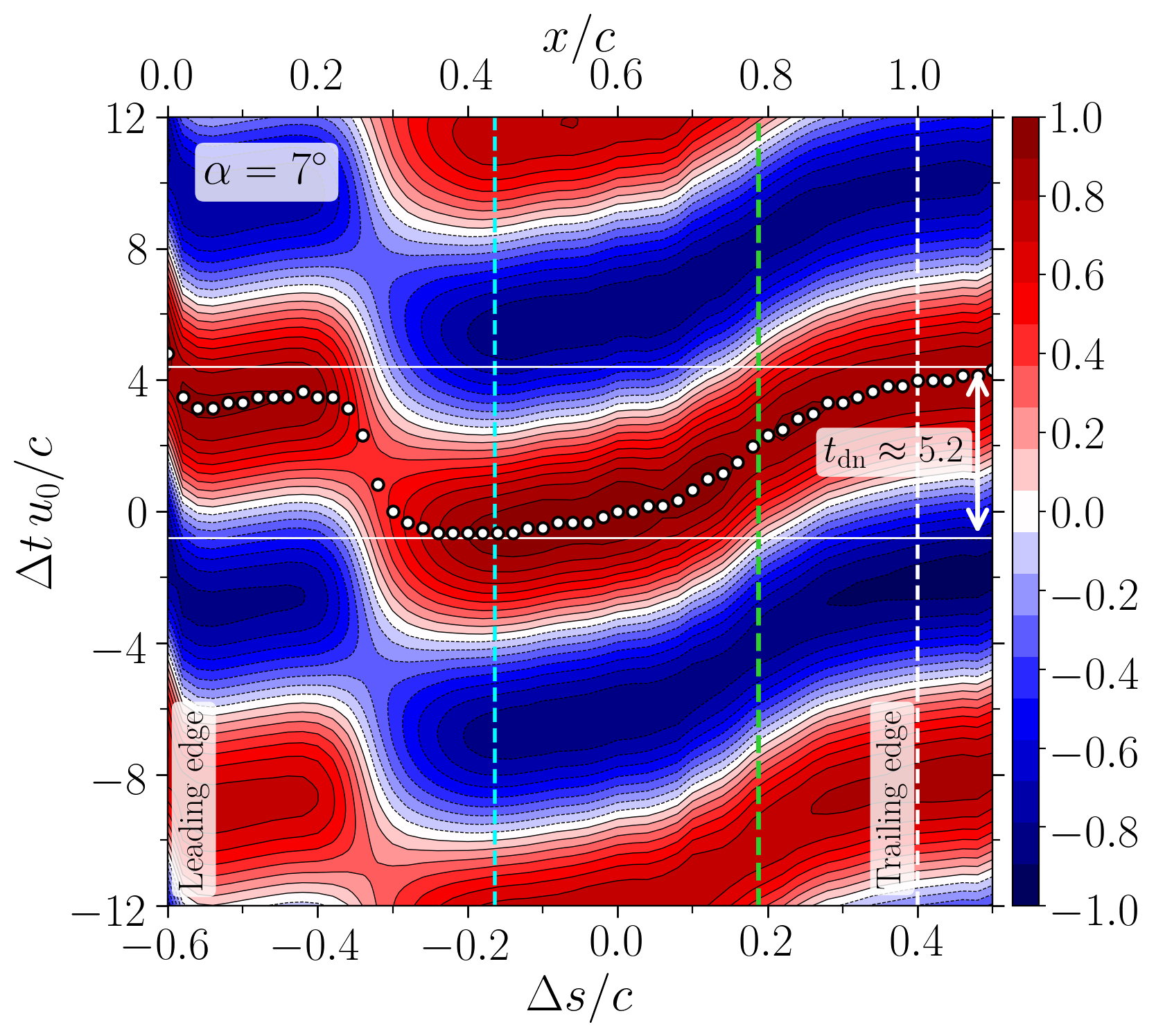}
\caption{
Assessment of the aeroacoustic feedback for the buffet case 
at $\Rey=3\times10^5$. 
(a) Wave fronts computed via geometrical acoustics. 
The red and blue fronts correspond to back- and front-impinging 
waves at the shock, with associated propagation times $t_b$ and $t_f$. 
(b) Space--time cross-correlation of the fluctuating pressure \eqref{eq:rpp}. 
White circles trace the locus of $\max_{\Delta t}(R_{pp})$; 
the cyan dashed line marks the mean shock position.
}
\label{fig:lowre}
\end{figure}

\begin{figure}
\centering
(a)\includegraphics[width=.47\textwidth]{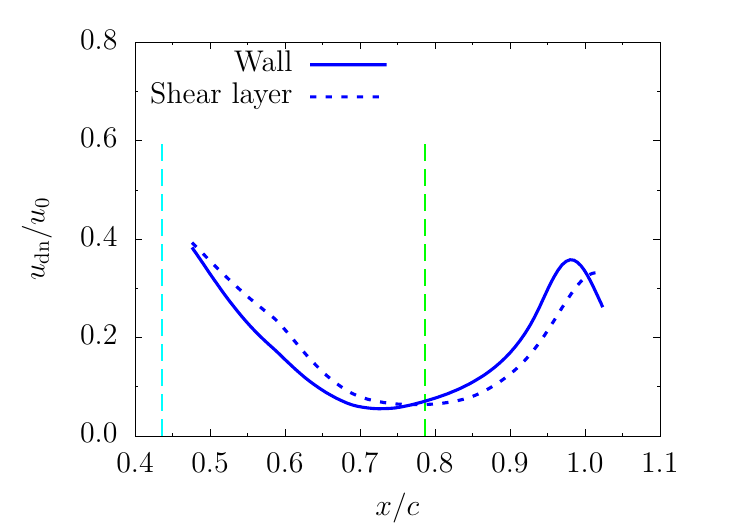}
(b)\includegraphics[width=.47\textwidth]{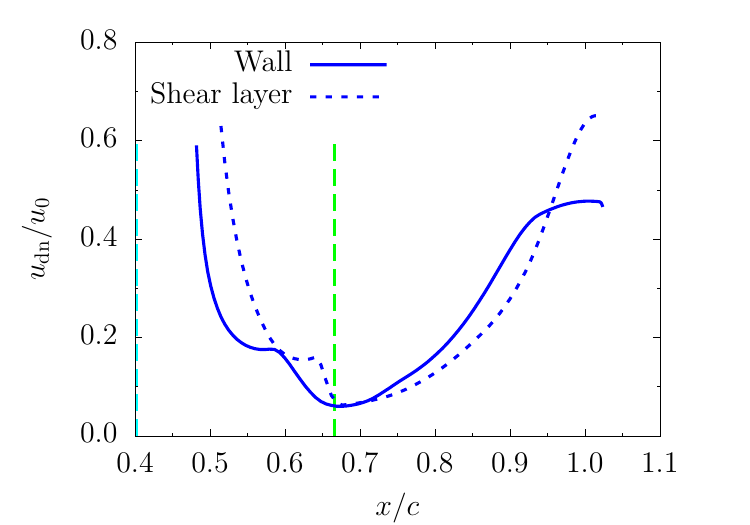}
\caption{
Downstream convection speed ($u_\mathrm{dn}$) estimated from the 
space--time pressure correlations for the fully developed buffet 
cases at $\Rey=3\times10^5$ (a) and $6\times10^5$ (b). 
Dashed and solid lines denote data sampled along 
the shear-layer path and at the wall. 
Cyan and green vertical lines mark the mean shock position 
and the mean separation point. 
}
\label{fig:conv} 
\end{figure}

\blue{
The low-Reynolds-number buffet case provides an additional test 
to assess the robustness of the identified aeroacoustic feedback. 
We thus apply the same geometrical acoustics approach to the flow case 
at $\Rey=3\times10^5$ and $\alpha=7^\circ$. The resulting ray paths, 
shown in figure \ref{fig:lowre}(a), are almost identical 
to those at $\Rey=6\times10^5$ in figure \ref{fig:GA}(b), 
giving similar propagation times of 
$t_b\approx3.5\,c/u_0$ and $t_f\approx5.5\,c/u_0$. 
}
\blue{
As for the downstream leg, we consider an analogous sampling 
path extending from the leading edge to 
$x=1.1\,c$ at a wall-normal offset of $n\approx0.02\,c$, 
corresponding to the peak of the mean shear at $x=0.6\,c$. 
The convective time inferred from the correlation ridge 
in figure \ref{fig:lowre}(b) is $t_{\mathrm{dn}}\approx5.2\,c/u_0$. 
Combined with the upstream propagation time, 
the predicted range is $8.7\text{ -- }10.7\,c/u_0$, 
compared with a DNS value of $T_B\approx11.1\,c/u_0$ 
(corresponding to $\St_B\approx0.09$, as listed in table \ref{tab:1}). 
As in the high Reynolds number case, the measured 
period falls close to the front-impinging upper bound. 
}

\blue{
Notably, the downstream path is slightly shorter at the lower 
Reynolds number, as the mean shock is located further 
downstream ($\overline{x}_s=0.436\,c$ against $0.403\,c$). 
This rules out a geometrical origin for the longer $t_{\mathrm{dn}}$, 
which instead arises from a lower convection speed ($u_{\mathrm{dn}}$). 
Figure \ref{fig:conv} compares the convection speed obtained by 
sampling the pressure field along the shear-layer path
and at the wall. The local convection speed, 
$u_{\mathrm{dn}}=\mathrm{d}(\Delta s)/\mathrm{d}(\Delta t)$, 
is obtained by differentiating a spline fit to the corresponding 
correlation ridge. The substantial overlap between the two profiles 
indicates that the inferred propagation speed is only slightly affected 
by the wall-normal offset of the sampling path.
Downstream of the mean shock location (cyan dashed line), 
$u_{\mathrm{dn}}$ decreases to a minimum close to the mean 
separation point (green dashed line), before increasing towards 
the near wake. While the two Reynolds numbers exhibit a comparable 
spatial variation, they differ in magnitude, with $u_{\mathrm{dn}}$ 
being consistently higher at $\Rey=6\times10^5$. 
This systematic difference, accumulated along the propagation path, 
accounts for the longer downstream propagation time at $\Rey=3\times10^5$.
}

%==============================================
\section{Conclusions}\label{sec:conclusions}
%==============================================
We have presented a fully resolved DNS database of turbulent transonic buffet 
over a supercritical airfoil at freestream Mach number $M=0.7$.  
\green{
The database covers the transition 
from stable flow ($\alpha=4^\circ$) to fully developed buffet 
($\alpha=7^\circ$) at chord-based Reynolds numbers of 
$\Rey=3\times10^5$ and $6\times10^5$. 
At both Reynolds numbers, buffet onset is preceded by a 
reversal in the direction of the mean shock motion as 
the incidence increases, confirming this reversal 
as a necessary precursor to the instability \citep{nitzsche2019fluid, accorinti2022experimental, schauerte2023experimental}.
}

\blue{
The Reynolds number controls the state of the incoming boundary layer.
At $\Rey=3\times10^5$, the momentum-thickness Reynolds number upstream
of the shock remains below the threshold for fully developed turbulence,
and the mean velocity profiles lack a logarithmic region.
At $\Rey=6\times10^5$, the velocity profiles collapse onto
canonical zero-pressure-gradient turbulent boundary-layer data.
This difference in the incoming boundary layer results in distinct
mean-flow separation. At the highest incidence, a shock-induced
separation bubble forms at the lower Reynolds number only, while the 
extent of trailing-edge separation increases in the fully turbulent case. 
}
\blue{
The buffet Strouhal number, $\St_B=fc/u_0$, remains close to 
$0.1$ across all configurations, showing a weak dependence on 
the incidence angle, despite the substantial growth of the separated 
region with increasing incidence.
The absence of a correspondingly large variation in frequency suggests 
that the separation length alone does not determine the characteristic 
time scale of the instability.  
}

\red{
Modal analysis (SPOD) reveals that a single 
eigenmode dominates the low-frequency dynamics from the pre-buffet
regime to the fully developed buffet. The energy of this mode 
increases by approximately
one order of magnitude per degree of incidence, whereas its spatial structure
and fundamental frequency remain nearly unchanged. 
This behaviour is consistent with
the growth and weakly nonlinear saturation of a single global instability
into a limit cycle 
\citep{crouch2009origin, crouch2024weakly, sansica2022system}. 
At the buffet frequency, the modal pressure fluctuations at the shock and
in the near wake are in antiphase, 
connecting the shock motion with the periodic growth and collapse 
of trailing-edge separation.
Two additional modes are identified across all flow regimes.
A high-frequency mode is dominated by von K\'arm\'an-type 
vortex shedding, and an intermediate-frequency mode is 
characterised by upstream-propagating acoustic disturbances 
associated with near-wake unsteadiness.
}

\red{
The DNS results allow us to examine each component of the 
aeroacoustic feedback loop proposed by \citet{lee1990oscillatory}.
Using the turbulent shear stress as a proxy for the acoustic source 
distribution indicates that the dominant source is located in the 
near wake rather than at the trailing edge, where large-scale separation 
limits the applicability of a compact dipole representation. 
Using geometrical acoustics, 
we identify two distinct paths connecting this
source region to the shock. A shorter path, which is typically 
invoked in aeroacoustic feedback models, approaches the shock from the 
downstream side, whereas a longer path traverses the supersonic 
region before impinging on the shock foot from the front. 
}
\blue{
This front-impinging path, which is associated with a propagation time 
of $t_{\mathrm{up}}\approx5.5\,c/u_0$ and is almost independent of 
both angle of attack and Reynolds number, tracks the 
characteristic line that has previously 
been identified as the strongest receptivity region 
\citep{sartor2015stability, paladini2019transonic, iwatani2023identifying}. 
}
Using a linearised shock-response theory, we find that a normal 
shock responds to acoustic forcing as a low-pass filter, 
with amplitude inversely proportional to frequency. 
The theory also predicts a stronger response to front-impinging 
than to back-impinging disturbances. 

\blue{
We examine the downstream leg of the loop using the space--time 
correlation coefficient of the fluctuating pressure field. 
In the incipient buffet regime, no continuous path 
connecting the shock to the near wake is observed. 
In the fully developed buffet regime, where the boundary layer 
separates completely during the upstream shock motion, 
coherent disturbances originate near the shock foot and 
convect downstream. Depending on the Reynolds number, 
the characteristic propagation time decreases from 
$t_{\mathrm{dn}}\approx5.2\,c/u_0$ to $3.7\,c/u_0$,
due to marked variations in the convection speed.
Adding the downstream convective time to the upstream 
acoustic time yields a predicted period of approximately 
$10.7\,c/u_0$ at $\Rey=3\times10^5$ and 
$9.2\,c/u_0$ at $\Rey=6\times10^5$, compared with the DNS 
values of $11.1\,c/u_0$ and $9.1\,c/u_0$, respectively.
}

\blue{
This agreement is consistent with the interpretation of 
buffet instability as an aeroacoustic feedback loop.
Within this framework, the upstream path appears to be 
almost insensitive to the angle of attack, suggesting that 
the onset of self-sustained oscillations is closely associated 
with the establishment of the downstream path. The latter 
emerges when a fully separated boundary layer intermittently 
develops, allowing pressure disturbances generated at the shock 
foot to propagate into the near wake. There, the scattering of 
large-scale shear-layer disturbances is accompanied by the generation of 
low-frequency, long-wavelength acoustic waves, which can excite the unstable 
global mode via the upstream path. Furthermore, the dependence of 
the downstream propagation time on the incoming boundary-layer 
state can provide a physical explanation for the observed 
Reynolds-number dependence of the buffet frequency.
}

\blue{
Taken together, these results connect several aspects of transonic 
buffet that have so far been considered separately. The persistence of 
a global mode across onset as described by stability analyses, 
the stronger receptivity to front-impinging perturbations 
as indicated by resolvent analyses, and the emergence of 
the downstream path with a fully separated boundary layer 
consistent with studies emphasising the key role of 
separation, can be identified as different elements of the 
same cycle.
}

\red{
We acknowledge that the present assessment of the aeroacoustic 
feedback relies on several approximations: 
1) geometrical acoustics 
is applied to the mean flow, 2) the acoustic source is localised 
through a proxy rather than from a direct evaluation of the 
Lighthill source term, and 3) the two characteristic propagation 
times are simply summed, without accounting for the time lag 
between a shear-layer disturbance reaching the near wake and the 
subsequent emission of an acoustic pulse. Each of these approximations 
introduces an uncertainty that is not quantified here. Therefore, 
the resulting agreement with the DNS results does not, 
by itself, determine which path, or combination of paths, 
governs the buffet period. 
A parallel study based on URANS simulations 
and linear stability analysis is currently underway to 
complement the present DNS database and investigate 
aspects of the buffet mechanism that remain beyond 
the scope of the present study.
}

\blue{
To the best of our knowledge, this is the first 
fully-resolved DNS database of 
turbulent transonic buffet, spanning the transition across the onset 
point at multiple Reynolds numbers. 
This database has enabled a detailed investigation of the 
physical mechanisms of buffet and provides a high-fidelity 
reference for calibrating and evaluating lower-order approaches, 
as well as a controlled two-dimensional 
baseline for future studies of buffet on swept wings.
}

\appendix 
%==============================================
\section{Numerical validation}\label{app:val}
%==============================================
\begin{figure}
\centering
(a)\includegraphics[width=.47\textwidth]{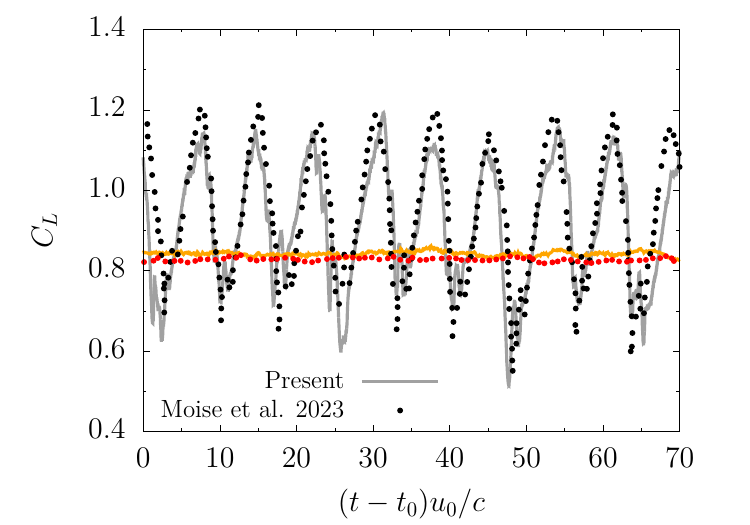}
(b)\includegraphics[width=.47\textwidth]{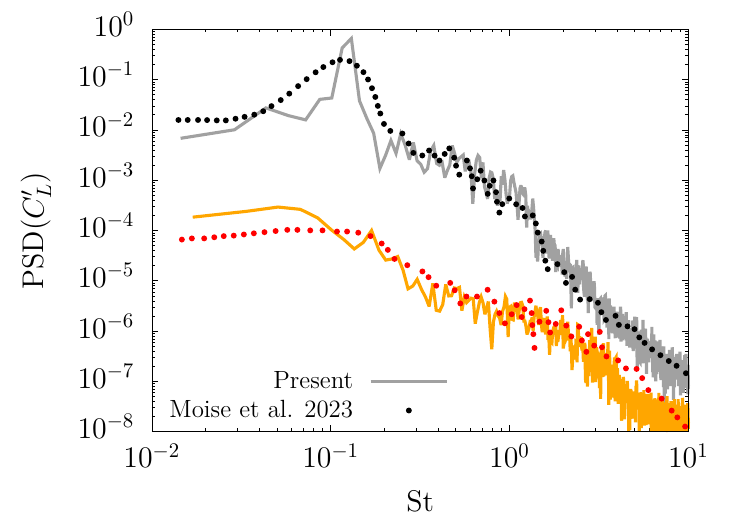}
\caption{  
Time-history of lift coefficient (a) and fluctuating lift spectra (b)  for flow cases at $\Rey = 6 \times 10^5$, $\alpha=5^\circ$, $M=0.7$, tripped boundary layer (orange line) and $\Rey = 5 \times 10^5$, $\alpha=7^\circ$, $M=0.7$, untripped boundary layer (grey line). The results are compared with reference data from \citet{moise2023transonic} at matching flow conditions (except for the Reynolds number of $\Rey = 5 \times 10^5$ in the turbulent case). 
}
\label{fig:val} 
\end{figure}

\red{
The present DNS results are compared with the implicit LES results of \citet{moise2023transonic} 
to validate the computational approach. The comparison 
considers two distinct flow regimes at $M=0.7$, including a fully developed buffet 
case with an untripped boundary layer and a tripped pre-buffet configuration. 
The laminar case was specifically designed for validation purposes and matches the 
flow conditions of the reference study ($\Rey = 5 \times 10^5$, $\alpha=7^\circ$). 
For the turbulent case, there is a slight difference in the Reynolds number between 
the setups ($\Rey = 6 \times 10^5$ in the present DNS versus $5 \times 10^5$ in the 
reference). Furthermore, both airfoil sides are tripped in the present setup, 
whereas in the reference study only the suction side is tripped.
}
\red{
Despite these minor differences, the time histories of the lift 
coefficient in figure \ref{fig:val}(a) show comparable oscillation amplitudes, 
and the spectra in figure \ref{fig:val}(b) confirm that the same dominant 
frequencies are captured in both regimes.
}

%==============================================
\section{Span-width sensitivity}\label{app:span}
\begin{figure}
\centering
(a)\includegraphics[width=.47\textwidth]{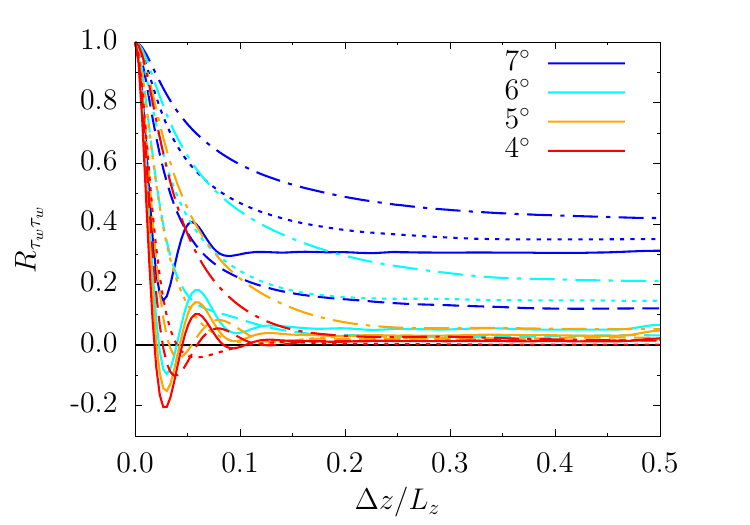}
(b)\includegraphics[width=.47\textwidth]{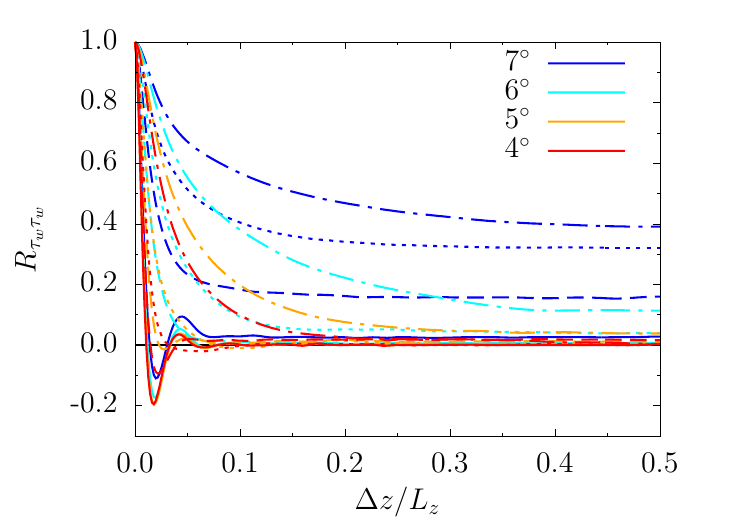}
\caption{
Spanwise two-point correlation of the wall shear stress at four streamwise 
locations along the suction side: $x/c=0.25$ (solid), $0.45$ (dashed), 
$0.65$ (dotted) and $0.85$ (dash-dotted), at $\Rey=3\times10^{5}$ (a) and 
$6\times10^{5}$ (b) at various angles of attack.
}
\label{fig:tpc} 
\end{figure}

%==============================================
\blue{
The adequacy of the computational span across all flow regimes is assessed in 
figure~\ref{fig:tpc}, which shows the spanwise two-point correlation of the 
wall shear stress, $R_{\tau_w\tau_w}(\Delta z)$. Four streamwise stations are 
considered along the suction side, spanning the upstream boundary layer 
(solid line), the shock-interaction region (dashed line), the downstream boundary 
layer (dotted line) and the trailing edge area (dash-dotted line). 
}

\blue{
In the pre-buffet regimes ($\alpha=4^\circ,\,5^\circ$), spanwise correlation decays 
to near zero across all stations for both $\Rey=3\times10^{5}$ (figure~\ref{fig:tpc}a) 
and $6\times10^{5}$ (figure~\ref{fig:tpc}b), demonstrating the complete decorrelation 
of the three-dimensional structures. As the incidence increases to $\alpha=6^\circ$, 
residual spanwise coherence emerges downstream of the interaction, driven by intermittent 
shock-induced and massive trailing-edge separations that sustain a quasi-two-dimensional 
footprint on the wall. In fully developed buffet conditions, the upstream boundary layer 
only achieves full decorrelation at the higher Reynolds numbers, whereas at the lower 
Reynolds numbers a residual correlation extends to the incoming flow. Although the 
magnitude of the residual correlation increases with incidence, reflecting the enlarged 
separation region, it consistently plateaus before reaching the spanwise boundary. 
This confirms that the chosen span width adequately captures the shock-buffet dynamics 
without imposing artificial confinement.
}

%==============================================
\section{Interaction of acoustic waves with a plane shock}\label{app:lin}
%==============================================
%===========================================================
\subsection{Linearised Euler equations}\label{app:euler}
%===========================================================
We consider the one-dimensional flow of an inviscid ideal gas, and 
we adopt a linear perturbation approach, whereby the instantaneous
flow variables are decomposed into a steady, uniform mean state and 
small unsteady fluctuations,
$q(x,t) = \bar{q}_i + q'_i(x,t)$,
where the subscript $i$ denotes the upstream ($i=1$) or downstream ($i=2$) region, 
the overline denotes mean quantities, and the prime 
denotes perturbations of order $O(\varepsilon)$ with $\varepsilon \ll 1$.
Substituting this decomposition into the Euler equations yields
\begin{subequations}
\label{eq:euler}
\begin{align}
\frac{\partial \rho'_i}{\partial t} + \bar{u}_i\frac{\partial \rho'_i}{\partial x}
+ \bar{\rho}_i\frac{\partial u'_i}{\partial x} &= 0, \label{eq:mass}\\
\bar{\rho}_i \left(\frac{\partial u'_i}{\partial t} 
+ \bar{u}_i\frac{\partial u'_i}{\partial x}\right)
+ \frac{\partial p'_i}{\partial x} &= 0, \label{eq:momentum}\\
\frac{\partial s'_i}{\partial t} 
+ \bar{u}_i\frac{\partial s'_i}{\partial x} &= 0. \label{eq:entropy}
\end{align}
\end{subequations}

%==============================================
\subsection{Linearised unsteady Rankine--Hugoniot conditions}\label{app:RH}
%==============================================
Let $x(t)$ denote the instantaneous position of a normal shock and 
$u_s(t)=\mathrm{d} x / \mathrm{d} t$ the shock velocity. 
The instantaneous Rankine--Hugoniot conditions for mass, momentum, and energy are
\begin{subequations}
\label{eq:RH}
\begin{align}
\rho_1\left(u_1-u_s\right) &= \rho_2\left(u_2-u_s\right),\\[4pt]
p_1+\rho_1\left(u_1-u_s\right)^2&= p_2+\rho_2\left(u_2-u_s\right)^2,\\
h_1+\frac{1}{2}\left(u_1-u_s\right)^2 &= h_2+\frac{1}{2}\left(u_2-u_s\right)^2. 
\end{align}
\end{subequations}
Linearising these relations about the steady normal shock state and evaluating
at the mean shock position ($x=0$) yields the first-order jump conditions:
\begin{subequations}
\label{eq:RHlin}
\begin{align}
u'_2 + \frac{\bar{u}_2}{\bar{\rho}_2\bar{c}_2^2}\,p'_2
+ \frac{\bar{u}_2-\bar{u}_1}{\bar{u}_1}\,u_s
- \frac{\bar{u}_2}{c_p}\,s'_2
&= \frac{\bar{u}_2}{\bar{u}_1}\,u'_1
+ \frac{\bar{u}_1}{\bar{\rho}_2\bar{c}_1^2}\,p'_1
- \frac{\bar{u}_2}{c_p}\,s'_1, \label{eq:RHmass}\\[8pt]
u'_2+\frac{1+{M}_2^2}{2\bar{\rho}_2\bar{u}_2}\,p'_2
- \frac{1}{2}\frac{\bar{u}_2}{c_p}\,s'_2
&= u'_1+\frac{1+{M}_1^2}{2\bar{\rho}_2\bar{u}_2}\,p'_1
- \frac{1}{2}\frac{\bar{u}_1}{c_p}\,s'_1, \label{eq:RHmomentum}\\[8pt]
u'_2 + \frac{p'_2}{\bar{\rho}_2\bar{u}_2}
- \frac{\bar{u}_2-\bar{u}_1}{\bar{u}_2}\,u_s
+ \frac{\bar{T}_2}{\bar{u}_2}\,s'_2
&= \frac{p'_1}{\bar{\rho}_1\bar{u}_2}
+ \frac{\bar{u}_1}{\bar{u}_2}\,u'_1 
+ \frac{\bar{T}_1}{\bar{u}_2}\,s'_1, \label{eq:RHenergy}
\end{align}
\end{subequations}
where ${M}_i=\bar{u}_i/\bar{c}_i$ denotes the mean Mach number on side $i$ 
of the shock. 

%==============================================
\subsection{Acoustic disturbance incident from downstream}\label{app:back}
%==============================================
We seek monochromatic plane-wave perturbations in each uniform region. 
By replacing the {\em ansatz} 
\begin{equation}
\phi(x,t)=\widehat{\phi}\,e^{i(\omega t - k x)}
\label{eq:plane_wave}
\end{equation}
in the convected wave equation \eqref{eq:convected} yields the dispersion 
relation
\begin{equation}
(\omega-{u}k)^2-{c}^2 k^2=0 \quad\Longrightarrow\quad \omega = k({u}\pm {c}).
\label{eq:dispersion}
\end{equation}
We consider an incident acoustic plane wave coming from downstream 
(subsonic region \(2\)) and a reflected acoustic wave:
\begin{equation}
\phi_2(x,t)=\big(A_i e^{-ik_i x} + A_r e^{-ik_r x}\big)e^{i\omega t},
\label{eq:phi2a}
\end{equation}
where
\begin{equation}
k_i=\frac{\omega}{\bar{u}_2-\bar{c}_2}=\frac{\omega}{\bar{c}_2({M}_2-1)},\qquad
k_r=\frac{\omega}{\bar{u}_2+\bar{c}_2}=\frac{\omega}{\bar{c}_2({M}_2+1)}.
\label{eq:ki_kr}
\end{equation}
The shock interaction produces 
entropy perturbations that are convected with the mean flow: 
\begin{equation}
s'_2(x,t)=c_p\,B_b\,e^{i(\omega t - k_s x)},\qquad k_s=\frac{\omega}{\bar{u}_2},
\label{eq:entropy_convected}
\end{equation}
where the subscript $b$ denotes the back impact scenario. 
Using \(u'=\partial_x\phi\) and \eqref{eq:bernoulli} we obtain the velocity and 
pressure perturbations at the mean shock location (\(x=0\)) 
\begin{equation}
u'_2 = -i\big(k_i A_i + k_r A_r\big), \qquad
p'_2 = i\bar{\rho}_2\bar{c}_2\big(k_i A_i - k_r A_r\big).
\end{equation}
For compactness introduce the coefficients
\begin{equation}
\alpha_i^{(u)}\equiv -ik_i,\quad \alpha_r^{(u)}\equiv -ik_r,\quad
\alpha_i^{(p)}\equiv i\bar{\rho}_2\bar{c}_2 k_i,\quad 
\alpha_r^{(p)}\equiv -i\bar{\rho}_2\bar{c}_2 k_r,
\label{eq:alpha_def}
\end{equation}
so that \(u'_2=\alpha_i^{(u)}A_i+\alpha_r^{(u)}A_r\) 
and \(p'_2=\alpha_i^{(p)}A_i+\alpha_r^{(p)}A_r\).
For a supersonic upstream flow there are no transmitted acoustic 
or entropy perturbations upstream, hence
\begin{equation}
u'_1=p'_1=s'_1=0.
\label{eq:upstream_zero}
\end{equation}
Setting \(u_s=i\omega X_b\) and substituting the downstream expressions 
into the linearised jump conditions \eqref{eq:RHlin} 
yields a linear system for the unknowns \((A_r,B_b,X_b)\) forced by \(A_i\). 
After collecting terms the system may be written as
\begin{equation}
\mathbf{A}\begin{bmatrix}A_r\\[2pt] B_b\\[2pt] X_b\end{bmatrix} + \mathbf{b}\,A_i=\mathbf{0},
\label{eq:matrix_system}
\end{equation}
with the coefficient matrix and forcing vector obtained directly 
from \eqref{eq:RHlin} and \eqref{eq:alpha_def}. 
For the algebraic simplifications we use \(\bar{u}_2={M}_2\bar{c}_2\) 
and the modal identities \(\omega-\bar{u}_2 k_i=-\bar{c}_2 k_i\), 
\(\omega-\bar{u}_2 k_r=+\bar{c}_2 k_r\).
Carrying out the algebra yields the reduced linear system
\begin{equation}
\begin{bmatrix}
- \dfrac{i\omega}{\bar{c}_2} 
& -{{M}_2\bar{c}_2} 
& \dfrac{\bar{u}_2-\bar{u}_1}{\bar{u}_1}i\omega\\[12pt]
- \dfrac{i\omega}{2\bar{c}_2}\dfrac{{M}_2+1}{{M}_2} 
& -\dfrac{1}{2}{{M}_2\bar{c}_2} 
& 0\\[12pt]
- \dfrac{i\omega}{{M}_2\bar{c}_2} 
& \dfrac{\bar{T}_2c_p}{\bar{u}_2} 
& -\dfrac{\bar{u}_2-\bar{u}_1}{\bar{u}_2}i\omega
\end{bmatrix}
\begin{bmatrix}A_r\\[8pt]B_b\\[8pt]X_b\end{bmatrix}
+
\begin{bmatrix}
\;\dfrac{i\omega}{\bar{c}_2}\;\\[12pt]
\;\dfrac{i\omega}{2\bar{c}_2}\dfrac{{M}_2-1}{{M}_2}\;\\[12pt]
- \dfrac{i\omega}{{M}_2\bar{c}_2}
\end{bmatrix} A_i
= \mathbf{0}.
\label{eq:reduced_system}
\end{equation}
From the second row of \eqref{eq:reduced_system} one obtains \(B_b\) 
explicitly in terms of \(A_r\) and \(A_i\). After simplification,
\begin{equation}
B_b = \frac{i\omega}{\bar{u}_2^2}\Big[({M}_2-1)A_i-({M}_2+1)A_r\Big], 
\label{eq:B_expr}
\end{equation}
Substituting \eqref{eq:B_expr} into the first row and solving for \(X_b\) gives
\begin{equation}
X_b = -\,\frac{\bar{u}_1}{(\bar{u}_2-\bar{u}_1)\,\bar{u}_2}\,(A_r + A_i).
\label{eq:X_expr}
\end{equation}
Finally, substituting both \(B_b\) and \(X_b\) into the third row produces 
a linear relation between \(A_r\) and \(A_i\). 
Using the perfect-gas identity \(c_p\bar{T}_2=\bar{c}_2^2/(\gamma-1)\) 
and simplifying yields the acoustic reflection 
coefficient $\mathcal{R}=A_r/A_i$
\begin{equation}
\mathcal{R} =
\frac{M_2\left[2 \gamma M_1^2-(\gamma-1)\right]
-\left[2 M_1^2+(\gamma-1)\right]}
{M_2\left[2 \gamma M_1^2-(\gamma-1)\right]
+\left[2 M_1^2+(\gamma-1)\right]}.
\label{eq:Ar_to_Ai}
\end{equation}

%==============================================
\subsection{Acoustic disturbance incident from upstream}\label{app:front}
%==============================================
We consider a monochromatic acoustic plane wave incident from the 
upstream region 1. Since the upstream flow is supersonic, acoustic 
disturbances cannot propagate upstream against the flow. 
Consequently, there is no reflected wave in region 1.
The incident potential is given by
\begin{equation}
\phi_1(x,t) = A_i\, e^{i(\omega t - k_1 x)},
\label{eq:phi1a}
\end{equation}
where the wavenumber corresponds to the downstream-propagating mode
\begin{equation}
k_1 = \frac{\omega}{\bar{u}_1+\bar{c}_1} = \frac{\omega}{\bar{c}_1({M}_1+1)}.
\label{eq:k1_def}
\end{equation}
In the subsonic region downstream, we seek a transmitted acoustic wave
and an entropy wave:
\begin{equation}
\phi_2(x,t) = A_t\, e^{i(\omega t - k_t x)}, \qquad
s'_2(x,t) = c_p\, B_f\, e^{i(\omega t - k_s x)},
\label{eq:downstream_ansatz}
\end{equation}
where the subscript $f$ denotes the front impact scenario. 
Here \(k_t\) is identical to the reflected wavenumber \(k_r\) defined 
in \eqref{eq:ki_kr}, as both waves propagate downstream relative to the shock, 
while \(k_s=\omega/\bar{u}_2\).
At the mean shock location \(x=0\), the downstream perturbations are
\begin{equation}
u'_2 = -ik_t A_t \equiv \alpha_t^{(u)} A_t, \qquad
p'_2 = -i\bar{\rho}_2\bar{c}_2 k_t A_t \equiv \alpha_t^{(p)} A_t, 
\end{equation}
while the upstream perturbations are determined by the incident wave:
\begin{equation}
u'_1 = -ik_1 A_i, \qquad
p'_1 = \bar{\rho}_1\bar{c}_1 u'_1.
\label{eq:upstream_pert}
\end{equation}
Setting \(u_s=i\omega X_f\) and substituting these expressions
into the linearised jump conditions \eqref{eq:RHlin}
yields a linear system for the unknowns \((A_t,B_f,X_f)\) forced by \(A_i\).
Using the identity \(\bar{\rho}_1\bar{u}_1=\bar{\rho}_2\bar{u}_2\) and simplifying,
the system may be written as
\begin{equation}
\begin{bmatrix}
- \dfrac{i\omega}{\bar{c}_2}
& -{{M}_2\bar{c}_2}
& \dfrac{\bar{u}_2-\bar{u}_1}{\bar{u}_1}i\omega\\[12pt]
- \dfrac{i\omega}{2\bar{c}_2}\dfrac{{M}_2+1}{{M}_2}
& -\dfrac{1}{2}{{M}_2\bar{c}_2}
& 0\\[12pt]
- \dfrac{i\omega}{{M}_2\bar{c}_2}
& \dfrac{\bar{T}_2 c_p}{\bar{u}_2}
& -\dfrac{\bar{u}_2-\bar{u}_1}{\bar{u}_2}i\omega
\end{bmatrix}
\begin{bmatrix}A_t\\[8pt]B_f\\[8pt]X_f\end{bmatrix}
+
\begin{bmatrix}
\;\dfrac{i\omega}{\bar{c}_1}\dfrac{\bar{u}_2}{\bar{u}_1}\;\\[12pt]
\;\dfrac{i\omega}{2\bar{c}_1}\dfrac{1+{M}_1}{{M}_1}\;\\[12pt]
\;\dfrac{i\omega}{\bar{u}_2}\;
\end{bmatrix} A_i
= \mathbf{0}.
\label{eq:matrix_system_trans}
\end{equation}
From the second row of \eqref{eq:matrix_system_trans} 
one obtains \(B_f\) explicitly in terms of the acoustic amplitudes.
After simplification,
\begin{equation}
B_f = \frac{i\omega}{\bar{u}_2^2}\left[ \bar{u}_2\frac{1+{M}_1}{\bar{u}_1} A_i 
- (1+{M}_2)A_t \right].
\label{eq:B_trans}
\end{equation}
Substituting \eqref{eq:B_trans} into the first row leads 
to the shock displacement amplitude 
\begin{equation} 
X_f = \frac{1}{\bar{u}_1-\bar{u}_2} \left[ \frac{\bar{u}_1}{\bar{u}_2} A_t 
- \left( 1 + M_1 - M_2 C\right) A_i \right], 
\label{eq:X_trans} 
\end{equation}
where $C = \bar{c}_2/\bar{c}_1$ is the speed of sound ratio. Finally, 
substituting both $B_f$ and $X_f$ into the third row gives 
the transmission coefficient $\mathcal{T}=A_t/A_i$
\begin{equation}
\mathcal{T} =
\frac{M_2 C}{M_1}
\left[
\frac{C(M_1+1) - M_1 M_2 (\gamma-1)(M_1 - M_2 C)}
{C(M_2+1) - M_2 (\gamma-1)(M_1 - M_2 C)}
\right].
\label{eq:At_to_Ai}
\end{equation}

\bibliographystyle{jfm}
\bibliography{Bibliography}

@article{pirozzoli2013generalized,
  author = {Pirozzoli, S. and Colonius, T.},
  journal = {Journal of Computational Physics},
  pages = {109--126},
  publisher = {Elsevier},
  title = {Generalized characteristic relaxation boundary conditions for unsteady compressible flow simulations},
  volume = {248},
  year = {2013}
}

@article{salvadore2025streams,
  author = {Salvadore, F. and Soldati, G. and Ceci, A. and Rossi, G. and Memmolo, A. and Della Posta, G. and Modesti, D. and Sathyanarayana, S. and Bernardini, M. and Pirozzoli, S.},
  journal = {Computer Physics Communications},
  pages = {109652},
  publisher = {Elsevier},
  title = {{STREAmS}-2.1: {S}upersonic turbulent accelerated {N}avier-{S}tokes solver version 2.1},
  year = {2025}
}

@article{schlatter2010assessment,
  author = {Schlatter, P. and {\"O}rl{\"u}, R.},
  journal = {Journal of Fluid Mechanics},
  pages = {116},
  publisher = {Cambridge University Press},
  title = {Assessment of direct numerical simulation data of turbulent boundary layers},
  volume = {659},
  year = {2010}
}

@article{pirozzoli2011stabilized,
  author = {S. Pirozzoli},
  journal = {Journal of Computational Physics},
  number = {8},
  pages = {2997--3014},
  publisher = {Elsevier},
  title = {Stabilized non-dissipative approximations of Euler equations in generalized curvilinear coordinates},
  volume = {230},
  year = {2011}
}

@online{construct2d,
        title={Construct2D},
        date={2018},
        author={D. Prosser},
        url={https://sourceforge.net/projects/construct2d/},
        year={2018}
}

@article{soldati2024flew,
  author = {Soldati, G. and Ceci, A. and Pirozzoli, S.},
  journal = {Aerotecnica Missili \& Spazio},
  pages = {1--13},
  publisher = {Springer},
  title = {{FLEW}: A {DNS} Solver for Compressible Flows in Generalized Curvilinear Coordinates},
  year = {2024}
}

@article{spalart1993experimental,
  author = {Spalart, P.R. and Watmuff, J.H.},
  journal = {Journal of Fluid Mechanics},
  pages = {337--371},
  publisher = {Cambridge University Press},
  title = {Experimental and numerical study of a turbulent boundary layer with pressure gradients},
  volume = {249},
  year = {1993}
}

@article{lee1990transonic,
  title={Transonic buffet on a supercritical aerofoil},
  author={Lee, B.H.K.},
  journal={The Aeronautical Journal},
  volume={94},
  number={935},
  pages={143--152},
  year={1990},
  publisher={Cambridge University Press}
}

@inproceedings{crouch2009global,
  title={Global structure of buffeting flow on transonic airfoils},
  author={Crouch, J.D. and Garbaruk, A. and Magidov, D. and Jacquin, L.},
  booktitle={IUTAM Symposium on Unsteady Separated Flows and their Control: Proceedings of the IUTAM Symposium `Unsteady Separated Flows and their Control', Corfu, Greece, 18--22 June 2007},
  pages={297--306},
  year={2009},
  organization={Springer}
}

@article{crouch2009origin,
  title={Origin of transonic buffet on aerofoils},
  author={Crouch, J.D. and Garbaruk, A. and Magidov, D. and Travin, A.},
  journal={Journal of Fluid Mechanics},
  volume={628},
  pages={357--369},
  year={2009},
  publisher={Cambridge University Press}
}

@article{crouch2019global, 
  title={Global instability in the onset of transonic-wing buffet}, 
  volume={881}, 
  DOI={10.1017/jfm.2019.748}, 
  journal={Journal of Fluid Mechanics}, 
  author={Crouch, J.D. and Garbaruk, A. and Strelets, M.}, 
  year={2019}, 
  pages={3–22}
}

@article{crouch2024weakly,
  title={Weakly nonlinear behaviour of transonic buffet on airfoils},
  author={Crouch, J.D. and Ahrabi, B.R. and Kamenetskiy, D.S.},
  journal={Journal of Fluid Mechanics},
  volume={999},
  pages={A8},
  year={2024},
  publisher={Cambridge University Press}
}

@article{lee1990oscillatory,
  title={Oscillatory shock motion caused by transonic shock boundary-layer interaction},
  author={Lee, B.H.K.},
  journal={AIAA Journal},
  volume={28},
  number={5},
  pages={942--944},
  year={1990}
}

@article{jacquin2009experimental,
  title={Experimental study of shock oscillation over a transonic supercritical profile},
  author={Jacquin, L. and Molton, P. and Deck, S. and Maury, B. and Soulevant, D.},
  journal={AIAA Journal},
  volume={47},
  number={9},
  pages={1985--1994},
  year={2009}
}

@article{d2021experimental,
  title={Experimental investigation of the transonic buffet cycle on a supercritical airfoil},
  author={D’Aguanno, A. and Schrijer, F.F.J. and van Oudheusden, B.W.},
  journal={Experiments in Fluids},
  volume={62},
  number={10},
  pages={214},
  year={2021},
  publisher={Springer}
}

@techreport{pearcey1962simple,
  author = {Pearcey, H.H. and Holder, D.W.},
  title = {{Simple Methods for the Prediction of Wing Buffeting Resulting from Bubble Type Separation}},
  institution = {{National Physical Laboratory}},
  year = {1962},
  number = {{Aero Rept. 1024}},
  address = {Teddington, UK}
}

@article{sartor2015stability,
  title={Stability, receptivity, and sensitivity analyses of buffeting transonic flow over a profile},
  author={Sartor, F. and Mettot, C. and Sipp, D.},
  journal={AIAA Journal},
  volume={53},
  number={7},
  pages={1980--1993},
  year={2015},
  publisher={American Institute of Aeronautics and Astronautics}
}

@article{sansica2023global,
  title={Global stability analysis of full-aircraft transonic buffet at flight Reynolds numbers},
  author={Sansica, A. and Hashimoto, A.},
  journal={AIAA Journal},
  volume={61},
  number={10},
  pages={4437--4455},
  year={2023},
  publisher={American Institute of Aeronautics and Astronautics}
}

@article{timme2020global,
  title={Global instability of wing shock-buffet onset},
  author={Timme, S.},
  journal={Journal of Fluid Mechanics},
  volume={885},
  pages={A37},
  year={2020},
  publisher={Cambridge University Press}
}

@article{plante2021link,
  title={Link between subsonic stall and transonic buffet on swept and unswept wings: from global stability analysis to nonlinear dynamics},
  author={Plante, F. and Dandois, J. and Beneddine, S. and Laurendeau, E. and Sipp, D.},
  journal={Journal of Fluid Mechanics},
  volume={908},
  pages={A16},
  year={2021},
  publisher={Cambridge University Press}
}

@article{paladini2019various,
  title={Various approaches to determine active regions in an unstable global mode: application to transonic buffet},
  author={Paladini, E. and Marquet, O. and Sipp, D. and Robinet, J.C. and Dandois, J.},
  journal={Journal of Fluid Mechanics},
  volume={881},
  pages={617--647},
  year={2019},
  publisher={Cambridge University Press}
}

@article{lee2001self,
  title={Self-sustained shock oscillations on airfoils at transonic speeds},
  author={Lee, B.H.K.},
  journal={Progress in Aerospace Sciences},
  volume={37},
  number={2},
  pages={147--196},
  year={2001},
  publisher={Elsevier}
}

@techreport{pearcey1958method,
  author       = {Pearcey, H.H.},
  title        = {A Method for the Prediction of the Onset of Buffeting and Other Separation Effects from Wind Tunnel Tests on Rigid Models},
  institution  = {Advisory Group for Aerospace Research and Development (AGARD)},
  type         = {Technical Report},
  number       = {TR-223},
  address      = {Neuilly-sur-Seine, France},
  year         = {1958},
}

@inproceedings{stanewsky1990experimental,
  author       = {Stanewsky, E. and Basler, D.},
  title        = {Experimental Investigation of Buffet Onset and Penetration on a Supercritical Airfoil at Transonic Speeds},
  booktitle    = {Aircraft Dynamic Loads Due to Flow Separation},
  series       = {Advisory Group for Aerospace Research and Development (AGARD) Conference Proceedings},
  number       = {CP-483},
  organization = {Advisory Group for Aerospace Research and Development (AGARD)},
  address      = {Sorrento, Italy},
  pages        = {4.1--4.11},
  year         = {1990},
}

@article{accorinti2022experimental,
  title={Experimental investigation of transonic shock buffet on an {OAT15A} profile},
  author={Accorinti, A. and Baur, T. and Scharnowski, S. and K{\"a}hler, C.J.},
  journal={AIAA Journal},
  volume={60},
  number={11},
  pages={6289--6300},
  year={2022},
  publisher={American Institute of Aeronautics and Astronautics}
}

@techreport{mabey1981oscillatory,
  author       = {Mabey, D.G.},
  title        = {Oscillatory Flows from Shock-Induced Separations on Biconvex Aerofoils of Varying Thickness in Ventilated Wind Tunnels},
  institution  = {Advisory Group for Aerospace Research and Development (AGARD)},
  type         = {Technical Report},
  number       = {TR-296},
  address      = {Aix-en-Provence, France},
  year         = {1981},
}

@article{iovnovich2012reynolds,
  title={Reynolds-averaged {N}avier-{S}tokes study of the shock-buffet instability mechanism},
  author={Iovnovich, M. and Raveh, D.E.},
  journal={AIAA Journal},
  volume={50},
  number={4},
  pages={880--890},
  year={2012}
}

@article{jackson1987finite,
  title={A finite-element study of the onset of vortex shedding in flow past variously shaped bodies},
  author={Jackson, C.P.},
  journal={Journal of Fluid Mechanics},
  volume={182},
  pages={23--45},
  year={1987},
  publisher={Cambridge University Press}
}

@article{raghunathan1998transonic,
  title={Transonic shock oscillations on {NACA}0012 aerofoil},
  author={Raghunathan, S. and Mitchell, R.D. and Gillan, M.A.},
  journal={Shock Waves},
  volume={8},
  number={4},
  pages={191--202},
  year={1998},
  publisher={Springer}
}

@article{goncalves2004turbulence,
  title={Turbulence model and numerical scheme assessment for buffet computations},
  author={Goncalves, E. and Houdeville, R.},
  journal={International Journal for Numerical Methods in Fluids},
  volume={46},
  number={11},
  pages={1127--1152},
  year={2004},
  publisher={Wiley Online Library}
}

@article{sansica2022system,
  title={System identification of two-dimensional transonic buffet},
  author={Sansica, A. and Loiseau, J.C. and Kanamori, M. and Hashimoto, A. and Robinet, J.C.},
  journal={AIAA Journal},
  volume={60},
  number={5},
  pages={3090--3106},
  year={2022},
  publisher={American Institute of Aeronautics and Astronautics}
}

@article{xiao2006numerical,
  title={Numerical study of transonic buffet on a supercritical airfoil},
  author={Xiao, Q. and Tsai, H.M. and Liu, F.},
  journal={AIAA Journal},
  volume={44},
  number={3},
  pages={620--628},
  year={2006}
}

@article{deck2005numerical,
  title={Numerical simulation of transonic buffet over a supercritical airfoil},
  author={Deck, S.},
  journal={AIAA Journal},
  volume={43},
  number={7},
  pages={1556--1566},
  year={2005}
}

@article{iwatani2023identifying,
  title={Identifying the self-sustaining mechanisms of transonic airfoil buffet with resolvent analysis},
  author={Iwatani, Y. and Asada, H. and Yeh, C.A. and Taira, K. and Kawai, S.},
  journal={AIAA Journal},
  volume={61},
  number={6},
  pages={2400--2411},
  year={2023},
  publisher={American Institute of Aeronautics and Astronautics}
}

@article{hartmann2013interaction,
  title={On the interaction of shock waves and sound waves in transonic buffet flow},
  author={Hartmann, A. and Feldhusen, A. and Schr{\"o}der, W.},
  journal={Physics of Fluids},
  volume={25},
  number={2},
  year={2013},
  publisher={AIP Publishing}
}

@article{memmolo2018scrutiny,
  title={Scrutiny of buffet mechanisms in transonic flow},
  author={Memmolo, A. and Bernardini, M. and Pirozzoli, S.},
  journal={International Journal of Numerical Methods for Heat \& Fluid Flow},
  volume={28},
  number={5},
  pages={1031--1046},
  year={2018},
  publisher={Emerald Publishing Limited}
}

@article{zauner2019direct,
  title={Direct numerical simulations of transonic flow around an airfoil at moderate Reynolds numbers},
  author={Zauner, M. and De Tullio, N. and Sandham, N.D.},
  journal={AIAA Journal},
  volume={57},
  number={2},
  pages={597--607},
  year={2019},
  publisher={American Institute of Aeronautics and Astronautics}
}

@article{placek2016wind,
  title={Wind Tunnel Tests of laminar-turbulent transition influence on basic aerodynamic characteristics of laminar airfoil in transonic flow regime},
  author={Placek, R. and Miller, M.},
  journal={Inst. of Aviation TR D},
  volume={5},
  year={2016}
}

@article{szubert2016numerical,
  title={Numerical study of the turbulent transonic interaction and transition location effect involving optimisation around a supercritical aerofoil},
  author={Szubert, D. and Asproulias, I. and Grossi, F. and Duvigneau, R. and Hoarau, Y. and Braza, M.},
  journal={European Journal of Mechanics-B/Fluids},
  volume={55},
  pages={380--393},
  year={2016},
  publisher={Elsevier}
}

@article{placek2018flow,
  title={The flow separation development analysis in subsonic and transonic flow regime of the laminar airfoil},
  author={Placek, R. and Rucha{\l}a, P.},
  journal={Transportation Research Procedia},
  volume={29},
  pages={323--329},
  year={2018},
  publisher={Elsevier}
}

@article{quadrio2022drag,
  title={Drag reduction on a transonic airfoil},
  author={Quadrio, M. and Chiarini, A. and Banchetti, J. and Gatti, D. and Memmolo, A. and Pirozzoli, S.},
  journal={Journal of Fluid Mechanics},
  volume={942},
  pages={R2},
  year={2022},
  publisher={Cambridge University Press}
}

@article{berizzi2025aerodynamic,
  title={Aerodynamic performance of a transonic airfoil with spanwise forcing},
  author={Berizzi, N. and Gatti, D. and Soldati, G. and Pirozzoli, S. and Quadrio, M.},
  journal={Journal of Fluid Mechanics},
  volume={1010},
  pages={A18},
  year={2025},
  publisher={Cambridge University Press}
}

@article{hosseini2016direct,
  title={Direct numerical simulation of the flow around a wing section at moderate {R}eynolds number},
  author={Hosseini, S.M. and Vinuesa, R. and Schlatter, P. and Hanifi, A. and Henningson, D.S.},
  journal={International Journal of Heat and Fluid Flow},
  volume={61},
  pages={117--128},
  year={2016},
  publisher={Elsevier}
}

@article{schlatter2012turbulent,
  title={Turbulent boundary layers at moderate {R}eynolds numbers: inflow length and tripping effects},
  author={Schlatter, P. and {\"O}rl{\"u}, R.},
  journal={Journal of Fluid Mechanics},
  volume={710},
  pages={5--34},
  year={2012},
  publisher={Cambridge University Press}
}

@article{pirozzoli2021natural,
  title={Natural grid stretching for {DNS} of wall-bounded flows},
  author={Pirozzoli, S. and Orlandi, P.},
  journal={Journal of Computational Physics},
  volume={439},
  pages={110408},
  year={2021},
  publisher={Elsevier}
}

@article{pope2001turbulent,
  title={Turbulent flows},
  author={Pope, S.B.},
  journal={Measurement Science and Technology},
  volume={12},
  number={11},
  pages={2020--2021},
  year={2001}
}

@article{moise2023transonic,
  title={Transonic buffet characteristics under conditions of free and forced transition},
  author={Moise, P. and Zauner, M. and Sandham, N. D and Timme, S. and He, W.},
  journal={AIAA Journal},
  volume={61},
  number={3},
  pages={1061--1076},
  year={2023},
  publisher={American Institute of Aeronautics and Astronautics}
}

@article{moise2024connecting,
  title={Connecting transonic buffet with incompressible low-frequency oscillations on aerofoils},
  author={Moise, P. and Zauner, M. and Sandham, N.D.},
  journal={Journal of Fluid Mechanics},
  volume={981},
  pages={A23},
  year={2024},
  publisher={Cambridge University Press}
}

@article{moise2022large,
  title={Large-eddy simulations and modal reconstruction of laminar transonic buffet},
  author={Moise, P. and Zauner, M. and Sandham, N.D.},
  journal={Journal of Fluid Mechanics},
  volume={944},
  pages={A16},
  year={2022},
  publisher={Cambridge University Press}
}

@article{towne2018spectral,
  title={Spectral proper orthogonal decomposition and its relationship to dynamic mode decomposition and resolvent analysis},
  author={Towne, A. and Schmidt, O.T. and Colonius, T.},
  journal={Journal of Fluid Mechanics},
  volume={847},
  pages={821--867},
  year={2018},
  publisher={Cambridge University Press}
}

@article{schmidt2020guide,
  title={Guide to spectral proper orthogonal decomposition},
  author={Schmidt, O.T. and Colonius, T.},
  journal={AIAA Journal},
  volume={58},
  number={3},
  pages={1023--1033},
  year={2020},
  publisher={American Institute of Aeronautics and Astronautics}
}

@book{lumley1970stochastic,
  title={Stochastic tools in turbulence},
  author={Lumley, J.L.},
  year={1970},
  publisher={Elsevier}
}

@article{fukushima2018wall,
  title={Wall-modeled large-eddy simulation of transonic airfoil buffet at high {R}eynolds number},
  author={Fukushima, Y. and Kawai, S.},
  journal={AIAA Journal},
  volume={56},
  number={6},
  pages={2372--2388},
  year={2018},
  publisher={American Institute of Aeronautics and Astronautics}
}

@book{pierce1981acoustics,
  title={Acoustics: an introduction to its physical principles and applications},
  author={Pierce, A.D.},
  year={1981},
  publisher={Springer}
}

@article{mahesh1995interaction,
  title={The interaction of an isotropic field of acoustic waves with a shock wave},
  author={Mahesh, K. and Lee, S. and Lele, S.K. and Moin, P.},
  journal={Journal of Fluid Mechanics},
  volume={300},
  pages={383--407},
  year={1995},
  publisher={Cambridge University Press}
}

@article{robinet2001critical,
  title={Critical interaction of a shock wave with an acoustic wave},
  author={Robinet, J.C. and Casalis, G.},
  journal={Physics of Fluids},
  volume={13},
  number={4},
  pages={1047--1059},
  year={2001},
  publisher={American Institute of Physics}
}

@techreport{ribner1954convection,
  title={Convection of a pattern of vorticity through a shock wave},
  author={Ribner, H.S.},
  year={1954},
  institution={National Advisory Committee for Aeronautics},
  number={NACA-TN-2864},
}

@techreport{moore1953unsteady,
  title={Unsteady oblique interaction of a shock wave with a plane disturbance},
  author={Moore, F.K.},
  year={1953},
  institution={National Advisory Committee for Aeronautics},
  number={NACA-TN-2879},
}

@article{mckenzie1968interaction,
  title={Interaction of linear waves with oblique shock waves},
  author={McKenzie, J.F. and Westphal, K.O.},
  journal={Physics of Fluids},
  volume={11},
  number={11},
  pages={2350--2362},
  year={1968},
  publisher={American Institute of Physics}
}

@inproceedings{garnier2010large,
  title={Large-eddy simulation of transonic buffet over a supercritical airfoil},
  author={Garnier, E. and Deck, S.},
  booktitle={Direct and Large-Eddy Simulation VII},
  series={ERCOFTAC Series},
  volume={13},
  pages={549--554},
  year={2010},
  publisher={Springer},
  address={Heidelberg}
}

@article{schauerte2023experimental,
  title={Experimental analysis of transonic buffet conditions on a two-dimensional supercritical airfoil},
  author={Schauerte, C.J. and Schreyer, A.-M.},
  journal={AIAA Journal},
  volume={61},
  number={8},
  pages={3432--3448},
  year={2023},
  publisher={American Institute of Aeronautics and Astronautics}
}

@inproceedings{nitzsche2019fluid,
  title={Fluid-mode flutter in plane transonic flows},
  author={Nitzsche, J. and Ringel, L. M. and Kaiser, C. and Hennings, H.},
  booktitle={International Forum on Aeroelasticity and Structural Dynamics 2019, IFASD 2019},
  year={2019}
}

@article{nitzsche2009numerical,
  title={A numerical study on aerodynamic resonance in transonic separated flow},
  author={Nitzsche, J.},
  year={2009}
}

@article{zauner2023co,
  title={On the co-existence of transonic buffet and separation-bubble modes for the {OALT25} laminar-flow wing section},
  author={Zauner, M. and Moise, P. and Sandham, N.D.},
  journal={Flow, Turbulence and Combustion},
  volume={110},
  number={4},
  pages={1023--1057},
  year={2023},
  publisher={Springer}
}

@article{molton2013control,
  title={Control of buffet phenomenon on a transonic swept wing},
  author={Molton, P. and Dandois, J. and Lepage, A. and Brunet, V. and Bur, R.},
  journal={AIAA Journal},
  volume={51},
  number={4},
  pages={761--772},
  year={2013},
  publisher={American Institute of Aeronautics and Astronautics}
}

@article{dandois2016experimental,
  title={Experimental study of transonic buffet phenomenon on a {3D} swept wing},
  author={Dandois, J.},
  journal={Physics of Fluids},
  volume={28},
  number={1},
  year={2016},
  publisher={AIP Publishing}
}

@article{iovnovich2015numerical,
  title={Numerical study of shock buffet on three-dimensional wings},
  author={Iovnovich, M. and Raveh, D.E.},
  journal={AIAA Journal},
  volume={53},
  number={2},
  pages={449--463},
  year={2015},
  publisher={American Institute of Aeronautics and Astronautics}
}

@article{paladini2019transonic,
  title = {Transonic buffet instability: from two-dimensional airfoils to three-dimensional swept wings},
  author = {Paladini, E. and Beneddine, S. and Dandois, J. and Sipp, D. and Robinet, J.-C.},
  journal = {Phys. Rev. Fluids},
  volume = {4},
  issue = {10},
  pages = {103906},
  numpages = {18},
  year = {2019},
  month = {Oct},
  publisher = {American Physical Society},
  doi = {10.1103/PhysRevFluids.4.103906},
  url = {https://link.aps.org/doi/10.1103/PhysRevFluids.4.103906}
}

@article{he2021triglobal,
  title={Triglobal infinite-wing shock-buffet study},
  author={He, W. and Timme, S.},
  journal={Journal of Fluid Mechanics},
  volume={925},
  pages={A27},
  year={2021},
  publisher={Cambridge University Press}
}

@inproceedings{davidson2016influence,
  title={Influence of transition on the flow downstream of normal shock wave--boundary layer interactions},
  author={Davidson, T.S. and Babinsky, H.},
  booktitle={54th AIAA Aerospace Sciences Meeting},
  pages={0044},
  year={2016}
}

@article{lusher2025implicit,
  title={Implicit large eddy simulations of three-dimensional turbulent transonic buffet on wide-span infinite wings},
  author={Lusher, D.J. and Sansica, A. and Hashimoto, A.},
  journal={Journal of Fluid Mechanics},
  volume={1007},
  pages={A26},
  year={2025},
  publisher={Cambridge University Press}
}

@article{zauner2020wide,
  title = {Wide domain simulations of flow over an unswept laminar wing section undergoing transonic buffet},
  author = {Zauner, M. and Sandham, N.D.},
  journal = {Phys. Rev. Fluids},
  volume = {5},
  issue = {8},
  pages = {083903},
  numpages = {27},
  year = {2020},
  month = {Aug},
  publisher = {American Physical Society},
  doi = {10.1103/PhysRevFluids.5.083903},
  url = {https://link.aps.org/doi/10.1103/PhysRevFluids.5.083903}
}

@article{lusher2024effect,
  title={Effect of tripping and domain width on transonic buffet on periodic {NASA-CRM} airfoils},
  author={Lusher, D. J and Sansica, A. and Hashimoto, A.},
  journal={AIAA Journal},
  volume={62},
  number={11},
  pages={4411--4430},
  year={2024},
  publisher={American Institute of Aeronautics and Astronautics}
}

@book{babinsky2011shock,
  title={Shock wave-boundary-layer interactions},
  author={Babinsky, H. and Harvey, J.K.},
  volume={32},
  year={2011},
  publisher={Cambridge University Press}
}

@article{dandois2018large, 
  title={Large-eddy simulation of laminar transonic buffet}, 
  volume={850}, 
  DOI={10.1017/jfm.2018.470}, 
  journal={Journal of Fluid Mechanics}, 
  author={Dandois, J. and Mary, I. and Brion, V.}, 
  year={2018}, 
  pages={156–178}}

@article{schauerte2025influences,
  title={Influences in Experimental Studies on the Characteristics of Transonic Shock Buffet},
  author={Schauerte, C.J. and Schreyer, A.-M.},
  journal={AIAA Journal},
  volume={63},
  number={4},
  pages={1278--1294},
  year={2025},
  publisher={American Institute of Aeronautics and Astronautics}
}

@article{davidson2018influence,
  title={Influence of boundary-layer state on development downstream of normal shock interactions},
  author={Davidson, T.S.C. and Babinsky, H.},
  journal={AIAA Journal},
  volume={56},
  number={6},
  pages={2298--2307},
  year={2018},
  publisher={American Institute of Aeronautics and Astronautics}
}

@article{preston1958the, 
  title={The minimum {R}eynolds number for a turbulent boundary layer and the selection of a transition device}, 
  volume={3}, 
  number={4}, 
  journal={Journal of Fluid Mechanics}, 
  author={Preston, J.H.}, 
  year={1958}, 
  pages={373–384}
}

@article{spalart1988direct,
  title={Direct simulation of a turbulent boundary layer up to ${R}_\theta=1410$.},
  author={Spalart, P.R.},
  journal={Journal of Fluid Mechanics},
  volume={187},
  pages={61--98},
  year={1988},
  publisher={Cambridge University Press}
}

@inproceedings{bandyopadhyay1987resonant,
  title={Resonant flow in a row of small transverse cavities submerged in a turbulent boundary layer},
  author={Bandyopadhyay, P.},
  booktitle={19th AIAA, Fluid Dynamics, Plasma Dynamics, and Lasers Conference},
  pages={1235},
  year={1987}
}

@article{wu2009direct,
  title={Direct numerical simulation of turbulence in a nominally zero-pressure-gradient flat-plate boundary layer},
  author={Wu, X. and Moin, P.},
  journal={Journal of Fluid Mechanics},
  volume={630},
  pages={5--41},
  year={2009},
  publisher={Cambridge University Press}
}

@article{simpson1989turbulent,
  title={Turbulent boundary-layer separation},
  author={Simpson, R.L.},
  journal={Annual Review of Fluid Mechanics},
  volume={21},
  number={1},
  pages={205--232},
  year={1989},
  publisher={Annual Reviews 4139 El Camino Way, PO Box 10139, Palo Alto, CA 94303-0139, USA}
}

@inproceedings{sandborn1974boundary,
  title={Boundary Layer Separation and Reattachment},
  author={Sandborn, V.A.},
  booktitle={Fluid Mechanics, Acoustics, and Design of Turbomachinery}, 
  volume={304},
  pages={279},
  year={1974},
  organization={Scientific and Technical Information Office, National Aeronautics and Space Administration}
}

@article{sansica2016instability,
  title={Instability and low-frequency unsteadiness in a shock-induced laminar separation bubble},
  author={Sansica, A. and Sandham, N.D. and Hu, Z.},
  journal={Journal of Fluid Mechanics},
  volume={798},
  pages={5--26},
  year={2016},
  publisher={Cambridge University Press}
}

@article{lusher2020shock,
  title={Shock-wave/boundary-layer interactions in transitional rectangular duct flows},
  author={Lusher, D.J. and Sandham, N.D.},
  journal={Flow, Turbulence and Combustion},
  volume={105},
  number={2},
  pages={649--670},
  year={2020},
  publisher={Springer}
}

@article{katzer1989lengthscales,
  title={On the lengthscales of laminar shock/boundary-layer interaction},
  author={Katzer, E.},
  journal={Journal of Fluid Mechanics},
  volume={206},
  pages={477--496},
  year={1989},
  publisher={Cambridge University Press}
}

@article{lusher2020effect,
  title={The effect of flow confinement on laminar shock-wave/boundary-layer interactions},
  author={Lusher, D.J. and Sandham, N.D.},
  journal={Journal of Fluid Mechanics},
  volume={897},
  pages={A18},
  year={2020},
  publisher={Cambridge University Press}
}

@book{young1989boundary,
  title={Boundary Layers},
  author={Young, A.D.},
  year={1989},
  publisher={American Institute of Aeronautics and Astronautics},
  series={AIAA Education Series},
  address={Washington, D.C.},
  number = {} 
}

@article{holder1964transonic,
  title={The transonic flow past two-dimensional aerofoils},
  author={Holder, D.W.},
  journal={The Aeronautical Journal},
  volume={68},
  number={644},
  pages={501--516},
  year={1964},
  publisher={Cambridge University Press}
}

@article{vinuesa2018turbulent,
  title={Turbulent boundary layers around wing sections up to ${R}e_c$ = 1,000,000},
  author={Vinuesa, R. and Negi, P.S. and Atzori, M. and Hanifi, A. and Henningson, D.S. and Schlatter, P.},
  journal={International Journal of Heat and Fluid Flow},
  volume={72},
  pages={86--99},
  year={2018},
  publisher={Elsevier}
}

@article{kojima2020resolvent,
  title={Resolvent analysis on the origin of two-dimensional transonic buffet},
  author={Kojima, Y. and Yeh, C.-. and Taira, K. and Kameda, M.},
  journal={Journal of Fluid Mechanics},
  volume={885},
  pages={R1},
  year={2020},
  publisher={Cambridge University Press}
}

@article{poplingher2019modal,
  title={Modal analysis of transonic shock buffet on {2D} airfoil},
  author={Poplingher, L. and Raveh, D. E and Dowell, E.H.},
  journal={AIAA Journal},
  volume={57},
  number={7},
  pages={2851--2866},
  year={2019},
  publisher={American Institute of Aeronautics and Astronautics}
}

@inproceedings{fukushima2019self,
  title={Self-sustained shock-wave oscillation mechanisms of transonic airfoil buffet},
  author={Fukushima, Y. and Kawai, S.},
  booktitle={AIAA Scitech 2019 Forum},
  pages={1846},
  year={2019}
}

@article{iorio2014direct,
  title={Direct and adjoint global stability analysis of turbulent transonic flows over a {NACA0012} profile},
  author={Iorio, M.C. and Gonzalez, L.M. and Ferrer, E.},
  journal={International Journal for Numerical Methods in Fluids},
  volume={76},
  number={3},
  pages={147--168},
  year={2014},
  publisher={Wiley Online Library}
}

@inproceedings{iwatani2022pod,
  title={{POD}, {DMD}, and resolvent analysis of transonic airfoil buffet},
  author={Iwatani, Y. and Asada, H. and Kawai, S.},
  booktitle={AIAA SCITECH 2022 Forum},
  pages={0461},
  year={2022}
}

@article{monkewitz1990local,
  title={Local and global instabilities in spatially developing flows},
  author={Monkewitz, P.},
  journal={Annual Review of Fluid Mechanics},
  year={1990}
}

@article{giannetti2007structural,
  title={Structural sensitivity of the first instability of the cylinder wake},
  author={Giannetti, F. and Luchini, P.},
  journal={Journal of Fluid Mechanics},
  volume={581},
  pages={167--197},
  year={2007},
  publisher={Cambridge University Press}
}

@article{lee1994role,
  title={Role of Kutta waves on oscillatory shock motion on an airfoil},
  author={Lee, B.H.K. and Murty, H. and Jiang, H.},
  journal={AIAA Journal},
  volume={32},
  number={4},
  pages={789--796},
  year={1994}
}

@article{gao2025optimal,
  title={Optimal position and angle of jet control for transonic shock buffet with resolvent analysis},
  author={Gao, C. and Ma, R. and Ren, K. and Yuan, H. and Zhang, W.},
  journal={AIAA Journal},
  volume={63},
  number={8},
  pages={3415--3433},
  year={2025},
  publisher={American Institute of Aeronautics and Astronautics}
}

@inproceedings{houtman2022resolvent,
  title={Resolvent analysis of large aircraft wings in edge-of-the-envelope transonic flow},
  author={Houtman, J. and Timme, S. and Sharma, A.},
  booktitle={AIAA SciTech 2022 Forum},
  pages={1329},
  year={2022}
}

@article{candel1977numerical,
  title={Numerical solution of conservation equations arising in linear wave theory: application to aeroacoustics},
  author={Candel, S.M.},
  journal={Journal of Fluid Mechanics},
  volume={83},
  number={3},
  pages={465--493},
  year={1977},
  publisher={Cambridge University Press}
}

@article{curle1955the,
    author = {Curle, N.},
    title = {The influence of solid boundaries upon aerodynamic sound},
    journal = {Proceedings of the Royal Society of London. A. Mathematical and Physical Sciences},
    volume = {231},
    number = {1187},
    pages = {505-514},
    year = {1955},
    month = {09},
    issn = {0080-4630},
}

@article{howe1978review,
  title={A review of the theory of trailing edge noise},
  author={Howe, M.S.},
  journal={Journal of Sound and Vibration},
  volume={61},
  number={3},
  pages={437--465},
  year={1978},
  publisher={Elsevier}
}

@techreport{rossiter1964wind,
  author      = {Rossiter, J.E.},
  title       = {Wind-tunnel experiments on the flow over rectangular cavities at subsonic and transonic speeds},
  institution = {Aeronautical Research Council},
  year        = {1964},
  type        = {R\&M},
  number      = {3438}
}

@article{lighthill1952on,
    author = {Lighthill, M.J.},
    title = {On sound generated aerodynamically I. General theory},
    journal = {Proceedings of the Royal Society of London. A. Mathematical and Physical Sciences},
    volume = {211},
    number = {1107},
    pages = {564-587},
    year = {1952},
    month = {03},
    issn = {0080-4630},
}

@article{amiet1976noise,
  title={Noise due to turbulent flow past a trailing edge},
  author={Amiet, R.K.},
  journal={Journal of Sound and Vibration},
  volume={47},
  number={3},
  pages={387--393},
  year={1976},
  publisher={Elsevier}
}

@article{williams1970aerodynamic,
  title={Aerodynamic sound generation by turbulent flow in the vicinity of a scattering half plane},
  author={Williams, J.E. Ffowcs and Hall, L.H.},
  journal={Journal of Fluid Mechanics},
  volume={40},
  number={4},
  pages={657--670},
  year={1970},
  publisher={Cambridge University Press}
}

@article{crighton1975basic,
  title={Basic principles of aerodynamic noise generation},
  author={Crighton, D.G.},
  journal={Progress in Aerospace Sciences},
  volume={16},
  number={1},
  pages={31--96},
  year={1975},
  publisher={Elsevier}
}

@article{lighthill1954sound,
  title={On sound generated aerodynamically. II. Turbulence as a source of sound},
  author={Lighthill, M.J.},
  journal={Proceedings of the Royal Society of London. Series A, Mathematical and Physical Sciences},
  pages={1--32},
  year={1954},
  publisher={JSTOR}
}

@article{arbey1983noise,
  title={Noise generated by airfoil profiles placed in a uniform laminar flow},
  author={Arbey, H. and Bataille, J.},
  journal={Journal of Fluid Mechanics},
  volume={134},
  pages={33--47},
  year={1983},
  publisher={Cambridge University Press}
}

@article{tam1974discrete,
  title={Discrete tones of isolated airfoils},
  author={Tam, C.K.W.},
  journal={The Journal of the Acoustical Society of America},
  volume={55},
  number={6},
  pages={1173--1177},
  year={1974},
  publisher={Acoustical Society of America}
}

@article{desquesnes2007numerical,
  title={Numerical investigation of the tone noise mechanism over laminar airfoils},
  author={Desquesnes, G. and Terracol, M. and Sagaut, P.},
  journal={Journal of Fluid Mechanics},
  volume={591},
  pages={155--182},
  year={2007},
  publisher={Cambridge University Press}
}

@inproceedings{mohan1991periodic,
  title={Periodic flows on rigid aerofoils at transonic speeds},
  author={Mohan, S.},
  booktitle={29th Aerospace Sciences Meeting},
  pages={598},
  year={1991}
}

@techreport{roos1977measurements,
  author      = {Roos, F.W. and Riddle, D.W.},
  title       = {Measurements of surface-pressure and wake-flow fluctuations in the flow field of a Whitcomb supercritical airfoil},
  institution = {National Aeronautics and Space Administration},
  type        = {NASA Technical Note},
  number      = {D-8443},
  year        = {1977}
}

@article{roos1980some,
  title={Some features of the unsteady pressure field in transonic airfoil buffeting},
  author={Roos, F.W.},
  journal={Journal of Aircraft},
  volume={17},
  number={11},
  pages={781--788},
  year={1980}
}
\end{document}